\documentclass{aa}  

\usepackage{graphicx}
\usepackage{xcolor,soul}
\usepackage{svg}
\usepackage{longtable}
\usepackage{array}
\usepackage{lscape}
\usepackage{adjustbox}
\usepackage[toc,page]{appendix}
\usepackage{txfonts}
\usepackage[breaklinks, colorlinks, citecolor=blue, linkcolor=blue]{hyperref}
\usepackage{amsmath,siunitx}

\providecommand{\mst}{\ensuremath{\,{\rm M_\odot}}}
\providecommand{\rst}{\ensuremath{\,{\rm R_\odot}}}

\begin{document}

   \title{Search for planets around stars with wide brown dwarfs}

   \subtitle{}

   \author{J.\ \v{S}ubjak\inst{1,2,3}
          \and
          N.\ Lodieu\inst{4,5}
          \and
          P. Kab\'{a}th\inst{1}
          \and
          H.\,M.\,J.\,Boffin\inst{3}
          \and
 G.\ Nowak\inst{4,5}
\and
 F.\ Grundahl\inst{6}
\and
 V.\ J.\ S.\ B\'ejar\inst{4,5}
\and
\newline
M.\ R.\ Zapatero Osorio\inst{7}
\and
V.\ Antoci\inst{6,8}
          }


   \institute{Astronomical Institute, Czech Academy of Sciences, Fri{\v c}ova 298, 251 65, Ond\v{r}ejov, Czech Republic \\
   \email{jan.subjak@asu.cas.cz}
         \and
         Astronomical Institute of Charles University, V Hole\v{s}ovi\v{c}k\'ach 2, 180 00, Prague, Czech Republic
         \and
         ESO, Karl-Schwarzschild-Stra{\ss}e 2, 85748 Garching bei M\"unchen, Germany
         \and
         Instituto de Astrof\'isica de Canarias (IAC), Calle V\'ia L\'actea s/n, E-38205 La Laguna, Tenerife, Spain
         \and
         Departamento de Astrof\'isica, Universidad de La Laguna (ULL), E-38206 La Laguna, Tenerife, Spain
        \and
        Stellar Astrophysics Centre, Department of Physics and Astronomy, Aarhus University, Ny Munkegade 120, DK-8000 Aarhus C, Denmark         
        \and
        Centro de Astrobiolog\'ia (CSIC-INTA), Ctra. Ajalvir km
        \and
        DTU Space, National Space Institute, Technical University of Denmark, Elektrovej 328, DK-2800 Kgs. Lyngby, Denmark
        }

   \date{\today{}; \today{}}

 
  \abstract
   {}
   {The project aims to understand better the role of wide brown dwarf companions on planetary systems.}
   {We obtained high-resolution spectra of six bright stars with co-moving wide substellar companions with the SONG, CARMENES, and STELLA high-resolution spectrographs. We used these spectra to derive radial velocities together with a complete set of stellar physical parameters. We then investigated radial velocities signals and discussed the fraction of planets in such systems. We also re-analyzed the ages of our targets, which were used to derive the physical parameters of wide brown dwarf companions. Finally, a compilation of systems with known planets from the literature is considered along with our sample to search for possible peculiarities in their parameter distributions.}
   {Based on the derived ages of six observed systems, we re-computed the masses of the wide companions, confirming their substellar nature. We confirmed planets in the HD\,3651 and HIP\,70849 systems and found a new planetary candidate in the HD\,46588 system. In our survey, which is sensitive mostly to Neptune-mass planets at short periods of a few days and Saturn-mass planets at longer periods of hundreds of days, we derived a frequency of planets orbiting stars with wide brown dwarf companions below 70\% with the uncertainties included. Comparing the parameter distributions of our sample with single stars, we observe the enhancement of planets with short periods below six days in systems with a wide stellar companion. Finally, planets in systems with wide BD companions follow their own eccentricity distribution with a maximum at $\sim0.65$ and have periods larger than 40 days, masses larger than $0.1\,M_J$, and eccentricities larger than 0.4\@.
   }
   {}


   \maketitle
%

\section{Introduction}



Thousands of brown dwarfs (BDs) have been discovered so far; many come from all-sky infrared surveys \citep[e.g.,][]{Martin99,Tsvetanov00,Cruz07,Lodieu07,Kirkpatrick11,Lodieu12,Deacon14,Rosell19}, such as the Two-Micron All-Sky Survey \citep[2MASS;][]{Skrutskie06}, the Wide-field Infrared Survey Explorer \citep[WISE;][]{Wright10}, the Deep Near-Infrared Survey of the Southern Sky \citep[DENIS;][]{Epchtein97}, the UKIRT Infrared Deep Sky Survey \citep[UKIDSS;][]{Lawrence07}, and the VISTA Hemisphere Survey \citep[VHS;][]{McMahon13}. Additional discoveries come from large-scale optical surveys, such as the Dark Energy Survey \citep[DES;][]{Abbott18}, the Sloan Digital Sky Survey \citep[SDSS;][]{York00}, and the Panoramic Survey Telescope and Rapid Response System \citep[Pan-STARRS;][]{Chambers16}.

Most of the discovered BDs are free-floating. Only a small fraction of them are confirmed as wide co-moving companions to stars \citep{Deacon14}. However, the recent study by \cite{Ponte20} provides almost 300 new candidate co-moving BDs using data from the Dark Energy Survey DR1 and $Gaia$ DR2 \citep{Gaia18}. \cite{Ponte20} observed a wide binary fraction of BDs about 2--4\% over most of the spectral types (L0--T8.5), more significantly for L dwarf companions with better statistics. We should also mention the rising group of transiting BDs \citep[e.g.,][]{Persson19, Subjak20, Carmichael20, Carmichael21, Palle21} discovered lately mainly thanks to the space missions such as the CoRoT mission \citep{Auvergne09}, the Kepler/K2 mission \citep{Borucki10}, and currently the TESS mission \citep{Ricker15}. The updated list of known transiting BDs can be found in \cite{Subjak20}. 
BDs can be commonly found also in binaries. Tens of them are listed in \cite{Liu2010}. Even two unique young systems with BDs in an eclipsing binary were reported by \citet{Stassun06} and \citet{Triaud20}. Another unique structure has systems of two BDs orbiting a primary star, such as HIP\,73990 \citep{Hinkley15} and $\upsilon$\,Oph \citep{Quirrenbach19}. We can hence find BDs in systems with different structures, which create opportunities to study different kinds of processes and interactions.


From this point of view are particularly interesting systems with wide BD companions around planet-host stars. To date, only nine such systems have been reported. If the mass of a companion is close to the stellar/substellar boundary, we include only a companion confirmed as a brown dwarf based on spectroscopic observations. The confirmed systems are: HD\,3651 \citep[e.g.,][]{Fischer03,Mugrauer06,Luhman07,Burgasser07,Brewer20}, HIP\,70849 \citep{Segransan11,Lodieu14}, HD\,168443 \citep{Marcy01,Pilyavsky11}, HD\,65216 \citep{Mayor04,Mugrauer07}, HD\,89744 \citep{Korzennik00,Mugrauer04}, HD\,4113 \citep{Tamuz08,Ednie18,Cheetham18}, GJ\,229 \citep{Geissler08,Tuomi14,Feng20}, HD\,41004 \citep{Santos02,Zucker03,Zucker04}, and $\epsilon$\,Indi \citep{Scholz03,Mccaughrean04,Feng19}. However, HD\,65216, HD\,41004 and $\epsilon$\,Indi have different architectures (see Appendix \ref{sec:family}).
 
Planets can be significantly affected by wide BD companions through different processes, such as gravitational instability, accretion, velocities of colliding planetesimals, dissipation, or Lidov-Kozai effects \citep[e.g.,][]{Boss06, Nelson03, Moriwaki04}. If we look at multiple stellar systems with planets, possible peculiarities in parameter distributions of inner planets were searched by many teams \citep[e.g.,][]{Butler97,Mugrauer05,Desidera07,Bonavita07}. 
Among them, we can highlight the peculiarities in the mass-period and eccentricity-period distributions \citep[e.g.,][]{Eggenberger04,Zucker02}. Such peculiarities reflect the interaction between inner planets and wide companions, placing constraints on various scenarios proposed for their formation and evolution \citep{Eggenberger04}. Thus, it is reasonable to search for peculiarities in systems with wide BD companions hosting planets.
   
The systems with wide BD companions and planets offer great potential and opportunities to enrich our knowledge about both groups of objects and test our planetary formation and evolution concepts. A large sample of such systems is needed to reveal possible peculiarities and provide a detailed discussion. It leads to fundamental questions like: What is the fraction of systems with wide BD companions that host planets? Are they common, or is there any process that disfavors this scenario? 
In this paper, we focus on a sub-sample of six stars with confirmed co-moving wide BD companions using high-resolution spectroscopic data. These systems are GJ\,504 \citep{Kuzuhara13}, HD\,46588 \citep{Loutrel11}, HD\,203030 \citep{Metchev06}, HD\,3651 \citep{Mugrauer06}, HN\,Peg \citep{Luhman07}, and HD\,118865 \citep{Burningham13}. We describe the target selection process in Section \ref{sec:targets}. We present new observations in Section \ref{sec:observations} and make use of archival datasets described in Section \ref{sec:support}. We derive the physical parameters of our targets in Section \ref{sec:analysis} and investigate the nature of radial velocity (RV) signals to search for planets in Section \ref{sec:frequency}. In Section \ref{sec:discussion} we compare the sample of systems with confirmed planets and wide BD companions previously published in the literature with single planet-host-stars and planet-host-stars with wide stellar companions to discuss possible peculiarities in their parameter distributions. Finally, we summarize final conclusions in Section \ref{sec:summary}.


%
%
\section{Target justification}\label{sec:targets}
%


Following the discovery of a T4.5 dwarf companion at 6.3$'$ ($\sim$9000 au) from HIP\,70849 \citep{Lodieu14}, a K7V star which hosts a 9 M$_{\rm Jup}$ planet with an eccentric orbit \citep{Segransan11}, 
we searched the literature for other wide co-moving systems with one substellar component and a bright ($V$\,$\leq$\,6 mag) host suitable for intensive spectroscopic follow-up with the 1-m robotic Stellar Observations Network Group (SONG) telescope \citep{Andersen19}. Among the brightest primaries of wide substellar companions visible from the Northern hemisphere, we recovered HD\,3651 \citep{Mugrauer06}. The other targets suitable for our project were at that time: HD\,46588 \citep{Loutrel11}, HN\,Peg \citep{Luhman07}, GJ\,504 \citep{Kuzuhara13,Janson13,Skemer16}. We subsequently added HD\,203030 \citep{Metchev06} when the magnitude limit of the SONG guiding was increased to $V$\,=\,8.5\,mag as well as HD\,118865 \citep{Burningham13} to gauge the performance of the STELLar Activity (STELLA) 1.2-m robotic telescopes \citep{Strassmeier04}. These six sources constituted our sample when we started the project in 2015. The parameters of the observed sample of stars are listed in Table \ref{tab1}. We emphasize that the population of wide L and T dwarf companions has increased significantly over the past decade, as compiled in \cite{Deacon12} and \cite{Ponte20}. Hence, this work can be considered as a pilot program and the statistical results should be considered as preliminary.

\begin{table*}
	\centering
	\caption{System parameters for stars in our sample.}
	\label{tab1}
	\scalebox{0.81}{
	\begin{tabular}{lccccccr} 
		\hline
		\hline
		System       & GJ\,504 & HD\,3651 & HD\,46588 & HD\,203030 & HN\,Peg & HD\,118865 & Source\\
		\hline
        RA$_{J2000}$ (hh:mm:ss.ss) & 13 16 46.52 & 00 39 21.81 & 06 46 14.15 & 21 18 58.22 & 21 44 31.33 & 13 39 34.33 & 2\\
        Dec$_{J2000}$ (d:':") & 09 25 26.97 & 21 15 01.72 & 79 33 53.32 & 26 13 49.96 & 14 46 18.98 & 01 05 18.13 & 2\\
        \smallskip\\
        TESS $T$ mag & $4.655 \pm 0.007$ & $5.124 \pm 0.006$ & $4.949 \pm 0.006$ & $7.766 \pm 0.006$ & $5.420 \pm 0.007$ & $7.462 \pm 0.006$ & 3\\
        $Gaia$ $G$ mag & $5.042 \pm 0.004$ & $5.653 \pm 0.001$ & $5.309 \pm 0.002$ & $8.247 \pm 0.001$ & $5.824 \pm 0.002$ & $7.831 \pm 0.001$ & 1\\
        Tycho $B_T$ mag & $5.768 \pm 0.021$ & $6.744 \pm 0.020$ & $5.946 \pm 0.014$ & $9.090 \pm 0.120$ & $ 6.533 \pm 0.023$ & $8.524 \pm 0.029$ & 4\\
        Tycho $V_T$ mag & $5.188 \pm 0.023$ & $5.884 \pm 0.023$ & $5.433 \pm 0.009$ & $8.450 \pm 0.030$ & $5.977 \pm 0.023$ & $7.970 \pm 0.030$ & 4\\
        2MASS $J$ mag & $4.392 \pm 0.284$ & $4.549 \pm 0.206$ & $4.512 \pm 0.212$ & $7.068 \pm 0.019$ & $4.793 \pm 0.035$ & $6.977 \pm 0.024$ & 5\\
        2MASS $H$ mag & $4.107 \pm 0.208$ & $4.064 \pm 0.240$ & $4.262 \pm 0.146$ & $6.737 \pm 0.018$ & $4.598 \pm 0.036$ & $6.729 \pm 0.044$ & 5\\
        2MASS $K_S$ mag & $4.033 \pm 0.238$ & $3.999 \pm 0.036$ & $4.141 \pm 0.034$ & $6.653 \pm 0.023$ & $4.559 \pm 0.038$ & $6.666 \pm 0.020$ & 5\\
        WISE1 mag & $3.787 \pm 0.357$ & $3.786 \pm 0.365$ & $4.179 \pm 0.280$ & $6.658 \pm 0.071$ & $4.506 \pm 0.221$ & $6.670 \pm 0.065$ & 6\\
        WISE2 mag & $3.552 \pm 0.253$ & $3.577 \pm 0.254$ & $3.910 \pm 0.128$ & $6.627 \pm 0.021$ & $4.294 \pm 0.133$ & $6.650 \pm 0.022$ & 6\\
        WISE3 mag & $3.831 \pm 0.015$ & $3.898 \pm 0.014$ & $4.184 \pm 0.016$ & $6.655 \pm 0.015$ & $4.558 \pm 0.015$ & $6.703 \pm 0.016$ & 6\\
        WISE4 mag & $3.757 \pm 0.022$ & $3.901 \pm 0.023$ & $4.147 \pm 0.028$ & $6.611 \pm 0.068$ & $4.488 \pm 0.029$ & $6.670 \pm 0.060$ & 6\\
        \smallskip\\
	    $\mu_\alpha\,cos(\delta)$ (mas/yr) & $-335.473 \pm 0.192$ & $-461.948 \pm 0.068$ & $-99.163 \pm 0.054$ & $133.810 \pm 0.022$ & $231.108 \pm 0.030$ & $-95.558 \pm 0.027$ & 1\\
	    $\mu_{\delta}$ (mas/yr) & $191.038 \pm 0.195$ & $-369.624 \pm 0.025$ & $-604.042 \pm 0.073$ & $9.245 \pm 0.018$ & $-113.200 \pm 0.027$ & $-48.191 \pm 0.020$ & 1\\
	    Parallax (mas) & $56.858 \pm 0.122$ & $90.025 \pm 0.048$ & $54.938 \pm 0.060$ & $25.459 \pm 0.021$ & $55.148 \pm 0.035$ & $16.504 \pm 16.504$ & 1\\
		\hline
		\hline
		
	\end{tabular}
	}
	\smallskip\\
References: 1 - $Gaia$ eDR3, \citet{Gaia21}; 2 - $Gaia$ DR2, \citet{Gaia18}; \\ 3 - TESS, \citet{Stassun18}; 4 - Tycho, \citet{Hog00}; 5 - 2MASS, \citet{Cutri03}; 6 - WISE, \citet{Wright10} \\
\end{table*}

%
%
\section{Spectroscopic observations}\label{sec:observations}

%
%
%

We performed high-resolution spectroscopy with several telescopes and instruments whose observations are described below. 

\subsection{SONG spectroscopy}
\label{GJ504_planet:spectro_SONG}

We observed our sample of stars with the robotic Stellar Observations Network Group (SONG) Hertzsprung telescope \citep{Andersen19} through a dedicated program with nightly observations when weather permitted for two years ($\geq$\,4 semesters) in most cases. The time interval of the observations, together with the number of acquired spectra, are reported in Table \ref{table:5}. We used the slit of 1.2\,arcsec, yielding a spectral resolution of 90,000\@. To ensure sufficient count over the full wavelength range of SONG spectra, we employed exposure times between 360 and 600s, leading to a signal-to-noise ratio of at least 50 per resolution element for all observations. The numbers of gathered spectra for each system, together with the range of dates of observations, are in Table \ref{table:5}.

SONG is a 1--m telescope located in the Observatorio del Teide in Tenerife, Canary Islands. It is equipped with a high-resolution echelle spectrograph operating in the visible band between 440 and 690\,nm with an average dispersion of 0.002\,nm/pixel. The detector is an Ikon-L commercial charged-coupled device camera developed by Andor with a 2k$\times$2k chip size. The SONG spectra and the RVs have been kindly provided by the SONG team as derived by the iSONG software \citep{antoci13} following the procedure described in \citep{Grundahl17}.

\subsection{Calar Alto/CARMENES spectroscopy}
\label{GJ504_planet:spectro_CARMENES}

We collected visible and near-infrared spectra for each system with the Calar Alto high-resolution search for M dwarfs with Exoearths with Near-infrared and optical Echelle Spectrographs \citep[CARMENES;][]{Quirrenbach14}. The time interval of the observations, together with the number of acquired spectra, are reported in Table \ref{table:5}. All observations were conducted in queue mode by the staff of the Calar Alto observatory, satisfying the requested conditions: seeing better than two arcsec and clear skies. We set the exposure times between 250-900 seconds depending on the star's brightness, targeting the minimum signal-to-noise ratio (S/N) of 150 based on the estimates of the exposure time calculator.

CARMENES consists of two independent high-resolution echelle spectrographs in the visible (520--960\,nm; R\,=\,93,400) and near-infrared (960--1710\,nm; R\,=\,81,800) which are simultaneously fed through fibres.

The data were reduced using CARACAL \citep{Caballero16}, and the visible and near-infrared radial velocities were obtained with {\tt SERVAL} \citep{Zechmeister18}. {\tt SERVAL} determines radial velocity by co-adding all available spectra of the target with an S/N higher than 10 and creating a high-quality template of the star used as a reference spectrum. The radial velocities were corrected for barycentric motion, secular perspective acceleration, instrumental drift, and nightly zero points \citep{Trifonov18}.

\subsection{STELLA spectroscopy}
We observed HD\,118865 system with the STELLar Activity (STELLA) two 1.2-m robotic telescopes located at Iza\~na Observatory in Tenerife \citep{Strassmeier04}. The fibre-fed Echelle Spectrograph of STELLA with an e2v 2k$\times$2k CCD detector covers the wavelength range between 390–880\,nm and has the resolving power of R\,=\,55,000\@. The spectra were automatically reduced using the IRAF-based STELLA data-reduction pipeline \citep{Weber08}, and RVs were kindly provided by the STELLA team. The time interval of the observations, together with the number of acquired spectra, are reported in Table \ref{table:5}.

%
%



\begin{table}                  
\caption{List of spectroscopic observations for our sample of stars.}             
\label{table:5}      
\resizebox{\columnwidth}{!}{%
\centering                          
\begin{tabular}{c c c}        
\hline\hline                 
Star & \# spectra & Date \\    
\hline  
\multicolumn{3}{|c|}{SONG spectrograph} \\
\hline
   GJ\,504 & 262 & 2016-Feb-7 to 2017-Aug-19 \\      
   HD\,3651 & 414 & 2014-Sep-03 to 2016-Feb-14 \\
   HD\,46588 & 352 & 2015-Apr-3 to 2017-Mar-6 \\
   HD\,203030 & 111 & 2017-Apr-17 to 2018-Sep-29 \\
   HN\,Peg & 250 & 2015-Apr-22 to 2017-Jan-15 \\
\hline
\multicolumn{3}{|c|}{CARMENES spectrograph} \\
\hline
   GJ\,504 & 39 & 2019-Jan-04 to 2019-Jun-28 \\      
   HD\,3651 & 19 & 2019-Jan-03 to 2019-Jun-29 \\
   HD\,46588 & 35 & 2019-Jan-04 to 2019-May-25 \\
   HD\,203030 & 32 & 2019-Jan-03 to 2019-Jun-29 \\
   HN\,Peg & 25 & 2019-Jan-03 to 2019-Jun-29 \\
   HD\,118865 & 54 & 2019-Jan-04 to 2019-Jun-28 \\
\hline                                   
\multicolumn{3}{|c|}{STELLA spectrograph} \\
\hline
   HD\,118865 & 101 & 2017-Feb-01 to 2017-Jul-29 \\
\\
\multicolumn{3}{c}{Archival data} \\
\hline
\multicolumn{3}{|c|}{GJ\,504} \\
\hline
SOPHIE & 38 & 2013-Apr-01 to 2016-May-23 \\
Lick & 57 & 1987-Jun-12 to 2009-Feb-2 \\
HARPS-N & 106 & 2016-Jun-9 to 2018-Apr-1 \\
\hline
\multicolumn{3}{|c|}{HD\,3651} \\
\hline
EXPRES & 61 & 2019-Aug-18 to 2020-Feb-1 \\
HIRES & 161 & 1996-Oct-10 to 2013-Dec-14 \\
LICK & 155 & 1987-Sep-9 to 2011-Oct-12 \\
HRS & 35 & 2005-Jul-29 to 2007-Nov-19 \\
Tull & 4 & 2005-Sep-20 to 2006-Oct-12 \\
\hline
\multicolumn{3}{|c|}{HD\,203030} \\
\hline
HIRES & 17 & 2002-Aug-28 to 2013-Aug-26 \\
\hline
\multicolumn{3}{|c|}{HN\,Peg} \\
\hline
LICK & 37 & 1987-Sep-9 to 2004-Aug-2 \\
HARPS & 22 & 2013-Oct-10 to 2017-Jun-3 \\
\hline
\hline
\end{tabular}
}
\end{table}

\section{Supporting observational data}\label{sec:support}

We searched for archival spectroscopic observations, which we used as supplemental data in our analyses. In addition, we also downloaded photometric data from the Transiting Exoplanet Survey Satellite (TESS) \citep{Ricker15} available for the stars from our sample.

\subsection{Archival spectroscopy}

GJ\,504, HD\,3651, HN\,Peg and HD\,203030 are systems that were already observed spectroscopically. We downloaded public archival data to complement our own observational datasets.

For GJ\,504, we 
checked the HARPS-N archive and noticed that additional spectra had been collected with the HARPS-N \citep{Cosentino12} by the GAPS team between 9 June 2016 and 01 April 2018, amounting to a total of 106 spectra. We re-processed all HARPS-N spectra with the {\tt SERVAL} pipeline to get RV measurements and activity indicators. Additionally, we found 58 spectra \citep{Fischer14} collected with the LICK spectrograph \citep{Vogt87}; however, their relatively low cadence over the long baseline do not make them very useful as the star shows a high level of activity (we do not see any significant signal in the periodograms). For the same reason, we did not use 38 RVs \citep{Bonnefoy18} collected with the SOPHIE spectrograph \citep{Bouchy06}. In their periodogram, we only see peaks close to the rotation period of the stars visible in all used datasets. 

For HD\,3651, we also used RVs published in \citet{Brewer20}. These authors report 61 spectra from the EXPRES spectrograph \citep{Jurgenson16} mounted on the 4.3-m Lowell Observatory Discovery Channel Telescope (DCT). We also included an additional 161 archival Keck HIRES data \citep{Vogt94} with the 17-year time baseline \citep{Butler17}, 155 archival LICK RVs \citep{Fischer14}, 35 archival RVs obtained with the High Resolution Spectrograph (HRS) \citep{Tull98} mounted at the 9.2m Hobby–Eberly Telescope (HET) \citep{Wittenmyer09}, and four archival RVs obtained with the Tull spectrograph \citep{Tull95} at the 2.7m Harlan J.Smith Telescope \citep{Wittenmyer09}.

For HD\,203030, we found 17 Keck HIRES RVs \citep{Butler17}. However, with the 17-year time baseline and the star's level of activity, these data are not helpful, and we do not observe any significant signal in the periodogram.

Finally, for HN\,Peg, we found 37 LICK RVs \citep{Fischer14} and 22 HARPS RVs. However, because of the RV scatter caused by the stellar activity, such small datasets are also not helpful compared to the SONG dataset. The complete list of available archival data is reported in Table \ref{table:5}.

%
%

\begin{figure*}[!ht]
\centering
\includegraphics[width=0.9\textwidth,height=1.05\textwidth]{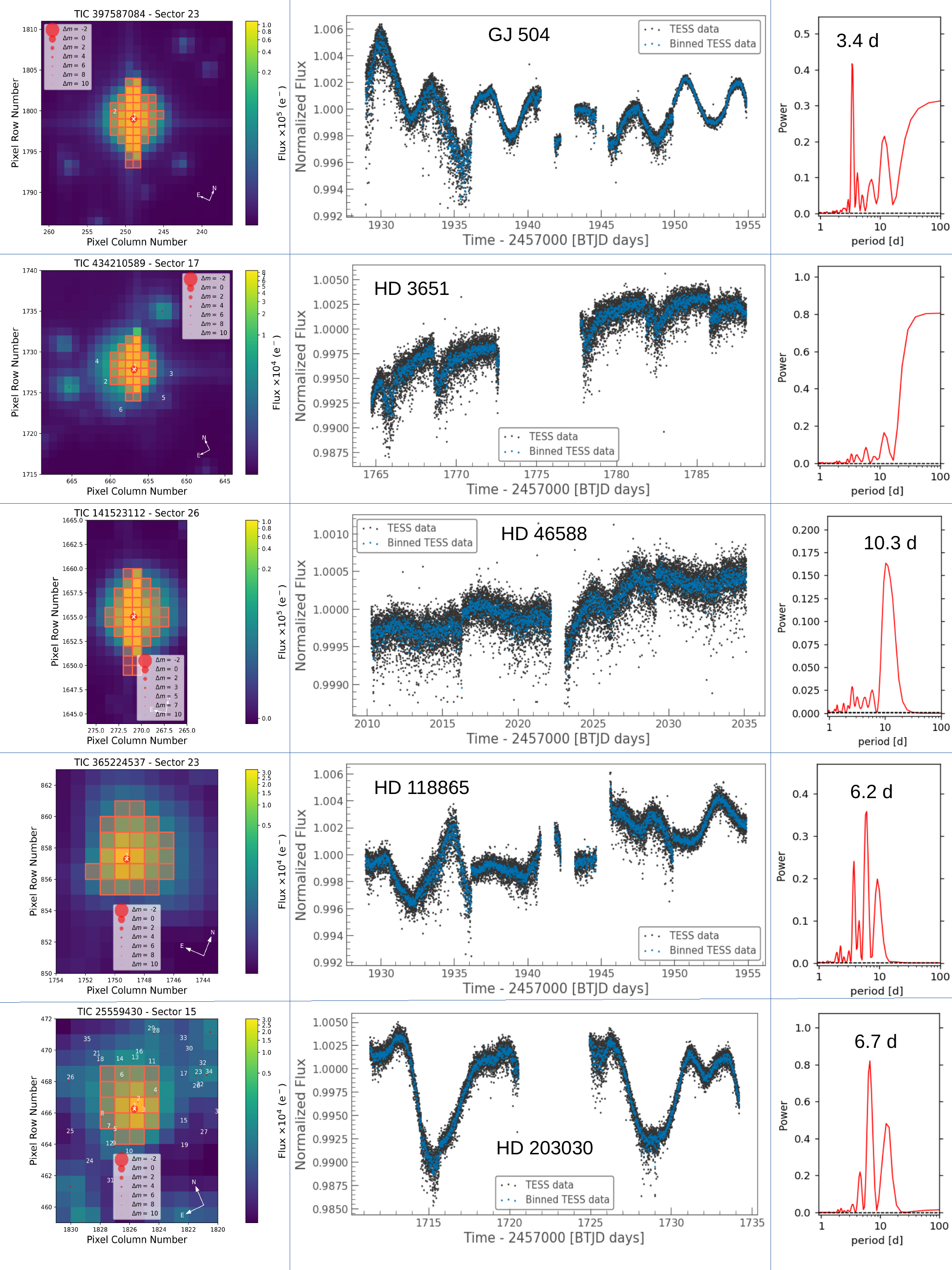}
\caption{Left: $Gaia$ DR2 catalog overplotted to the TESS TPF images. Middle: The PDCSAP light curves observed by TESS. Right: Marginalized posterior distributions of the rotation period from GP modeling. From top to bottom, we list GJ\,504/TIC\,397587084, HD\,3651/TIC\,434210589, HD\,46588/TIC\,141523112, HD\,118865/TIC\,365224537, and HD\,203030/TIC\,25559430.} 
\label{fig:tess_lc}
\end{figure*}

%
%

\subsection{TESS photometry}
\label{GJ504_planet:photo_TESS}

TESS observed each of our targets except HN\,Peg at least in one sector at a two-minute cadence mode during its prime mission. The full list of observations is summarized in Table \ref{table:1}. We used the {\tt lightkurve} package in Python \citep{Lightkurve18} to download TESS target pixel files from the Mikulski Archive for Space Telescopes (MAST)\footnote{\url{https://mast.stsci.edu/portal/Mashup/Clients/Mast/Portal.html}}. We then selected optimal aperture masks to extract light curves (LCs) for each system, which we corrected for outliers and normalized. On them, we performed the Pixel Level Decorrelation method \citep{Deming15} to remove systematics. The final TESS LCs are shown in the middle of Fig.\ \ref{fig:tess_lc}. None of our targets is recognized as TESS Object of Interest (TOI), meaning that no obvious transit-like feature has been identified in the LCs.

We used {\tt Tpfplotter} \citep{Aller20} to overplot the $Gaia$ DR2 catalogue to the TESS target pixel file (tpf) to investigate possible diluting sources in TESS photometry with the limiting difference in magnitude of 10. Tpf images for our sample of stars created with {\tt tpfplotter} can be seen on the left-hand side of Fig.\ \ref{fig:tess_lc}. There is only one additional source between TESS pixels used in SAP in the case of GJ\,504. This star has identification $Gaia$ID\,3732539649257672704 and $Gaia$ $G$ magnitude of 14. According to $Gaia$ colour bands, we classify this object as K7--K9 spectral type, with $G-R_P = 0.89$\,mag corresponding to K8--K9, and $B_P-R_P = 1.72$\,mag corresponding to K7--K8. To compute the spectral type we used the up-to-date version\footnote{\url{https://www.pas.rochester.edu/~emamajek/EEM_dwarf_UBVIJHK_colors_Teff.txt}} of the dwarf color sequence from \cite{Pecaut13}. For HD\,203030, there are five additional sources within the aperture of the TESS pixels. The full list of these sources with relevant information is displayed in Table \ref{table:2}. We conclude that these sources are too faint compared to GJ\,504, HD\,203030, respectively, to yield any significant dilution. We did not find any diluting sources in the TESS apertures for the rest of the systems in our sample.

\begin{table}
\caption{Dates of TESS observations for our sample of stars.}             
\label{table:1}      
\centering                          
\begin{tabular}{c c c}        
\hline\hline                 
Star & Sector & Date \\    
\hline                        
   GJ\,504 & 23 & 2020-Mar-20 to 2020-Apr-15 \\      
    & 50 & 2022-Mar-26 to 2022-Apr-22 \\
    \hline
   HD\,3651 & 17 & 2019-OCt-7 to 2019-Nov-2 \\
   \hline
   HD\,46588 & 19 & 2019-Nov-27 to 2019-Dec-24 \\
    & 20 & 2019-Dec-24 to 2020-Jan-21 \\
    & 26 & 2020-Jun-08 to 2020-Jul-04 \\
    & 40 & 2021-Jun-24 to 2021-Jul-23 \\
    & 47 & 2021-Dec-30 to 2022-Jan-28 \\
    & 53 & 2022-Jun-13 to 2022-Jul-09 \\
    \hline
   HD\,203030 & 15 & 2019-Aug-15 to 2019-Sep-11 \\
    & 55 & 2022-Aug-5 to 2022-Sep-1 \\
    \hline
   HD 118865 & 23 & 2020-Mar-18 to 2019-Apr-16 \\
    & 45 & 2021-Dec-02 to 2021-Dec-30 \\
    & 50 & 2022-Mar-26 to 2022-Apr-22 \\
     \hline
   HN\,Peg & 55 & 2022-Aug-5 to 2022-Sep-1 \\
    
\hline                                   
\hline
\end{tabular}
\end{table}

\begin{table}
 \centering
 \caption[]{The list of additional sources within the TESS apertures for our sample of stars.
 }
 \label{table:2}
\scalebox{0.9}{
 \begin{tabular}{@{\hspace{0mm}}l c c c@{\hspace{0mm}}}
 \hline
 \hline
$Gaia$ID & $Gaia$ $G$ mag  & Spectral type   &  close to \cr
 \hline
3732539649257672704 & 14.0 & K7--9 & GJ\,504  \cr
 \hline
1846882426008946304 & 15.2 & K3--4/K5--6 & HD\,203030  \cr
1846882219850515968 & 17.2 & K1--3 & HD\,203030  \cr
1846882430304186112 & 15.6 & G9--K0 & HD\,203030  \cr
1846882533383405696 & 17.7 & K4--6 & HD\,203030  \cr
1846882322929096704 & 18.1 & K3--4 & HD\,203030  \cr
 \hline
 \hline

 \end{tabular}
}
\end{table}

%
%

\section{Stellar and brown dwarf parameters}\label{sec:analysis}

In this section, we derive the stellar parameters of stars in our sample as well as the ages of individual systems, which are then used to infer the ages of wide companions and derive their parameters. We need to know the parameters of the host stars if we want to derive parameters of potential planets, and if we want later discuss peculiarities in the parameter distributions of these planets, it is crucial to describe every known object in these systems. We have done this for all studied systems regardless of whether we found any planetary candidate.
%
%

\subsection{Stellar parameters with {\tt iSpec} and {\tt VOSA}}
\label{sec:ispec_vosa}

We used the {\tt iSpec} framework \citep{Blanco14,Blanco19} to derive the parameters of the host stars from the co-added CARMENES spectra. {\tt iSpec} determines stellar parameters by minimalization of the $\chi^2$ value between the calculated synthetic spectrum and observed spectrum. To determine the effective temperature $T_{\rm eff}$, metallicity $\rm [Fe/H]$, surface gravity $\log{g}$, and the projected stellar equatorial velocity $v\sin{i}$ we followed the same procedure as in \cite{Fridlund17}. The whole procedure is described in detail in Appendix \ref{sec:iSpec}, and we report the stellar parameters for the stars of our sample in Table \ref{tab5}. 

As a sanity check, we also analyzed the spectral energy distribution with the Virtual Observatory SED Analyser \citep[{\tt VOSA}\footnote{\url{http://svo2.cab.inta-csic.es/theory/vosa/}};][]{Bayo08}. {\tt VOSA} performs the ${\chi}^2$ minimization procedure to compare theoretical models with the observed photometry. For our SED fitting, we use the Str{\"o}mgren-Crawford uvby{\ensuremath{\beta}} \citep{Paunzen15}, Tycho \citep{Hog00}, $Gaia$ DR2 \citep{Gaia18}, $Gaia$ eDR3 \citep{Gaia21}, 2MASS \citep{Cutri03}, AKARI \citep{Ishihara10}, and WISE \citep{Cutri14} photometry. The whole procedure is described in detail in Appendix \ref{sec:VOSA}, and we report the stellar parameters for the stars of our sample in Table \ref{tab5}.

%
%

\begin{table*}
	\centering
	\caption{Physical parameters for stars in our sample.}
	\label{tab5}
	\scalebox{1.1}{
	\begin{tabular}{lccccccr} 
		\hline
		\hline
		System       & GJ\,504 & HD\,3651 & HD\,46588 & HD\,203030 & HN\,Peg & HD\,118865 \\
		\hline
\multicolumn{7}{|c|}{iSpec \& PARAM 1.5 analysis} \\
\hline
$\rm T_{eff} (K)$ & $6081 \pm 102$ & $5284 \pm 78$ & $6230 \pm 115$ & $5578 \pm 96$ & $5945 \pm 106$ & $6347 \pm 109$ \\
$[{\rm Fe/H}]$ & $0.20 \pm 0.06$ & $0.22 \pm 0.05$ & $-0.06 \pm 0.12$ & $0.15 \pm 0.10$ & $-0.03 \pm 0.12$ & $0.20 \pm 0.06$ \\
$\log{g}$ & $4.29 \pm 0.08$ & $4.54 \pm 0.08$ & $4.36 \pm 0.08$ & $4.54 \pm 0.06$ & $4.44 \pm 0.08$ & $4.37 \pm 0.06$ \\
$v_{\rm rot} \sin{i_\star} $ (km/s) & $6.57 \pm 0.82$ & $1.22 \pm 0.87$ & $2.50 \pm 1.74$ & $6.10 \pm 0.61$ & $9.60 \pm 0.98$ & $7.1 \pm 1.06$\\
$EW_{Li}$ (\AA) & 0.087 & <0.010 & 0.051 & 0.050 & 0.103 & 0.076 \\
$M_\star$ ($\rm \mst$) & $1.16 \pm 0.04$ & $0.89 \pm 0.04$ & $1.09 \pm 0.06$ & $0.92 \pm 0.04$ & $1.00 \pm 0.06$ & $1.21 \pm 0.04$\\
$R_\star$ ($\rm \rst$) & $1.30 \pm 0.05$ & $0.86 \pm 0.03$ & $1.16 \pm 0.04$ & $0.85 \pm 0.03$ & $1.01 \pm 0.04$ & $1.19 \pm 0.03$\\
$P_{Rot}$ (days) & $3.4$ & $44.5$ & $10.3$ & $6.7$ & 4.59--5.17 (7) & $6.2$\\
\hline
\multicolumn{7}{|c|}{VOSA analysis} \\
\hline
$\rm T_{eff} (K)$ & 5900--6200 & 5200--5400 & 6100--6300 & 5400--5600 & 5900--6100 & 6000--6200\\
$[{\rm Fe/H}]$ & 0.0--0.5 & 0.0--0.5 & 0.0--0.5 & 0.0--0.5 & $-0.5$--0.5 & 0.0--0.5\\
$\log{g}$ & 4--5 & 4--5 & 4--5 & 4--5 & 4--5 & 4--5\\
$R_\star$ ($\rm \rst$) & 1.25--1.40 & 0.81--0.89 & 1.12--1.21 & 0.81--0.88 & 0.94--1.03 & 1.19--1.29\\
$L_\star$ ($L_{\odot}$) & 2.05--2.16 & 0.50--0.54 & 1.77--1.92 & 0.57--0.60 & 1.11--1.20 & 1.89--2.02\\
\hline
Sp. type & G0V (1) & K0V (2) & F8V (3) & G8V (4) & G0V (5) & F7V (6) \\
$log\,L_X$ (W) & 22.5 (8) & 20.2 (8) & 21.0 (8) & 21.8 (8) & 22.2 (8) & --- \\
\hline
\hline
	\end{tabular}
	}
{References: (1) - \citet{Anderson10}, (2) - \citet{Keenan89}, (3) - \citet{Abt09}, (4) - \citet{Frasca18} \\ 
(5) - \citet{Gray01}, (6) - \citet{Houk99}, (7) - \citet{Messina03}, (8) - \citet{Voges99}}
\end{table*}

\subsection{Analysing the surface rotation}
\label{surface_rotation}

To determine the rotation period from the TESS light curves (LCs), we apply generalised Lomb-Scargle (GLS) periodograms \citep{Zechmeister09} investigating the most dominant signals. Plots of periodograms for individual stars can be found in Fig.\ \ref{fig:tess_lc}. GJ\,504 and HD\,203030 show clear variations with the periods of 3.4\,days and 6.7 \,days, respectively, which we interpret as the rotation periods. HD\,3651 does not show clear variations but only a long trend; hence we interpret the rotation period to be larger than the TESS data baseline. It is consistent with the period of 44.5\,days derived through the calibration from the value of log\,$R'_{HK}$ by \citet{Fischer03}. \cite{Brewer20} then searched for photometric variations using 1192 photometric observations collected over the period of 25 yr with the 0.75m Automatic Photoelectric Telescope (APT) at Fairborn Observatory in southern Arizona. However, they did not find any significant variability within any observing season. Finally, after we removed long trends in the HD\,46588 and HD\,118865 datasets, we could identify variations with the periods of 10.3\,days and 6.2\,days that we interpret as the rotation periods. The variations in the HD\,118865 LC are not so clear as for other objects; however, we used this period only to derive the bottom limit for the age, as can be seen in the next section. We did not find any contradictions with rotation periods reported in the literature. For example, in the catalogue of rotation periods by \cite{Wright11} we can see the rotation period of 6.67\,d for HD\,203030, 4.86\,d for HN\,Peg and 48\,d for HD\,3651. \cite{Donahue96} then reported the rotation period of 3.33\,d for GJ\,504.

%
%

\subsection{Age analysis}\label{sec:age}
To determine the age of the sample stars, we provide an extensive analysis of several complementary age indicators. We include stellar isochrones fitting, gyrochronology analysis, lithium equivalent width, X-ray luminosity, and membership to young associations. Our effort is to examine each age indicator separately to provide the age intervals for each of them and to determine the final age as an overlap between these intervals. Detailed description of each age indicator can be found in Appendix \ref{sec:age_app} and the results are listed in Table \ref{tab:age}.

We found that all age indicators for our sample of stars are quite consistent with each other. Two inconsistencies are that the gyrochronology predicts a slightly lower age of HD\,203030 and the isochrone fitting for the GJ\,504, HD\,203030 and HD\,46588 gives older ages than the rest of the indicators. However, we found agreement in most cases when considering 95\% confidence intervals from isochrone fitting. The GJ\,504 age is intensively discussed in the literature. \cite{Kuzuhara13} estimated the age of GJ\,504 to be $160^{+350}_{-60}$ Myr using gyrochronology and chromosphere activity of the star. \cite{Fuhrmann15,Dorazi17} have found that the isochrones comparison suggests an older age between 1.8 and 3.5 Gyr. Authors speculated that the recent merging of hot Jupiter companion could explain the star's high activity level. \cite{Bonnefoy18} have recently revisited the system parameters. Instead of using an effective temperature as an input to isochrone fitting, they used an interferometric radius of $R_\star\,=\,1.35\pm0.04\rm \rst$ for GJ 504. However, the fact that we got, as an output from isochrone fitting, the radius with similar uncertainty as these authors measured suggests that both analyses should be equivalent in terms of age determination. They found two isochronal age scenarios: the young one with an age of $21\pm2$\,Myr and the old one with an age of $4\pm1.8$\,Gyr.

\begin{figure}[!ht]
\centering
\includegraphics[width=0.49\textwidth, trim= {0.0cm 0.0cm 0.0cm 0.0cm}]{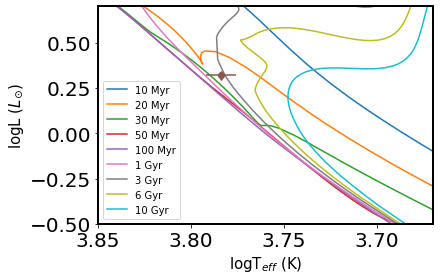}
\caption{Luminosity vs effective temperature plot. Curves represent MIST isochrones for ages: 10\,Myr (blue), 20\,Myr (orange), 30\,Myr (green), 50\,Myr (red), 100\,Myr (purple), 1\,Gyr (pink), 3\,Gyr (grey), 6\,Gyr (chartreuse), 10\,Gyr (celeste), and for [Fe/H]=0.25. Brown point represents the parameters of GJ\,504 with their error bars.} \label{fig:isoch_gj}
\end{figure}

We confirm that most age indicators lead toward the relatively young age of several hundred Myr. Similarly to \cite{Bonnefoy18}, we also found that isochrone fitting gives two possible age scenarios. It can be seen in Figure \ref{fig:isoch_gj}, where we plot GJ\, 504's luminosity from the SED analysis and effective temperature, together with the MIST stellar evolutionary tracks \citep{Choi16}. We also note that the age depends on the isochrone model used. Using the PARSEC isochrones \citep{Bressan12} we found the age of $3.8^{+1.5}_{-1.1}$\,Gyr, while for the MIST isochrones we found $2.6^{+1.2}_{-1.3}$\,Gyr. Similarly to other stars in our sample, in Table \ref{tab:age} we provide an age for GJ\,504 as an overlap of most of the activity indicators. However, we do not consider this age final, and we are open to both younger and older scenarios. That's why we keep this system in or sample as a system with the potential wide BD companion.


If we consider that BDs cool down with age, their detection becomes more challenging as they get older. In this context, it is not surprising that a lot of the stars in our sample are relatively young, i.e.\ 1\,Gyr or lower. Nonetheless, we found few systems older than 2.5 Gyr.

\begin{table*}
	\centering
	\caption{Age intervals for stars in our sample from different indicators together with the final adopted intervals.}
	\label{tab:age}
	\scalebox{0.75}{
	\begin{tabular}{lccccccr} 
		\hline
		\hline
		System age (Gyr)      & GJ\,504 & HD\,3651 & HD\,46588 & HD\,203030 & HN\,Peg & HD\,118865 \\
        \hline
		Isochrones & $3.8^{+1.5}_{-1.1}$ & $5.5^{+4.2}_{-3.5}$ & $2.9^{+1.8}_{-1.6}$ & $2.6^{+3.2}_{-1.9}$ & $3.8^{+2.8}_{-2.4}$ & $0.9^{+1.0}_{-0.6}$ \\
        Gyrochronology & 0.12--0.55 & $\geq$ 2.5 & 0.8--2.5 & 0.35--0.55 & 0.35--0.65 & 0.60--1.2 \\
        & (Pleiades--M37) & ($\geq$ Ruprecht 147) & (NGC 6811--Ruprecht 147) & (M37) & (M37--Praesepe) & (Praesepe--NGC 6811) \\
        Lithium EW & 0.20--0.65 & $\geq$ 0.20 & 0.6--4.0 & 0.60--0.65 & 0.20--0.65 & 0.6--4.0 \\
        & (M34--Prasepe) & ($\geq$ M34) & (Praesepe--M67) & (Praesepe) & (M34--Praesepe) & (Praesepe--M67) \\
        X-ray luminosity & $\leq$ 0.65 & $\geq$ 0.60 & $\geq$ 0.60 & 0.60--0.65 & $\leq$ 0.65 & --- \\
        & ($\leq$ Praesepe) & ($\geq$ Praesepe) & ($\geq$ Praesepe) & (Praesepe) & ($\leq$ Praesepe) & (---) \\
        Membership to YMGs & no & no & no & no & no & no \\
		\hline
		Final age & 0.20--0.55 & 2.5--10.0 & 0.8--2.5 & 0.35--0.65 & 0.35--0.65 & 0.6--1.2 \\
		& (M34--M37) & ($\geq$ Ruprecht 147) & (NGC 6811--Ruprecht 147) & (M37--Praesepe) & (M37--Praesepe) & (Praesepe--NGC 6811) \\
		\hline
		\hline
	\end{tabular}
	}
\end{table*}

\subsection{Revised parameters of the wide BD companions}
\label{phys_params_compan}

We used the SpeX Prism Library Analysis Toolkit ({\tt SPLAT}) \citep{Burgasser17} to derive the parameters of the wide BD companions in our sample. SPLAT is a python-based package designed to interface with the SpeX Prism Library ({\tt SPL}) \citep{Burgasser14}, which is an online repository of over 2000 low-resolution, near-infrared spectra of stars and BDs. The package enables conversion between observable (temperature, luminosity, surface gravity) and physical parameters (mass, radius, age) of BDs using published evolutionary model grids. A more detailed description can be found in Appendix \ref{phys_params_compan}, and results are summarized in Table \ref{tab3}. 

According to the derived mass intervals, we confirm all companions except GJ\,504b to be BDs. Using the adopted age from Table 3, we find GJ\,504b in the planetary regime. However, as previously discussed, we do not exclude older age from isochrone fitting, which places the companion in the brown dwarf regime.


\begin{table*}
	\centering
	\caption{Parameters of wide companions around the stars from our sample.}
	\label{tab3}
	\scalebox{0.83}{
	\begin{tabular}{lcccccccr} 
		\hline
		\hline
		System       & GJ\,504b & HD\,3651b & HD\,46588b & HD\,203030b & HN\,Pegb & HD\,118865b & Model\\
\hline \\
Age (Gyr) & 0.20$-$0.55 (4) & 2.5$-$10.0 (4) & 0.8$-$2.5 (4) & 0.35$-$0.65 (4) & 0.35$-$0.65 (4) & 0.6$-$1.2 (4) \\
Luminosity (log $L$/\(L_\odot\)) & $-6.15 \pm 0.15$ (5) & $-5.58 \pm 0.05$ (6) & $-4.68 \pm 0.05$ (7) & $-4.75 \pm 0.04$ (8) & $-4.77 \pm 0.03$ (9) & $-5.24 \pm 0.04$ (10) \\
\hline \\
& 440$-$544 & 738$-$854 & 1220$-$1377 & 1122$-$1211 & 1115$-$1191 & 854$-$932 & (1)\\
Temperature (K) & 450$-$553 & 746$-$862 & 1225$-$1383 & 1114$-$1225 & 1108$-$1203 & 864$-$941 & (2)\\
& 441$-$543 & 737$-$849 & 1212$-$1370 & 1115$-$1212 & 1108$-$1190 & 850$-$928 & (3a)\\
& 437$-$539 & 729$-$841 &  & 1100$-$1196 &  & 841$-$917 & (3b)\\
\hline
Adopted & 437$-$544 & 729$-$862 & 1212$-$1383 & 1100$-$1225 & 1108$-$1203 & 841$-$941 & \\
& 478$-$615 (*) & & & & & & \\
\hline \\
& 3.89$-$4.34 & 5.01$-$5.46 & 5.05$-$5.41 & 4.76$-$5.00 & 4.76$-$4.98 & 4.72$-$4.97 & (1)\\
Gravity ($cm/s^2$) & 3.85$-$4.32 & 4.97$-$5.40 & 5.01$-$5.37 & 4.71$-$4.95 & 4.71$-$4.94 & 4.69$-$4.94 & (2)\\
& 3.88$-$4.32 & 4.96$-$5.41 & 5.01$-$5.36 & 4.73$-$4.95 & 4.72$-$4.94 & 4.69$-$4.94 & (3a)\\
& 3.85$-$4.29 & 4.94$-$5.38 &  & 4.69$-$4.91 &  & 4.65$-$4.90 & (3b)\\
\hline
Adopted & 3.85$-$4.34 & 4.94$-$5.46 & 5.01$-$5.41 & 4.69$-$5.00 & 4.71$-$4.98 & 4.65$-$4.97 & \\
& 4.62$-$5.08 (*) & & & & & & \\
\hline \\
& 4.3$-$10.7 & 34.1$-$67.8 & 40.0$-$69.3 & 25.1$-$37.0 & 24.9$-$36.0 & 21.7$-$33.2 & (1)\\
Mass ({$\rm M_J$}) & 3.8$-$9.6 & 30.7$-$57.9 & 36.6$-$63.3 & 23.0$-$33.7 & 22.9$-$32.9 & 20.0$-$30.8 & (2)\\
& 4.2$-$10.1 & 31.0$-$61.9 & 37.2$-$64.1 & 23.7$-$34.3 & 23.5$-$33.5 & 20.7$-$31.2 & (3a)\\
& 4.1$-$9.9 & 30.3$-$60.4 &  & 22.9$-$33.0 &  & 20.0$-$30.2 & (3b)\\
\hline
Adopted & 3.8$-$10.7 & 30.3$-$67.8 & 36.6$-$69.3 & 22.9$-$37.0 & 22.9$-$36.0 & 20.0$-$33.2 & \\
& 16.5$-$37.1 (*) & & & & & & \\
\hline \\
& 1.10$-$1.18 & 0.76$-$0.91 & 0.82$-$0.94 & 0.96$-$1.04 & 0.96$-$1.04 & 0.94$-$1.02 & (1)\\
Radius ({$\rm R_J$}) & 1.07$-$1.15 & 0.75$-$0.90 & 0.82$-$0.94 & 0.96$-$1.05 & 0.96$-$1.05 & 0.93$-$1.01 & (2)\\
& 1.10$-$1.18 & 0.77$-$0.91 & 0.83$-$0.95 & 0.97$-$1.05 & 0.98$-$1.05 & 0.95$-$1.03 & (3a)\\
& 1.12$-$1.20 & 0.79$-$0.93 &  & 1.00$-$1.08 &  & 0.97$-$1.05 & (3b)\\
\hline
Adopted & 1.07$-$1.20 & 0.75$-$0.93 & 0.82$-$0.95 & 0.96$-$1.08 & 0.96$-$1.05 & 0.93$-$1.05 & \\
& 0.87$-$1.00 (*) & & & & & & \\
		\hline
		\hline \\
	\end{tabular}
	}
	{References: (1) - \citet{Burrows01}, (2) - \citet{baraffe03}, (3a) - case $[{\rm Fe/H}]$ = $0.0$ from \citet{saumon08} \\ 
	(3b) - case $[{\rm Fe/H}]$ = $+0.30$ from \citet{saumon08}, (4) - this work, (5) - \citet{Bonnefoy18}, (6) - \citet{Liu07}, \\ 
	(7) - \citet{Loutrel11}, (8) - \citet{Miles17}, (9) - \citet{Luhman07}, (10) - \citet{Burningham13}} \\
	(*) - using an age of $3.8^{+1.5}_{-1.1}$ from isochrone fitting
\end{table*}
%


\section{Frequency Analysis and Stellar Activity}\label{sec:frequency}


We performed a frequency analysis using GLS periodograms of RV measurements for each target to look for possible companions. We note that TESS LCs do not reveal obvious transits during the 27 days of observations per sector, implying that we cannot confirm any short-periodic Keplerian signal. Instead, we investigate various stellar activity indicators to discuss the nature of the most significant signals found in periodograms. The SONG, STELLA and CARMENES RV measurements for all stars in our sample are available at CDS.

The SONG datasets create the core of our project. The timescales and frequencies of observations make them the most suitable for frequency analysis. Unfortunately, we cannot provide many stellar activity indicators for SONG data as spectra are calibrated using the iodine cell technique, whose lines contaminate the spectrum. Furthermore, the SONG spectrograph's wavelength coverage does not cover the Ca\,H\&K doublet below 400\,nm (393.366 \& 396.847\,nm). However, we modified an in-house python pipeline designed to measure activity indicators in STELLA spectra \citep{Weber08} to be usable on SONG datasets. The pipeline corrects spectra for RVs, computes fluxes in selected wavelength regions, and compares the flux of the continuum with the flux in selected spectral features to track the changes in line profiles. We considered windows around well-known spectral features (e.g.,\ H$\alpha$ at 656.281\,nm, NaD at 589.592 \& 588.995\,nm, He at 587.564\,nm) and continuum regions from the literature \citep[e.g.,][]{Cincunegui07,Boisse09}, or set them manually. 
We also check the results with CARMENES activity indicators to verify the method. Based on the observed signals and the comparison with CARMENES activity indicators, the H$\alpha$ line appears to be the most reliable stellar activity tracker. 

Additionally, we used archival HARPS-N observations for GJ\,504, which we reduced with the {\tt SERVAL} pipeline \citep{Zechmeister18} to derive RVs and measure different activity indicators (H$\alpha$, CRX, dLW, NaD). We also collected CARMENES RVs for each star, also reduced with {\tt SERVAL}. These observations complement the SONG datasets in spite of the small number of epochs over a short baseline.

\subsection{HD3651}

\begin{figure}[!ht]
\centering
\includegraphics[width=0.51\textwidth, trim= {0.0cm 0.0cm 0.0cm 0.0cm}]{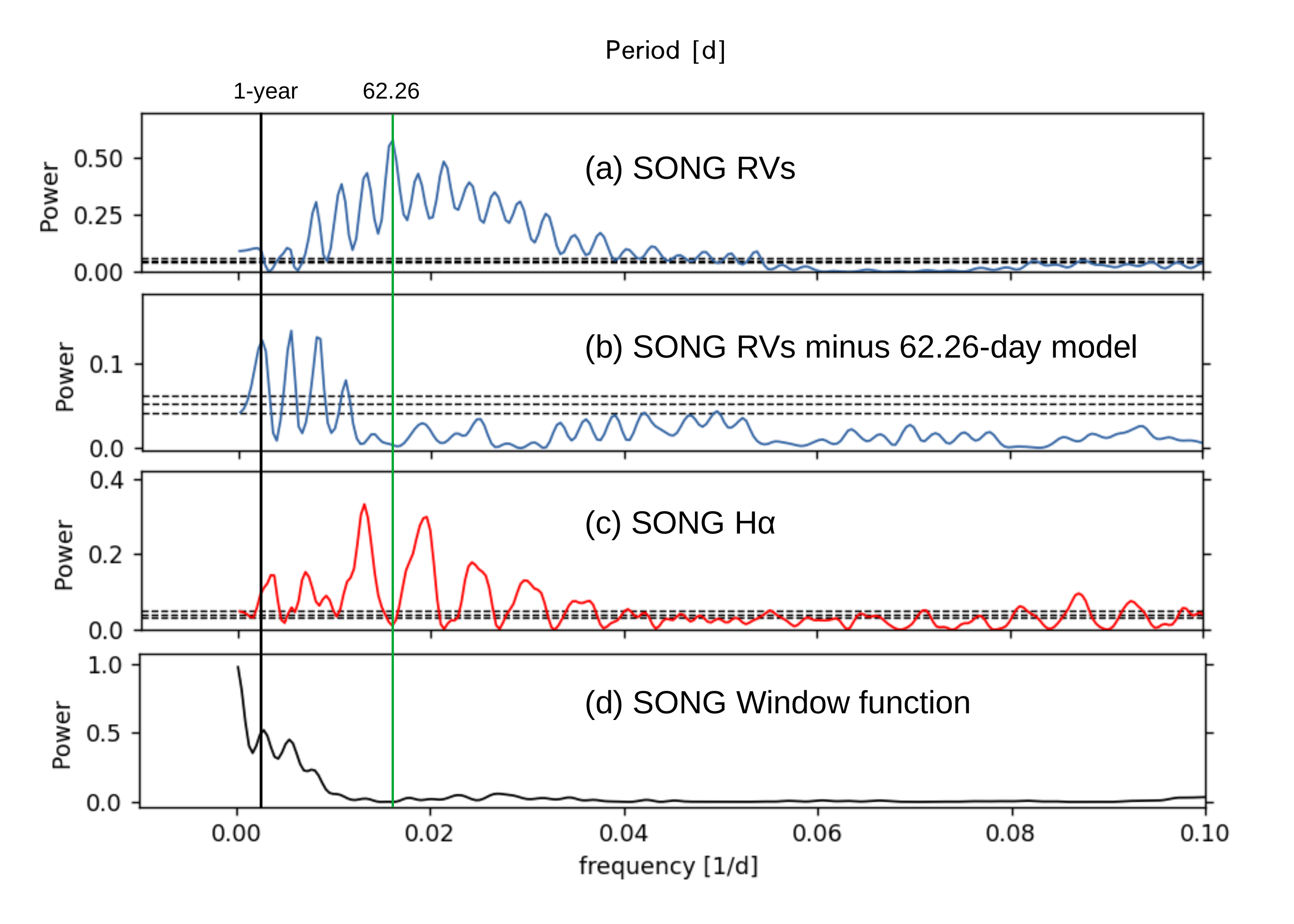}
\caption{Generalized Lomb-Scargle periodograms of the SONG RVs (blue) and the H$\alpha$ activity indicator (red) of HD\,3651:
(a) SONG RVs, (b) SONG RVs minus 62-day model, (c) H$\alpha$. The vertical green line represents the orbital period of the confirmed planet, and the black line the 1-year window function. Horizontal dashed lines show the theoretical FAP levels of 10\%, 1\%, and 0.1\% for each panel.} \label{fig:HD3651b}
\end{figure}

In the SONG periodogram of RVs, we observe the strongest signal at the period of the confirmed planet (see Fig.\ \ref{fig:HD3651b}). The periodogram of RVs shows a forest of peaks with a maximum at $\sim$62 days, consistent with the orbital period of the existing planet \citep{Fischer03,Brewer20}. To explore further the origin of this signal, we investigated the H$\alpha$ activity indicator derived specifically for the SONG dataset, whose periodograms are displayed in Fig.\ \ref{fig:HD3651b}. A similar forest as in the periodogram of RVs is also seen in the periodogram of H$\alpha$. However, there is no peak at $\sim$62 days, and we found only a weak correlation between RVs and H$\alpha$. These peaks can be linked with stellar rotation, which \citet{Fischer03} derived from the Ca II H and K line emission to be $\sim44.5$\,days. Fitting the orbital solution using SONG observations, we obtained a residual RMS of 5.87\,m\,s$^{-1}$, using the CARMENES dataset of 2.32\,m\,s$^{-1}$, and only of 58\,cm\,s$^{-1}$ using the EXPRES dataset \citep{Brewer20}, suggesting a low level of activity.

After subtracting the Doppler reflex motion of this planet, we can see significant peaks close to one year. We interpret this signal as the 1-year window function of the SONG dataset. We used a sinusoidal fit to derive the semi-amplitude of about 3 m/s, much smaller than those discussed below.

\begin{figure}[!ht]
\centering
\includegraphics[width=0.48\textwidth, trim= {0.0cm 0.0cm 0.0cm 0.0cm}]{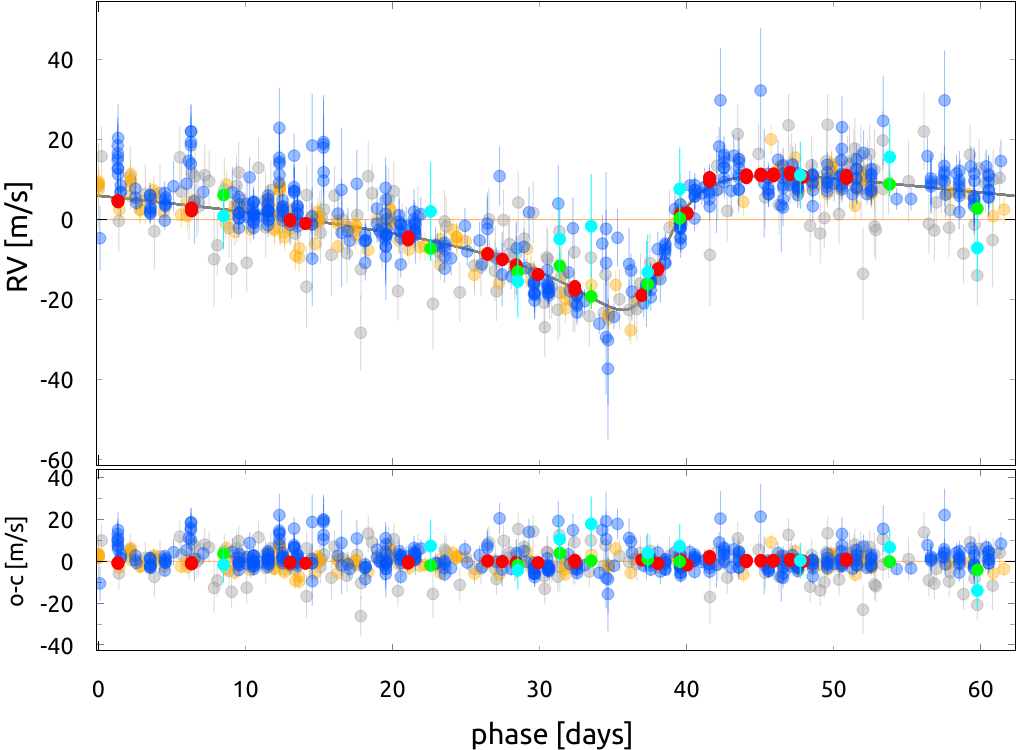}
\caption{Orbital solution for HD\,3651 showing the {\tt Exo-Striker} RV model in black. The orbital solution is derived by simultaneously fitting RVs from LICK (grey), KECK (orange), SONG (blue), VIS CARMENES (green), NIR CARMENES (cyan), and EXPRES (red).} \label{fig:HD3651a}
\end{figure}

We recovered the RV signal with LICK, KECK, SONG, CARMENES and EXPRES using the Monte Carlo-Markov Chain (MCMC) method implemented in the {\tt Exo-Striker} package \citep{Trifonov19}. We set 20 walkers and ran 1,000 burning phase steps and 10,000 MCMC phase steps in our analysis. We consider velocity offsets as the free parameters to fit simultaneously datasets originating from different spectrographs. We summarize the updated planetary parameters in Table \ref{table:hd3651b}. The orbital solution is plotted in Fig.\ \ref{fig:HD3651a} and the correlations between parameters together with the derived MCMC posterior probability distributions are presented in Fig.\ \ref{fig:HD3651-PPD} in the appendix.

In summary, we confirm the presence of a planet in this system. We agree with previous studies that the low levels of stellar activity \citep{Wright04} and photometric variability, along with the low scatter in RV measurements, make it very unlikely that the RV variations are caused by stellar activity. We did not find evidence of an additional companion in the system. \cite{Wittenmyer13} proposed a 2-planet solution for HD3651, with the second planet in 2:1 resonance. We can rule out such a solution and confirm that the solution with a single eccentric planet also leads to the lowest fitted $\chi^2$ value. We also checked if the transit would be visible in the TESS data. Unfortunately, transits are just outside of the observing window in both sectors 17 and 57.

\subsection{GJ\,504}

\begin{figure}[!ht]
\centering
\includegraphics[width=0.48\textwidth, trim= {0.0cm 0.0cm 0.0cm 0.0cm}]{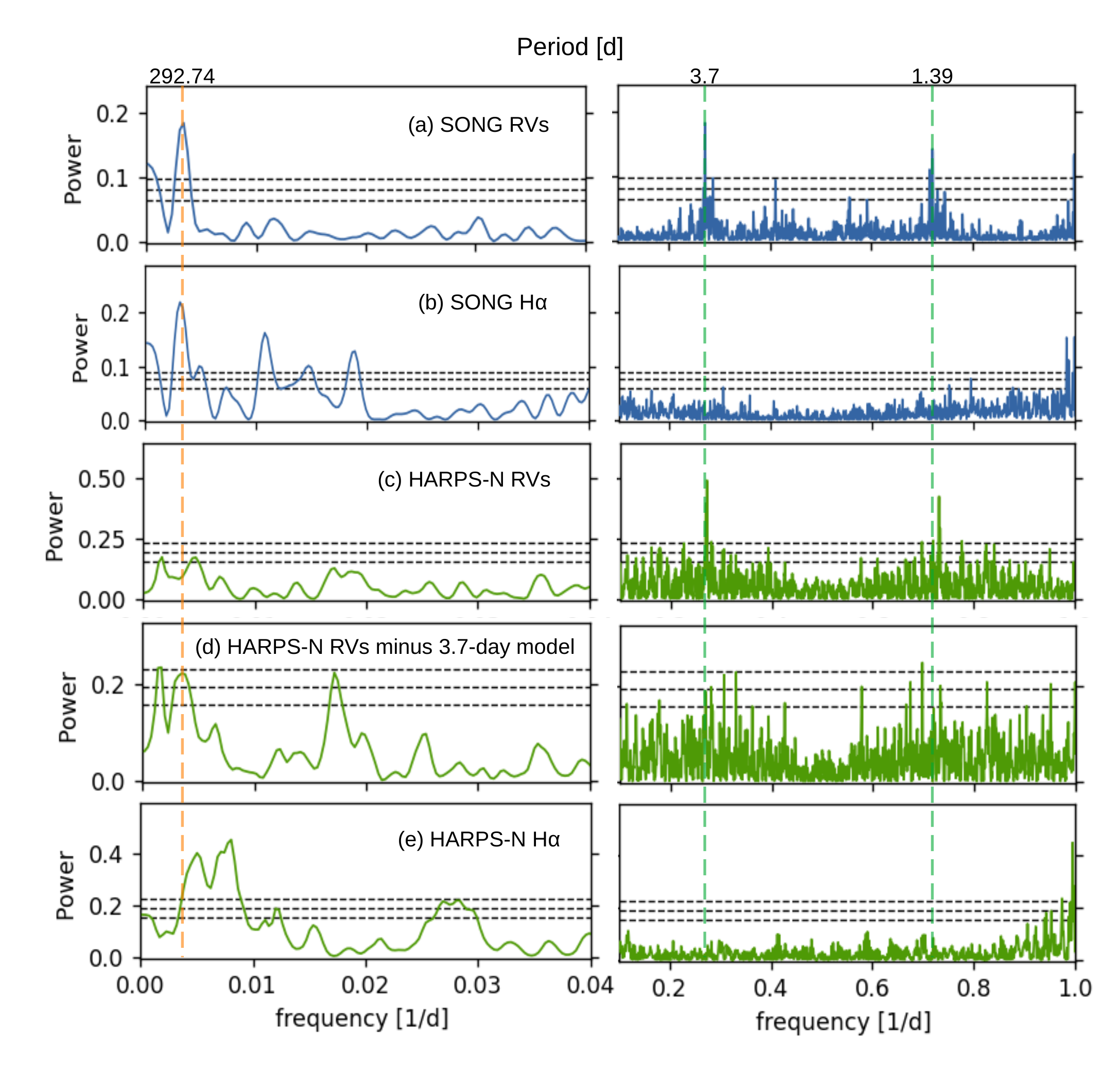}
\caption{GLS periodograms of RVs and activity indicators of GJ\,504: (a) SONG RVs, (b) SONG H$\alpha$ indicator, (c) HARPS-N RVs, (d) HARPS-N RVs after fitting a sinusoid of P\,=\,3.7 days, (e) HARPS-N H$\alpha$ indicator. Vertical green lines represent the stellar rotation period and its 1-day alias. The vertical orange line represents the most significant signal in the SONG RVs at 292.74 days. Horizontal dashed lines show the theoretical FAP levels of 10\%, 1\%, and 0.1\%.} 
\label{fig:GJ504}
\end{figure}

The GLS periodogram of the SONG data shows a significant peak at 3.67 days and another at 1.39 days (Fig.\ \ref{fig:GJ504}), which we assign to the rotation period and its 1-day alias. These two peaks also stand out in the periodograms of the HARPS-N and CARMENES datasets. 
In Section \ref{surface_rotation} we derive a rotation period of 3.4\,days from the TESS LC\@. 

The signal of the rotational period is not the most significant in SONG data. The highest peak is found at $\sim$300 days. We can see a similar peak even in the HARPS-N dataset despite the shorter timescale of observations. 
The short timescale makes the CARMENES dataset unsuitable for interpreting such a long signal; however, the observed trend looks to agree with the period of $\sim$300 days. To investigate this signal in more detail, we used the SONG H$\alpha$ activity indicator together with the HARPS-N H$\alpha$ activity indicator (Fig.\ \ref{fig:GJ504}). In the SONG H$\alpha$ we see the signal at $\sim$300 days, and a similar peak can also be seen in the HARPS-N H$\alpha$, suggesting that this signal is most likely associated with stellar activity.

\begin{figure}[!ht]
\centering
\includegraphics[width=0.46\textwidth, trim= {0.0cm 0.0cm 0.0cm 0.0cm}]{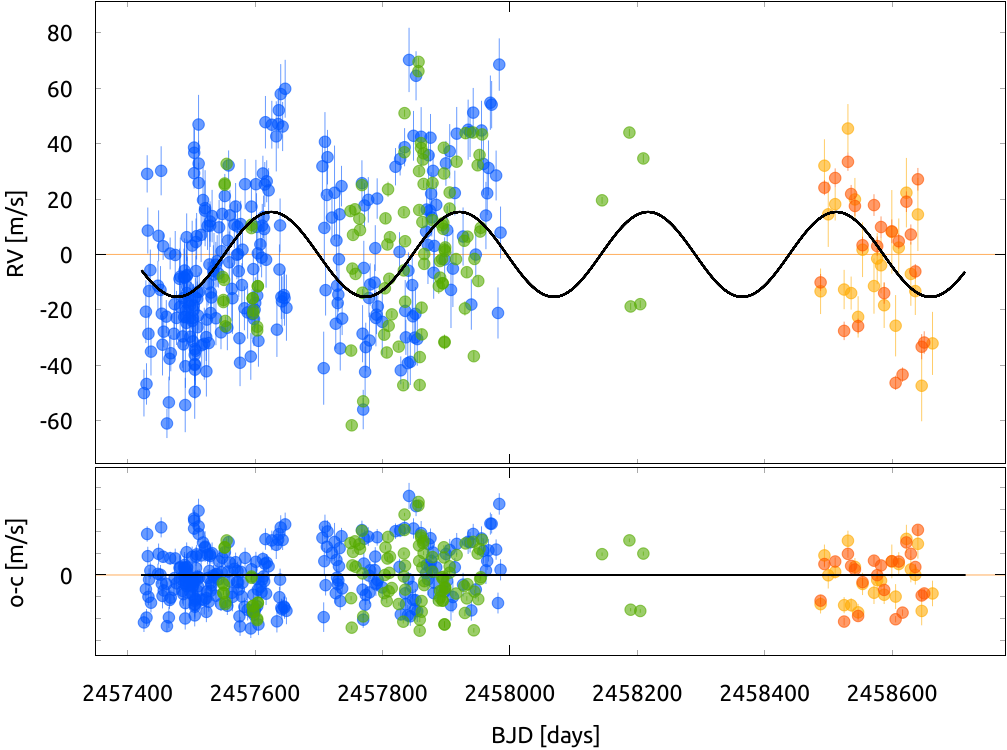}
\caption{SONG RVs of GJ\,504 (blue), HARPS-N RVs (green) and CARMENES RVs (orange) together with the inferred RV model of the 292.74-day signal (solid black line). The nominal error bars are in blue and red, hardly visible for the HARPS-N dataset. Bottom panel: residuals of the RV model.} \label{fig:GJ504_RVs}
\end{figure}

We plot the fit of the signal at $\sim$300 days for SONG, HARPS-N and CARMENES RVs in Fig.\ \ref{fig:GJ504_RVs}. Such a short activity cycle was previously observed for F-type stars (see Section \ref{subsec_hd46588}); however, it was not observed for a G-type star.

\subsection{HN\,Peg}

\begin{figure}[!ht]
\centering
\includegraphics[width=0.48\textwidth, trim= {0.0cm 0.0cm 0.0cm 0.0cm}]{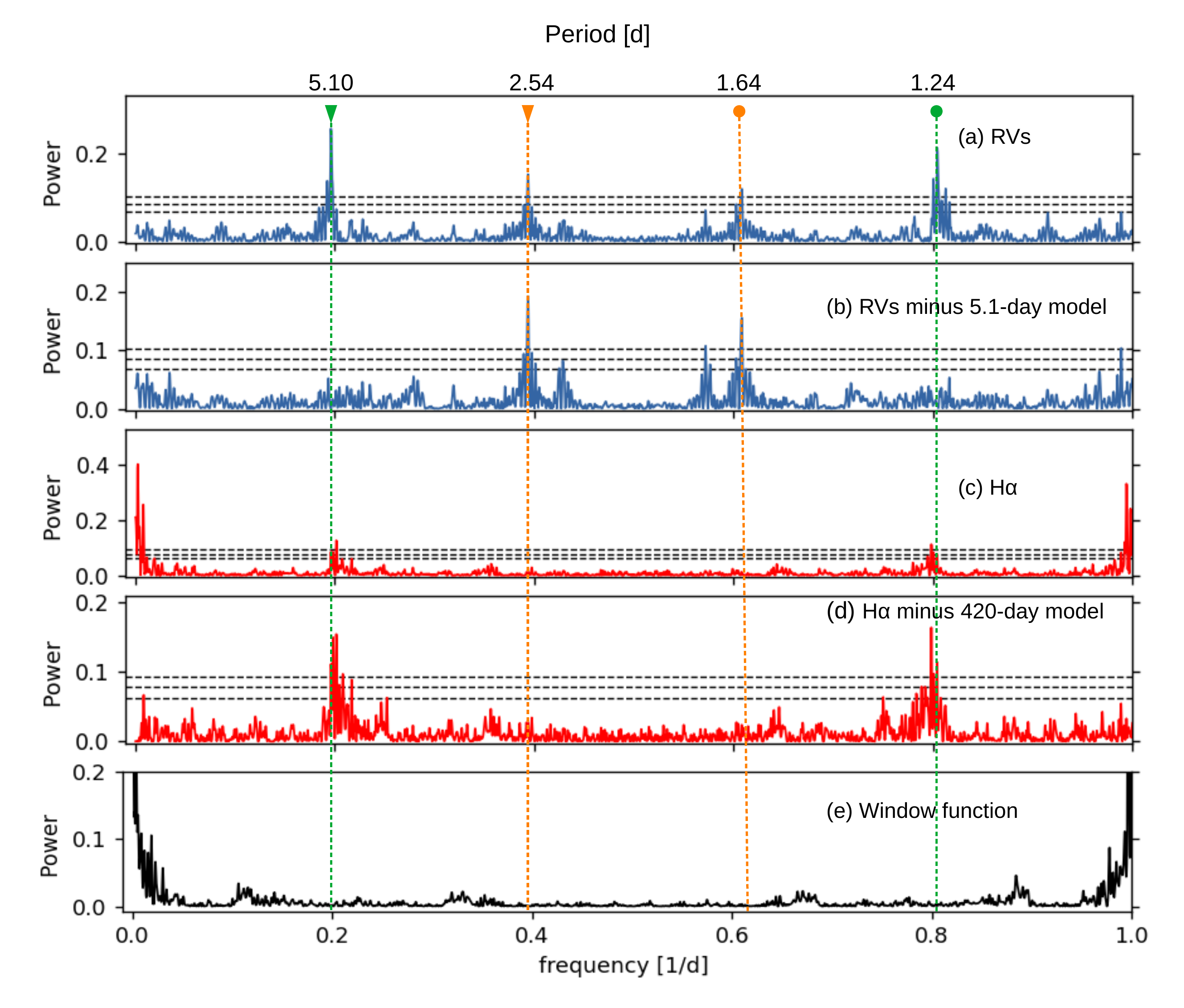}
\caption{GLS periodograms of RVs (blue) and activity indicators (red) of HN\,Peg:
(a) SONG RVs, (b) SONG RVs after fitting sinusoid with a period of 5.1 days, (c) SONG H$\alpha$  indicator, (d) SONG H$\alpha$ indicator after fitting sinusoid with P\,=\,420 days. Vertical green and orange lines represent the star's rotation period and the first harmonic of the rotation with their 1-day alias, respectively. Horizontal dashed lines show the theoretical FAP levels of 10\%, 1\%, and 0.1\% for each panel.} \label{fig:hnpeg}
\end{figure}

The GLS periodogram of the SONG RVs shows the most significant peak at 5.1 days and another peak at 1.24 days (Fig.\ \ref{fig:hnpeg}), which is a 1-day alias of the 5.1-day signal. The periodogram of RVs also shows a peak at 2.54 days and its 1-day alias at 1.64 days. The periodogram of the CARMENES data does not reveal any significant signal.

TESS did not observe this target yet; however, \citet{Messina03} determined the rotation period of the star. They performed photometric observations with three different Automatic Photoelectric Telescopes at Fairborn Observatory in southern Arizona and inferred a variable stellar rotation between 4.59 and 5.17 days. Hence, we interpret the 5.10-day signal in RVs as a rotational period of the star and the 2.54-day signal to be the first harmonic of the rotation period.

We can see the peak at 5.1 days with its 1-day alias in the H$\alpha$ activity indicator. This signal becomes more significant after removing a 420-day signal, which we interpret as systematic due to the window function. We confirm that the 5.1-day signal is linked with stellar activity. To conclude, we interpreted all signals to be linked to the stellar activity and did not find any evidence of a planetary companion.

\begin{figure}[!ht]
\centering
\includegraphics[width=0.46\textwidth, trim= {0.0cm 0.0cm 0.0cm 0.0cm}]{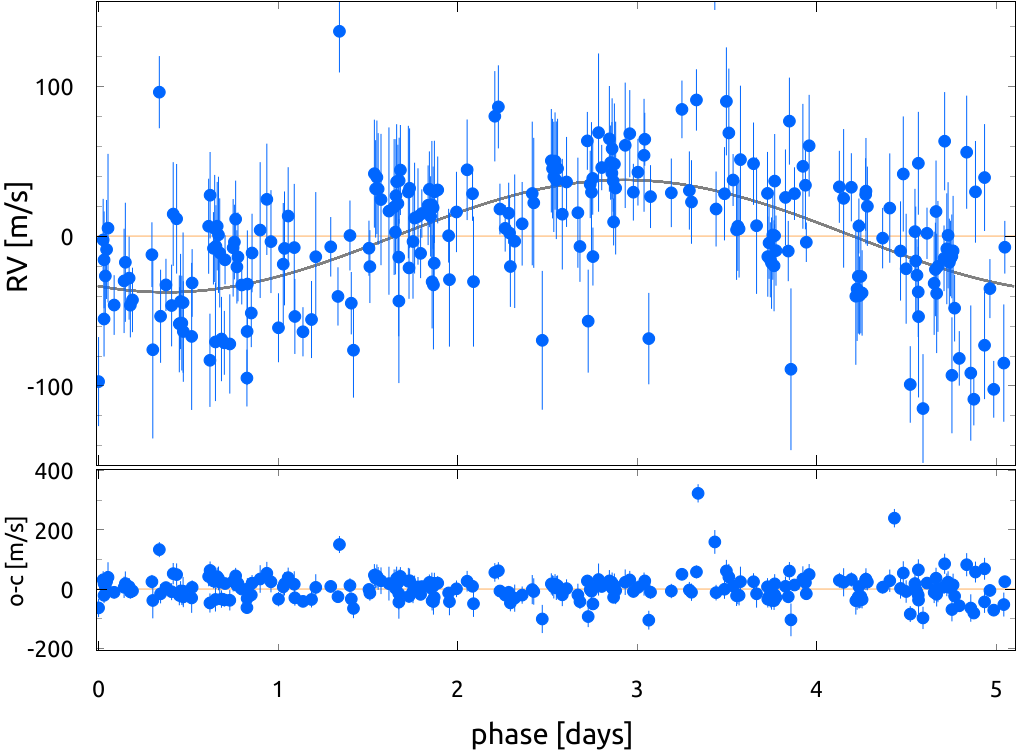}
\caption{SONG RVs of HN\,Peg (blue) together with the inferred RV model of the 5.1-day signal (solid black line). The nominal error bars are in the same colours.} \label{fig:HNPeg_RVs}
\end{figure}

\subsection{HD\,46588}
\label{subsec_hd46588}

\begin{figure}[!ht]
\centering
\includegraphics[width=0.48\textwidth, trim= {0.0cm 0.0cm 0.0cm 0.0cm}]{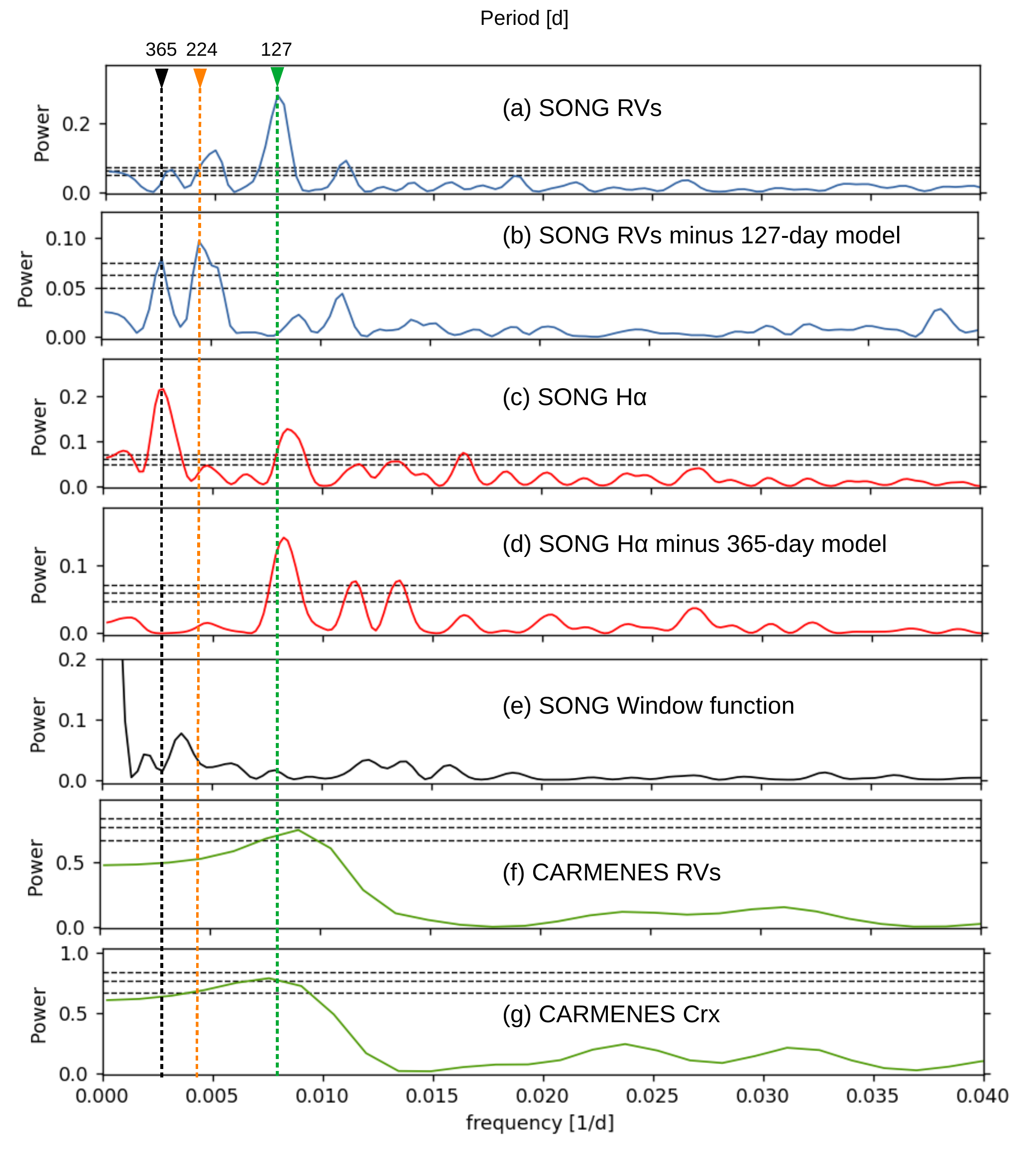}
\caption{GLS periodograms of SONG RVs (blue) and activity indicators (red), and CARMENES RVs and CRX (green) of HD\,46588: (a) SONG RVs, (b) SONG RVs after removing model with the period of 127 days, (c) SONG H$\alpha$ activity indicator, (d) SONG H$\alpha$ indicator after fitting sinusoid with P\,=\,365 days, (e) SONG window function, (f) CARMENES RVs (g) CARMENES chromatic index. The vertical green line represents short-term activity cycle, the orange line represents the possible planetary candidate, the black line represent the 1-year window function. Horizontal dashed lines show the theoretical FAP levels of 10\%, 1\%, and 0.1\% for each panel.} \label{fig:hd46588}
\end{figure}

\begin{figure}[!ht]
\centering
\includegraphics[width=0.46\textwidth, trim= {0.0cm 0.0cm 0.0cm 0.0cm}]{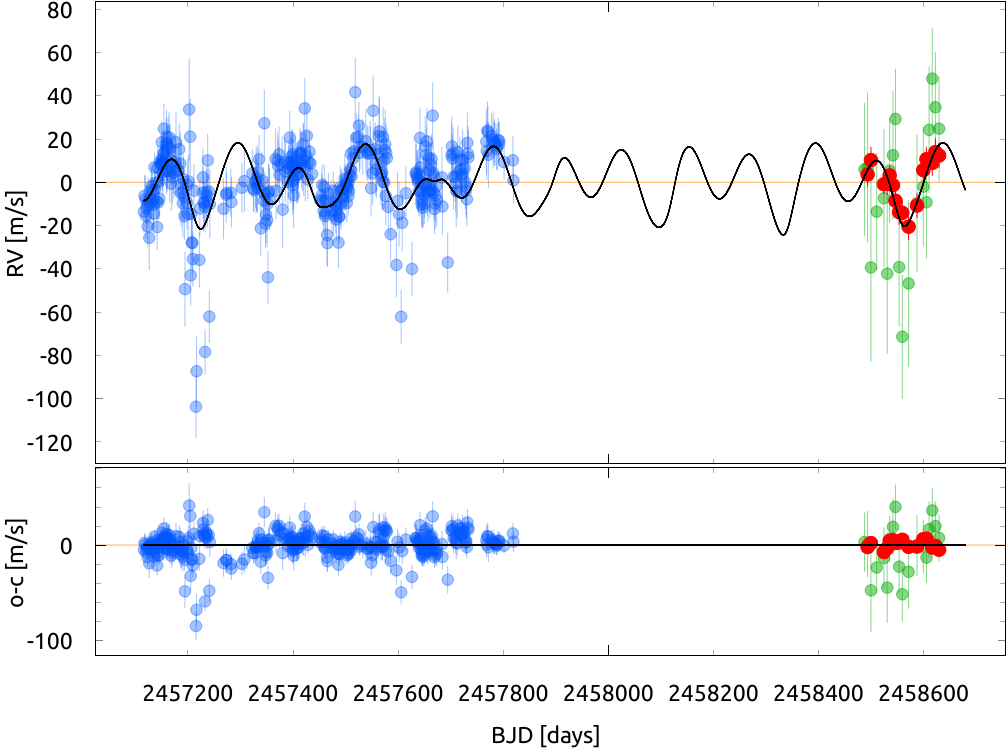}
\caption{SONG RVs of HD\,46588 (blue) and CARMENES RVs (red/green) together with the inferred RV model of the 127-day and 224-day signals (solid black line). The nominal error bars are in the same colours.} \label{fig:HD46588_RVs}
\end{figure}

The most dominant peak in the GLS periodogram of the SONG RVs is at 127 days (Fig.\ \ref{fig:hd46588}). In the CARMENES VIS RVs, we also detect a peak close to 127 days but poorly defined because of the short baseline. We find a strong correlation of 0.79 between the chromatic index (CRX) and the CARMENES RVs at 127 days, suggesting that this peak is associated with stellar activity. This interpretation is corroborated by the SONG H$\alpha$ indicator of the SONG dataset. We interpret this signal as the magnetic cycle of the star. We plot the fit of this signal for the SONG and CARMENES RVs in Fig.\ \ref{fig:HD46588_RVs}.

\begin{figure}[!ht]
\centering
\includegraphics[width=0.46\textwidth, trim= {0.0cm 0.0cm 0.0cm 0.0cm}]{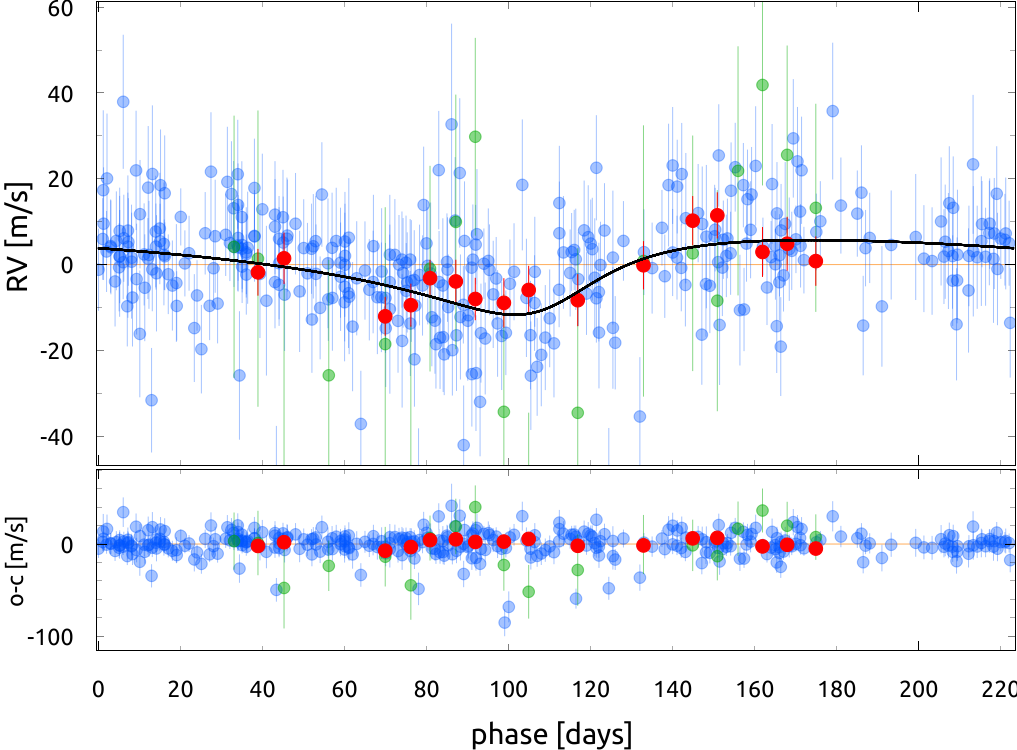}
\caption{The orbital solution for HD\,46588 showing the RV model in black. Blue points represent the SONG RVs and red/green points  represent the CARMENES RVs. The nominal error bars are in the same colours.} \label{fig:HD46588_plnt}
\end{figure}

In the SONG periodogram, we observe two other significant peaks at 224 days and 1 yr. The 1-yr peak is visible in the H$\alpha$ indicator, representing the 1-year window function of the SONG dataset. However, this is not the case for the peak at 224 days; hence, there might be a chance it might have a Keplerian origin. We computed the statistical significance of the peak at $\sim224$ days via the bootstrap randomisation process with $10^6$ realisations \citep{Murdoch93}. We found a false-alarm probability lower than one per cent suggesting that it is unlikely that this peak is due to random noise. However, we cannot say whether this signal is due to a planetary companion or connected to the stellar magnetic cycle, as the SONG data does not always behave as the simple sinusoidal model predicts, making the magnetic activity more complex. Further monitoring is needed to understand this system better; however, we report this system to have a possible planetary candidate. Fitted planetary parameters are summarized in Table \ref{table:hd3651b}. The fit was performed on the RV residuals after removal of the activity cycle with a period of 127 days. The orbital solution is plotted in Fig. \ref{fig:HD46588_plnt} and the correlations between parameters together with the derived MCMC posterior probability distributions are presented in Fig.\ \ref{fig:HD46588-PPD} in the appendix. We also performed a multidimensional GP approach \citep{Rajpaul15} to characterise the stellar and planetary signals in the SONG data using the PYANETI code \citep{Barragan22}. We used the H$\alpha$ indicator to constrain the stellar signal. We set wide priors for planet parameters based on the periodogram and the initial fit (see Figure \ref{fig:HD46588_plnt}). We found planetary parameters consistent in $1\sigma$ with the previous fit. Derived planetary parameters are summarized in Table \ref{table:hd3651b}. The plot of H$\alpha$ and RV time series and inferred stellar and planetary models is shown in Figure \ref{fig:gp_hd46588}. We can see that the stellar+planetary model can describe the observed RVs quite well.


HD\,46588 is extremely interesting in terms of stellar activity. The Sun has a 22-yr magnetic cycle. However, the physics underlying this phenomenon is still not well understood, similarly to activity cycles in other Sun-like stars. Long-term observations and monitoring of stars' chromospheric activity can be used to study activity cycles and determine their periods \citep{Wilson78,Baliunas95,Hall07,Isaacson10,Hempelmann16}. \cite{Baliunas95} measured activity cycles for more than 40 stars, with periods ranging between 2.5 and 20 years.

\cite{Baliunas95} observed the star $\tau$Boo (HD\,120136), which first appears to have a typical chromospheric cycle of 11.6$\pm$0.5\,years. $\tau$Boo is an F7 spectral type star with an M2 type companion on a close orbit of 3.3 days. The rotation period is determined to be close to this value, indicating synchronization. However, what makes this star interesting is evidence of additional variability in the Ca{\small{II}} line with a significantly shorter period than the original chromospheric cycle. \citet{Baliunas97} determined a period of 116 days, later confirmed by \citet{Mengel16} and \citet{Mittag17} investigating the S-index with the NARVAL and TIGRE instruments, respectively. \citet{Mittag17} also showed that the X-ray data support a periodicity of about 120 days, addressing an interesting question: if the fast activity period of $\tau$Boo is representative of main-sequence F stars? \citet{Mittag19} used analysis of S-index time series of F-type stars taken with the TIGRE telescope to discuss this question. They detect three more short-term cycles and one candidate between 0.5--1 year. They do not provide the number of investigated F-type stars, preventing us from discussing the statistics. However, $\tau$Boo is still the star with the fastest magnetic cycle ever observed.

In this context, HD\,46588 is a twin of $\tau$Boo. The star has the same F7 spectral type and hosts an L9 type companion at the projected physical separation of 1420\,au \citep{Loutrel11}. We found evidence for a chromospheric cycle at 127 days. New discoveries indeed support that short-term activity cycles can be typical in F-type stars. Furthermore, we derived the age of HD\,46588 about 2.5\,Gyr, which indicates that such short-term activity cycles can cause strong RV variations also for relatively old stars. HD\,118865 is the other F7 star in our sample. Our observations are only about 150 days for HD\,118865, and we observe just a linear trend in RVs (Section \ref{subsec_hd118865}). Further monitoring of F stars will bring more light to the discussion. However, the population of such stars with short-term activity cycles is starting to grow. 

\begin{table}
 \centering
 \caption[]{Planetary parameters.
 }
 \label{table:hd3651b}
\scalebox{0.9}{
 \begin{tabular}{@{\hspace{0mm}}l c c c@{\hspace{0mm}}}
 \hline
 & HD\,3651 & HD\,46588 & HD\,46588 (GP) \cr
 \hline
 $msini$ ($\ensuremath{\,{\rm M_J}}$) & $0.24\pm0.01$ & $0.25\pm_{-0.04}^{+0.06}$ & $0.30\pm0.04$ \cr
 $K$ (m/s) & $16.8\pm0.2$ & $8.6_{-1.4}^{+2.2}$ & $10.2\pm1.3$ \cr
  $T_0$ (days) & $2457029.7\pm0.3$ & $2457171_{-11}^{+10}$ & $2457165_{-2.0}^{+1.5}$ \cr
 $P$ (days) & $62.247\pm0.002$ & $223\pm3$ & $223\pm3$ \cr
$e$ & $0.624\pm0.005$ & $0.42_{-0.14}^{+0.19}$ & $0.35\pm0.05$ \cr
$\omega$ (deg) & $240.5\pm1.2$ & $218_{-18}^{+21}$ & $212_{-14}^{+13}$ \cr
 \hline
$P_{activity}$ (days) & - & $122.0\pm0.4$ & - \cr
$K_{activity}$ (m/s) & - & $13\pm1$ & - \cr
$P_{GP}$ (days) & - & - & $124_{-3}^{+4}$ \cr
$\lambda_P$ & - & - & $2.3_{-0.4}^{+0.5}$ \cr
$\lambda_e$ (days) & - & - & $52_{-6}^{+12}$ \cr
 \end{tabular}
}
\end{table}

\subsection{HD\,203030}
\label{subsec_hd203030}

In the periodogram of the TESS LC, we observed two strong peaks at 6.5 and 13 days. Typically we would attribute peaks to be the rotation period with the first harmonic. However, visual investigation of the LC reveals strong variations at 6.5 days and the gap in the data so that the middle period is missing. Hence, the 13-day peak is an artefact of data sampling. We can confirm such a hypothesis with the vsini value derived from spectra. Assuming an inclination of 90 degrees, we can derive the upper limit for the rotation period, which is 7.2 days, close to 6.5 days but much lower than 13 days.

\begin{figure}[!ht]
\centering
\includegraphics[width=0.48\textwidth, trim= {0.0cm 0.0cm 0.0cm 0.0cm}]{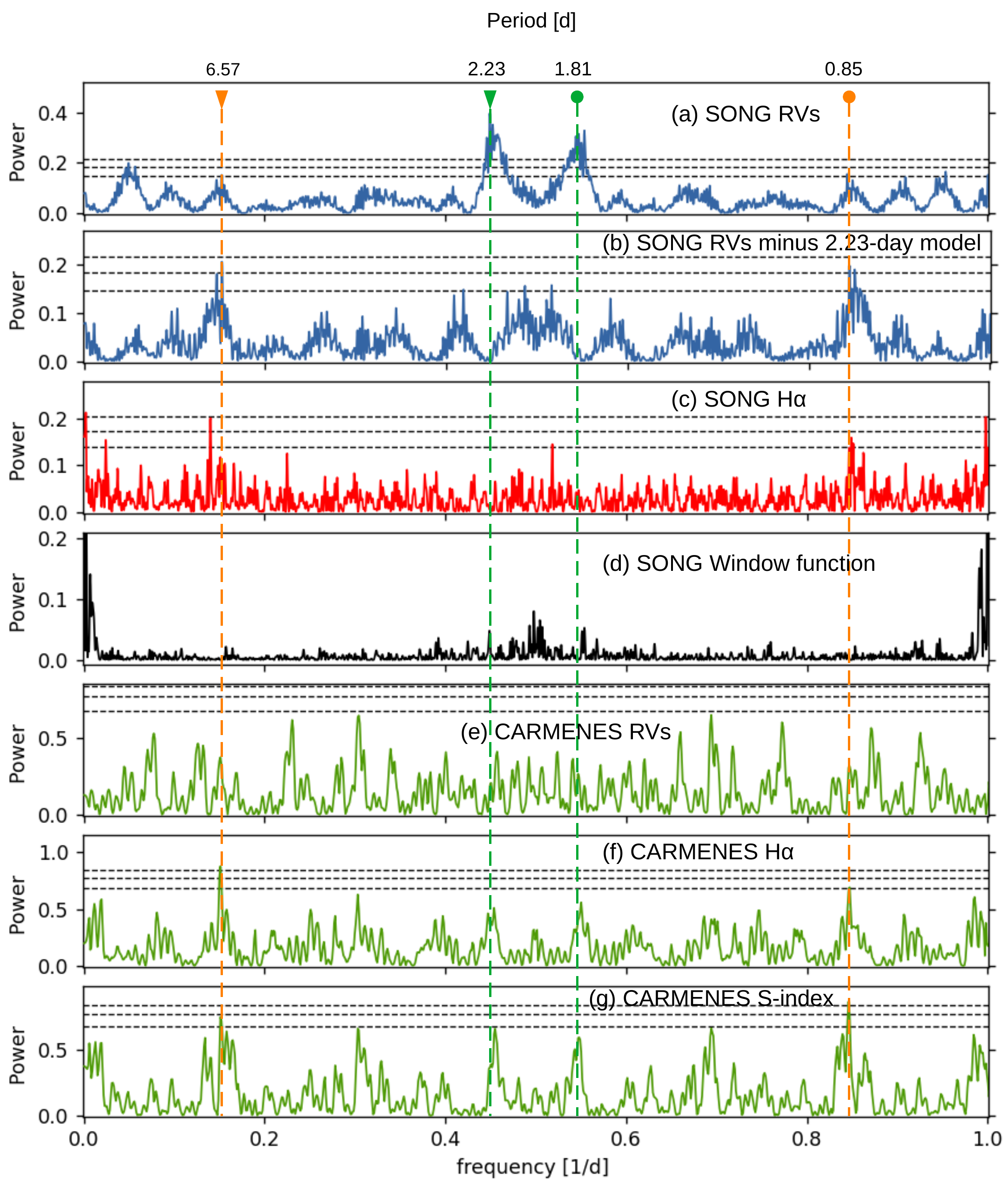}
\caption{GLS periodograms of SONG RVs (blue) and H$\alpha$ activity indicator (red) of HD\,203030: (a) SONG RVs, (b) SONG RVs after fitting sinusoid with P\,=\,2.23 days, (c) SONG H$\alpha$ indicator. Vertical orange lines represent the star's rotation period and its 1-day alias, and green lines represent the planetary candidate of the rotation period with its 1-day alias. Horizontal dashed lines show the theoretical FAP levels of 10\%, 1\%, and 0.1\% for each panel.} \label{fig:HD203030}
\end{figure}

The GLS periodograms for the SONG RVs show a significant peak at 2.23 days, with its 1-day alias at 1.81 days. After fitting this signal with a simple sinusoidal, we observe a peak at 6.5 days, representing the stellar rotation. We found an RMS scatter due to activity to be about 35 m\,s$^{-1}$, higher than the semi-amplitude of signals in the periodogram of HD\,203030 and much higher than 5.87 m\,s$^{-1}$ inferred for the quiet star HD\,3651\@. Hence, we conclude that stellar activity plays a dominant role in RVs, and the 2.2-day signal is probably just the second harmonic of the rotation period. In the periodogram of CARMENES RVs, we do not see any significant peak. SONG H$\alpha$ indicator reveals a signal at 7.14 days, while the CARMENES H$\alpha$ and S-index indicators both reveal a signal at 6.62 days, confirming the star's rotation period. We can also see a peak close to 2.23 days in S-index. All in all, we report that the variations in RVs are probably associated with stellar activity.

\begin{figure}[!ht]
\centering
\includegraphics[width=0.46\textwidth, trim= {0.0cm 0.0cm 0.0cm 0.0cm}]{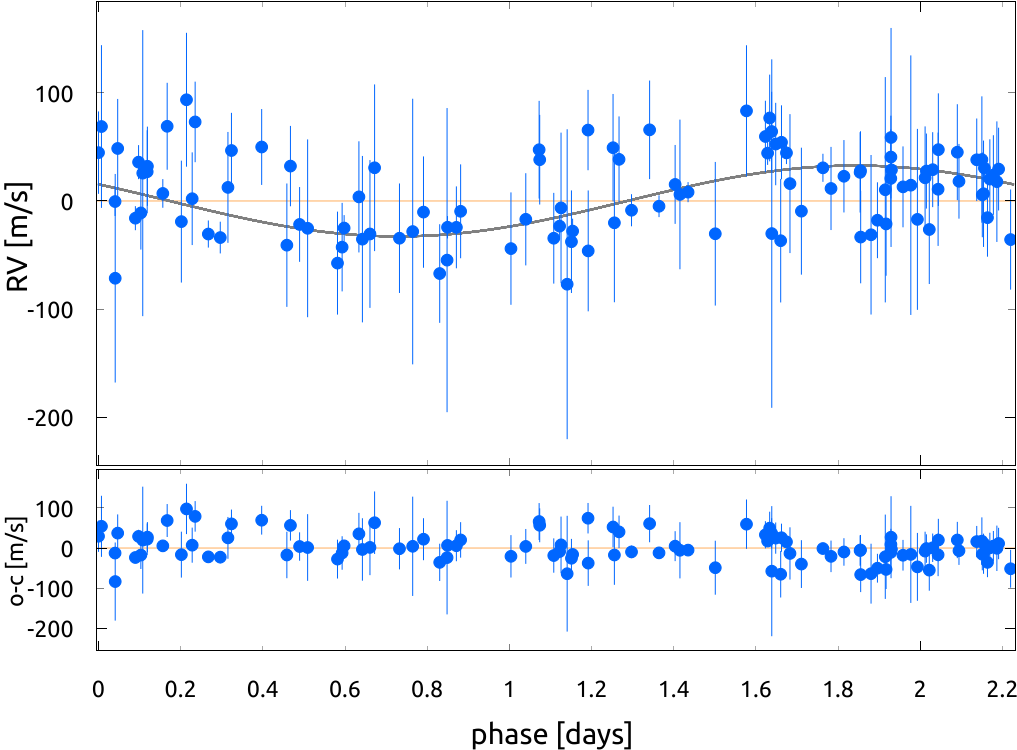}
\caption{SONG RVs of HD\,203030 (blue) together with the inferred RV model of the 2.23-day signal (solid black line). The nominal error bars are in the same colours.} \label{fig:HD203030_RVs}
\end{figure}

\subsection{HD\,118865}
\label{subsec_hd118865}

\begin{figure}[!ht]
\centering
\includegraphics[width=0.48\textwidth, trim= {0.0cm 0.0cm 0.0cm 0.0cm}]{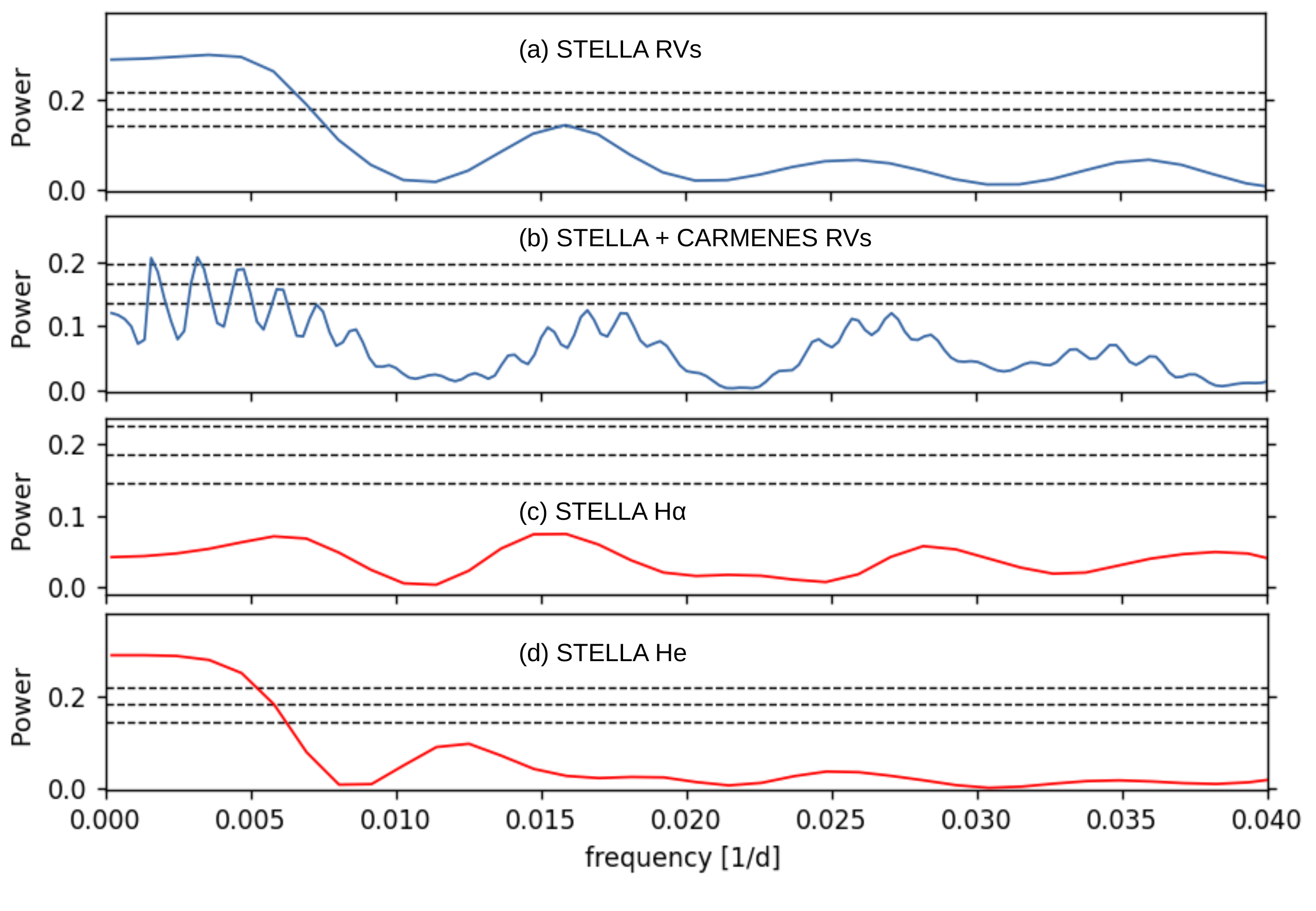}
\caption{GLS periodograms of STELLA RVs (blue) and the He activity indicator (red) of HD\,118865: (a) STELLA RVs, (b) STELLA He activity indicator. Horizontal dashed lines show the theoretical FAP levels of 10\%, 1\%, and 0.1\% for each panel.} \label{fig:HD118865_per}
\end{figure}

\begin{figure}[!ht]
\centering
\includegraphics[width=0.48\textwidth, trim= {0.0cm 0.0cm 0.0cm 0.0cm}]{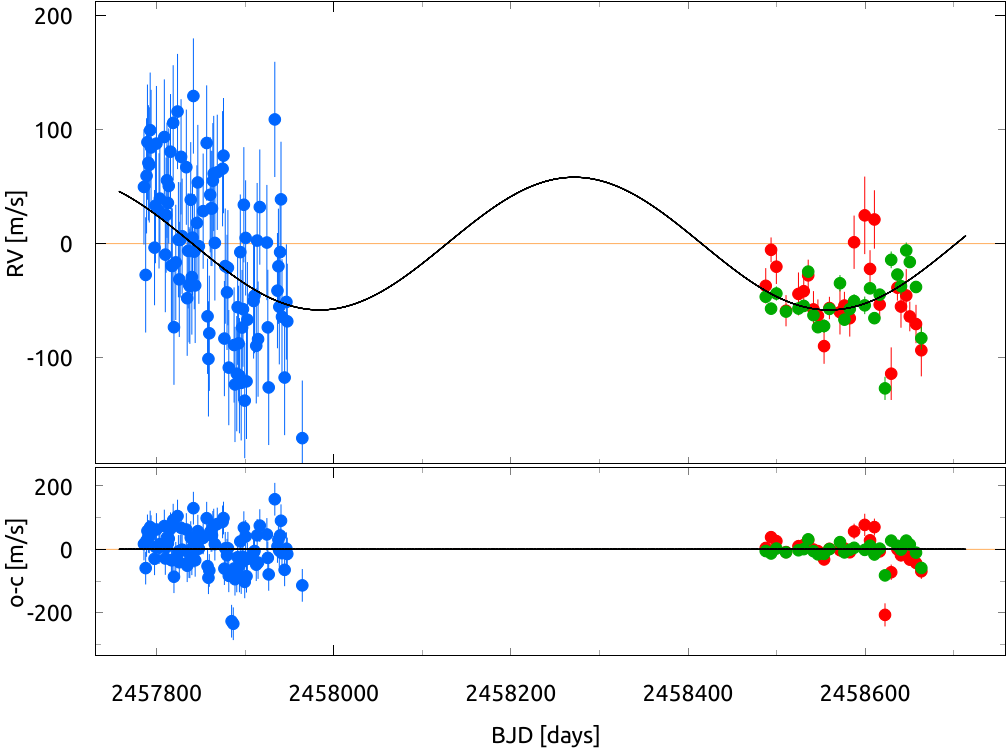}
\caption{STELLA RVs (blue), CARMENES VIS RVs (green), and CARMENES NIR RVs (red) of HD\,118865 together with the inferred RV model of the $\sim$590-day signal (solid black line). The nominal error bars are blue, red, and green. Bottom panel: residuals of the RV model.} \label{fig:HD118865}
\end{figure}

In the STELLA RVs, we observe a decreasing trend, which is not observed in the CARMENES dataset. It leads to a variety of possible solutions. In Fig.\ \ref{fig:HD118865} we illustrate one solution at a period of $\sim$575 days. We observe a similar trend in the STELLA He activity indicator (He I line at 587.56\,nm) (Fig.\ \ref{fig:HD118865_per}). Hence, we prefer the stellar activity scenario over the planetary companion even though the relatively large semi-amplitude and the fact that the star has an age of about 1\,Gyr or older. The star has an F spectral type, and as we discussed above, the magnetic cycle with a period of hundreds of days would not be surprising even for relatively old stars. Further spectroscopic monitoring is needed to constrain the period of this signal.

\section{Discussion}\label{sec:discussion}

\subsection{The role of wide companions on planet formation and evolution}

Wide stellar companions can significantly affect the physical parameters of inner planets. Peculiarities in their parameter distributions were predicted \citep[e.g.,][]{Mayer05,Kraus12,Kaib13} and observed \citep[e.g.,][]{Zucker02,Eggenberger04,Fontanive21}. \cite{Eggenberger04} compared orbital period, minimum planetary mass, and eccentricity for a sample of planets in systems with and without a wide stellar companion. They discussed two possible peculiarities identified in the distributions of short periodic planets. First, planets with periods below 40 days appear to have lower eccentricities when part of multi-stellar systems compared to planets around single stars. Second, the most massive short-periodic planets in their sample all belong to multi-stellar systems. However, at that time, only about a dozen multiple stellar systems with an exoplanet were known. \cite{Fontanive21} performed an extensive search for visual co-moving binary companions to exoplanet-host stars using $Gaia$ DR2 data in addition to known systems from literature. They compiled 938 stars, from which 186 are binaries, and 32 belong to hierarchical triples. They used all systems to study peculiarities in their parameter distributions similarly to \cite{Eggenberger04}. Specifically, they studied the mass and RV semi-amplitudes of planets, masses of primaries, and masses and separations of wide companions in these systems. To highlight some of their results, they found that (i) more massive planet-host stars are more often in multi-stellar systems, (ii) only close binaries ($<$\,1000\,au) seem to influence the formation or evolution of inner planets and (iii) the mass of the stellar companions has no significant effect on planetary properties.

We did not find any planets in our sample of stars with wide BD companions (only one planetary candidate in HD\,46588), so we used a different approach to discuss possible peculiarities in parameter distributions. We downloaded the catalogues of systems of single planet-host stars and planet-host stars with wide stellar binaries from \cite{Fontanive21} to produce similar plots as in \cite{Eggenberger04} and \cite{Fontanive21}. We also consider the planet's eccentricity distribution as it was not discussed in \cite{Fontanive21}. To these plots, we added for comparison the HD\,3651 planet, HD\,46588b planetary candidate and the four planets from systems with known wide BD companions from the literature (red points in Fig.\ \ref{fig:family_mass}) discussed in detail in Appendix \ref{sec:family}. We excluded HD\,65216, HD\,41004 and $\epsilon$\,Indi, as these systems have different architectures. We also excluded GJ\,229 because we are sceptic about the planetary nature of the reported signals. 

To better understand our data, we may firstly compare the separations and masses of wide BD companions from the sample of the systems with known planets (discussed in Appendix \ref{sec:family}) and our sample of five systems (we put HD\,3651 to the first group). Both samples are small; however, we did not notice any crucial differences. The first sample spans separations from 3\,au up to 9000\,au (3, 23, 476, 2500, 9000) while the second from 487\,au to 9200\,au (487, 784, 1420, 9200). With the current mass uncertainties, both samples practically span companion masses through the whole range typical for BDs. 

The plot of the minimum planetary mass versus the orbital period of planets can be seen in Fig.\ \ref{fig:family_mass}. We also divided systems of single planet-host stars into two categories: Group\,1 contains systems with one planet orbiting a star (black points in Fig.\ \ref{fig:family_mass}), and Group\,2 is made of systems with multiple planets orbiting a star (grey points)\footnote{We need to note, that such division reflects our current information about these systems. For example, systems from Group\,1 can be, in fact, members of Group\,2 with an undetected additional companion(s).}. Similarly, we divided multi-stellar systems into two categories: Group\,3 contains systems with one planet and a stellar companion orbiting a primary star (blue points in Fig.\ \ref{fig:family_mass}), and Group\,4 is made of systems with a stellar companion and multiple planets (green points). The final sample consists of 504 systems in Group\,1, 479 systems in Group\,2, 173 systems in Group\,3, and 103 systems in Group\,4\@. 

\begin{figure}[!t]
\centering
\includegraphics[width=0.48\textwidth, trim= {0.0cm 0.0cm 0.0cm 0.0cm}]{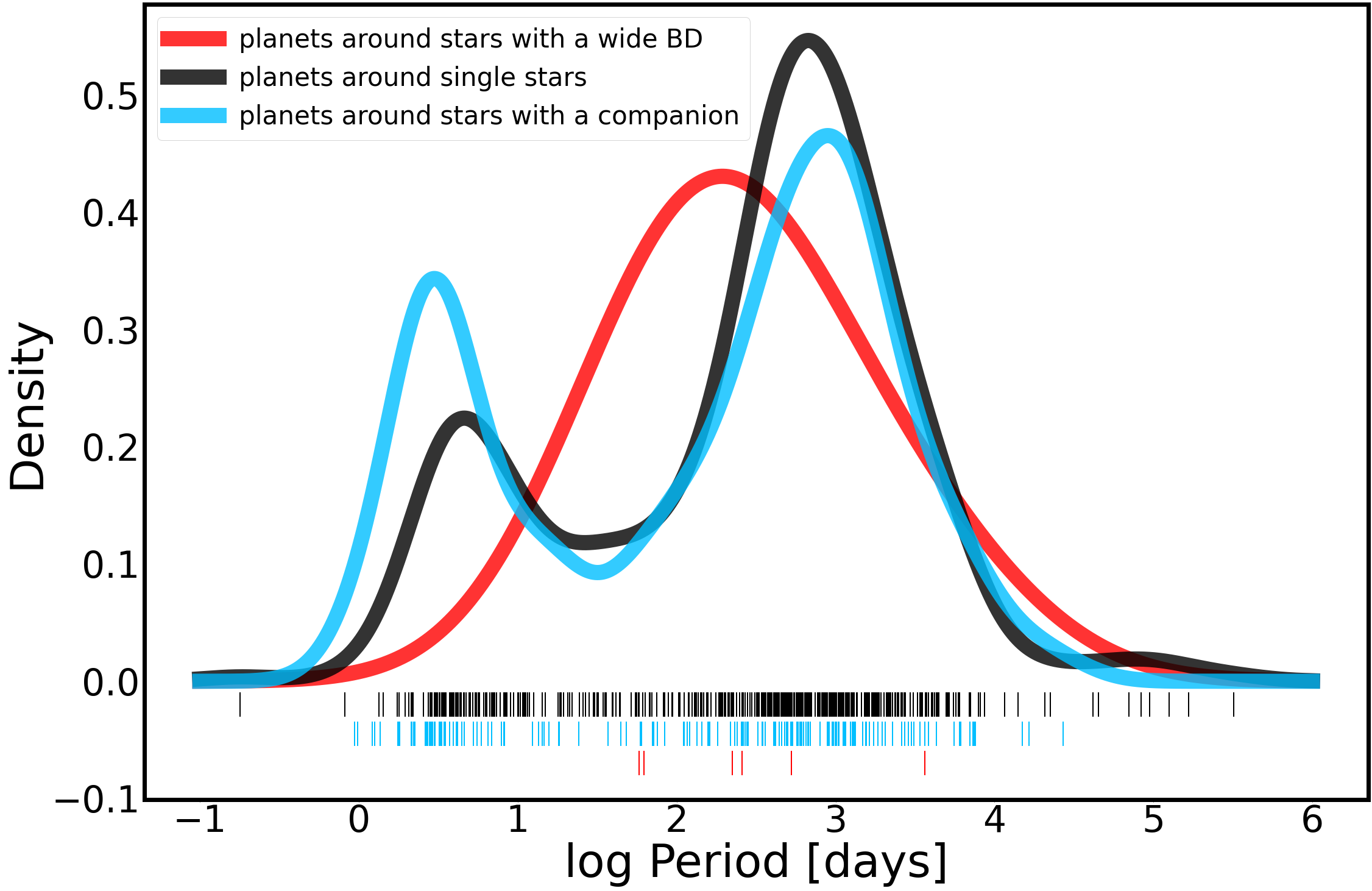}
\caption{KDEs of planetary orbital period comparing planets around single stars (black), planets around stars with stellar companion (blue), and planets around stars with BD companion (red).} \label{fig:KDE_period}
\end{figure}

\begin{figure}[!t]
\centering
\includegraphics[width=0.48\textwidth, trim= {0.0cm 0.0cm 0.0cm 0.0cm}]{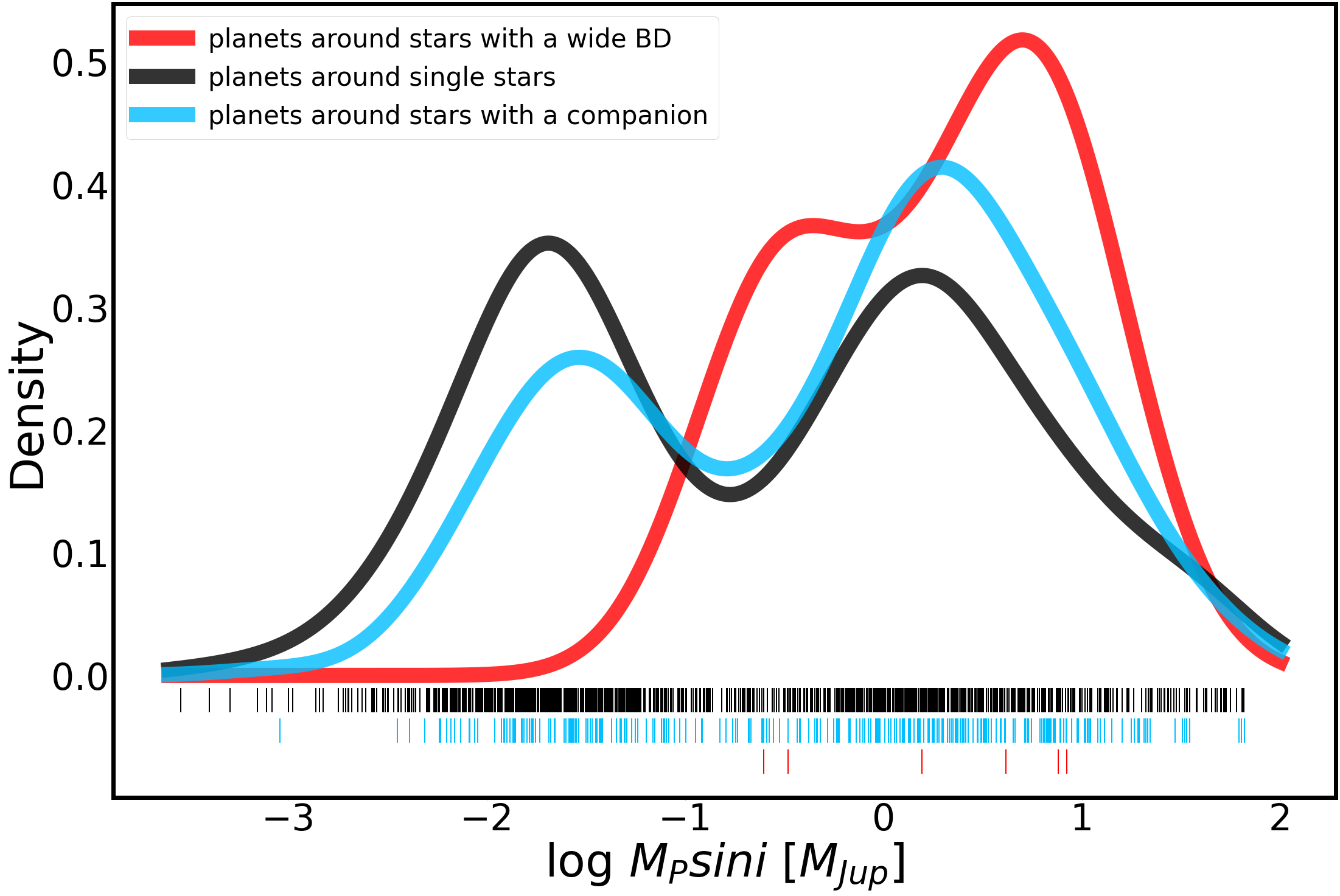}
\caption{KDEs of minimum planetary mass comparing planets around single stars (black), planets around stars with stellar companion (blue), and planets around stars with BD companion (red).} \label{fig:KDE_mass}
\end{figure}

\begin{figure*}[!ht]
\centering
\includegraphics[width=0.9\textwidth,height=0.8\textwidth]{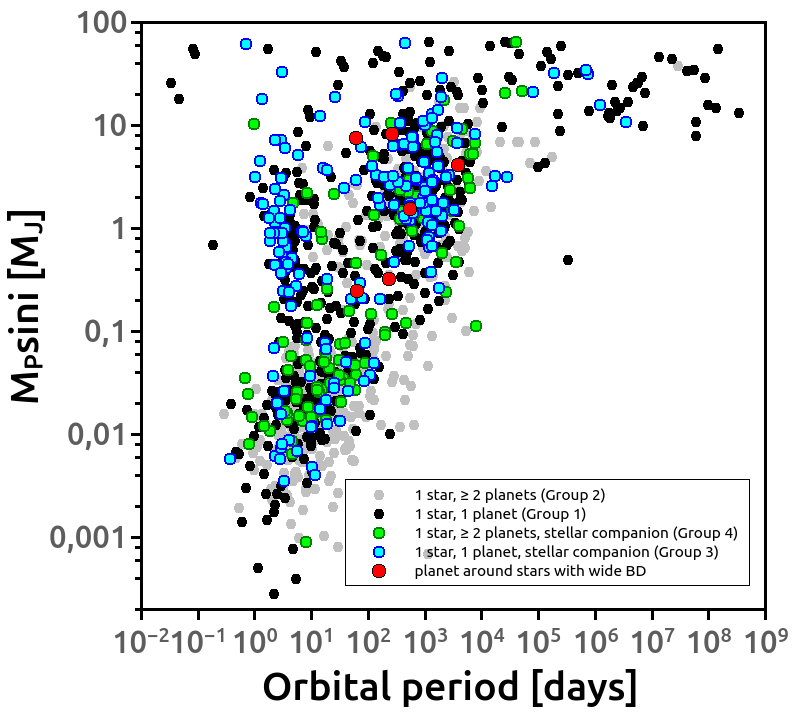}
\caption{Planetary minimum mass versus orbital period for the known single planet-host stars (black and grey points), planet-host stars with one planet and one stellar companion (blue points), planet-host stars with one stellar companion but more than one planet (green points), and finally, planet-host stars with wide BD companion (red points).} 
\label{fig:family_mass}
\end{figure*}

We do not see many planets more massive than $0.1\,M_J$ from Group\,2 and Group\,4 at close orbital periods. Multi-planetary systems with a planet's orbital periods below $\sim100$\,days mostly host less massive planets. The inward migration of more massive planets would disrupt the orbits of inner planets, which would lead to collisions or their ejection \citep{Mustill15,Saffe17}. We can also see that the upper end of the minimum planetary mass actually represents the population of BDs. While this paper focuses on the wide BD companions, here, we can also see BDs in closer orbits. Most of the systems in this mass region are from Group\,1 and Group\,3; hence, no additional planets are detected, suggesting that close BDs might prevent their existence. However, one would need to check how many of these systems were spectroscopically followed up with sufficient precision to detect planets. We also observe that the most distant companions in the plot are BDs, which is not so surprising because they are brighter than planets, hence, easier to detect by direct imaging.

As the majority of planets in systems with wide BDs lies in the mass region of Jovian planets between $0.1-15\,M_J$ \citep{Mordasini18,Spiegel2011}, we took only planets from this region to create kernel density estimates (KDEs) of the distributions of planet period. We chose a Gaussian kernel and used the cross-validation approach to derive the KDE bandwidths providing good insights into potential underlying trends. The only planetary system with a wide BD that does not lie in this mass region is GJ\,229, and we reported the RV signal in GJ\,229 as a planetary candidate rather than a confirmed planet (Appendix \ref{sec:family}). We created the KDEs for three groups: planets in systems with a one star (Group\,1\,+\,Group\,2), planets in systems with multiple stars (Group\,3\,+\,Group\,4), and planets in systems with known wide BD companions. The distributions of the planetary period can be seen in Fig. \ref{fig:KDE_period}. We observe the bimodal distribution of giant planets in the first two groups, which have two maxima around $\sim$\,5\,days and $\sim$\,1000\,days. However, both maxima are slightly shifted, and we observe an enhancement of short periodic planets in systems with a wide stellar companion. Giant planets are thought to form via core accretion in a proto-planetary disk at larger distances from stars than are often observed. Subsequent inward migration thus plays an important role in the existence of planets across a large variety of observed orbital periods \citep[e.g.][]{Lin96,Bodenheimer01}. Hence, enhanced migration seems to work for a large number of short-period planets. We do not observe such bimodal behaviour in the distribution of planets in the systems with a wide BD, as no short periodic planet with an orbital period below ten days is known so far. Our survey is sensitive to such planets, so we should have easily detected them (Section \ref{freq_p}). To investigate the significance of these results, we performed two-sided Kolmogorov-Smirnov tests comparing each sub-population of planets using the python package {\tt Scipy}, specifically the {\tt scipy.stats.ks\_2samp}\footnote{\url{https://docs.scipy.org/doc/scipy/reference/generated/scipy.stats.ks_2samp.html}} function. In other words, we tested the null hypothesis that the samples are drawn from the same distribution and used a p-value to support or reject this hypothesis. 
Comparing the distribution of planets around single stars with planets around stars with a stellar companion, we obtained the p-value of 0.029. However, comparing the group of planets around stars with a wide BD with planets around single stars, we obtained a p-value of 0.59, and with planets around stars with a stellar companion, we obtained a p-value of 0.41. Hence, we cannot confirm that the group of planets around stars with a wide BD follows its own distribution.


The distribution of minimum planetary mass can be seen in Fig. \ref{fig:KDE_mass}. Here we used all planets and divided them into the same three groups. Except for the population of planets around stars with a wide BD, we observe the bimodal distributions with two maxima around $\sim\,0.01\,M_J$ and $\sim\,1\,M_J$. Using the two-sided Kolmogorov-Smirnov test, we obtained the p-value of 0.08 comparing the planets around stars with a BD companion to planets around single stars and the p-value of 0.34 comparing the planets around stars with a BD companion to planets around stars with a stellar companion. It can suggest that a wide BD influences the mass distribution of planets in a similar way as a wide stellar companion, and these planets follow a similar distribution. Or there may be even a gap against low-mass planets in systems with a wide BD. However, these results can be affected by several factors. As BDs cool in time and become less luminous, their detection is easier when the system is young. The stellar activity of such young stars then influences the detection limits of RV campaigns and can cause bias in the non-detection of low-massive planets. In Section \ref{freq_p} we show our detection limits, which are often not sufficient to detect such low-massive planets. These insufficient detection limits are also caused by typically relatively massive primary stars in systems with a wide BD. We also observe that sub-Jovian planets ($M_Psini \leq 0.1 M_J$) are more often around single stars than around stars with a stellar companion, the trend already mentioned in \cite{Fontanive21}. The trend is probably caused by the fact that low-massive planets are easier to detect around low-mass M stars causing deeper transits and larger RV semi-amplitudes, and such stars have a lower rate of multiplicity in comparison to more massive stars \citep{Salama22}.

\begin{figure}[!t]
\centering
\includegraphics[width=0.48\textwidth, trim= {0.0cm 0.0cm 0.0cm 0.0cm}]{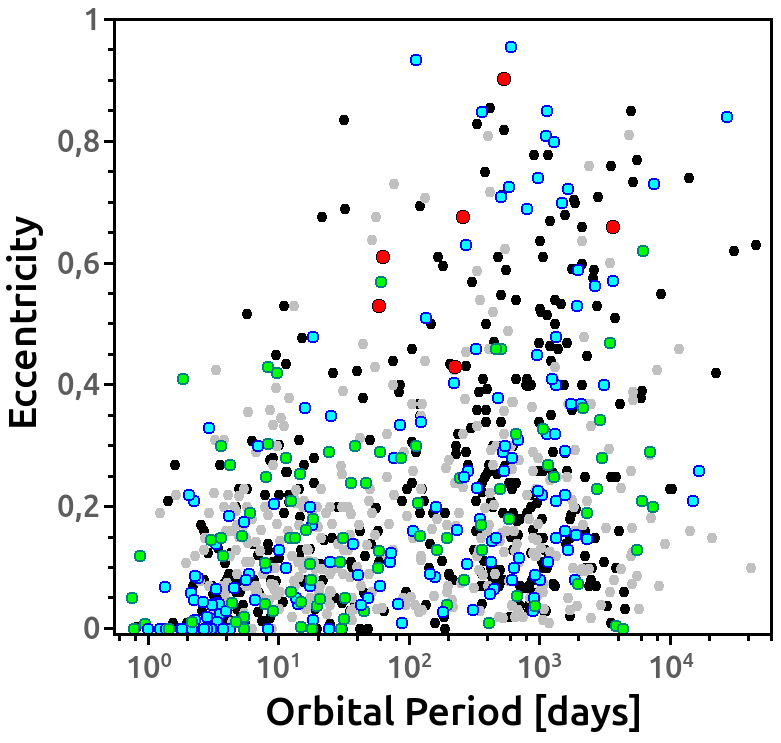}
\caption{Eccentricity versus orbital period for the known single planet-host stars (black and grey points), planet-host stars with one planet and one stellar companion (blue points), planet-host stars with one stellar companion but more than one planet (green points), and finally, planet-host stars with wide BD companion (red points).} \label{fig:family_ecc}
\end{figure}

\begin{figure}[!t]
\centering
\includegraphics[width=0.48\textwidth, trim= {0.0cm 0.0cm 0.0cm 0.0cm}]{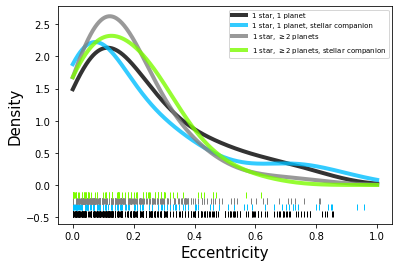}
\caption{KDEs of eccentricity comparing planets around single stars (black/grey) and planets around stars with stellar companion (blue/green).} \label{fig:KDE_ecc}
\end{figure}

The plot of the eccentricity versus the orbital period of planets can be seen in Fig.\ \ref{fig:family_ecc}. Colours represent the previous division into four groups plus planets around stars with a wide BD companion (red). To compare different groups, we used the same approach as before, creating KDEs for each group (see Fig. \ref{fig:KDE_ecc}). Group\,1 has visually a slightly different distribution than Group\,3 with an enhancement of small ($\leq 0.1$) and high ($\geq 0.8$) eccentricities. Comparing these two samples, we obtained a p-value of 0.012 from the two-sided Kolmogorov-Smirnov test, which means that these differences have a relatively high confidence level. 
Such enhancement of low eccentric planets is not surprising if we link it with the enhancement of short-period planets in systems with a wide stellar companion discussed above. We also compared Group\,1 with Group\,2 and found a p-value of 0.0008, which mean that planets in multi-planetary systems have a different distribution than single planets. We can see that planets in multi-planetary systems have lower eccentricities, and we observe only a few planets with an eccentricity larger than 0.5. Comparing Group\,2 with Group\,4, we obtained a p-value of 0.28; hence, we cannot rule out that these groups follow the same distribution. 

\begin{figure}[!t]
\centering
\includegraphics[width=0.48\textwidth, trim= {0.0cm 0.0cm 0.0cm 0.0cm}]{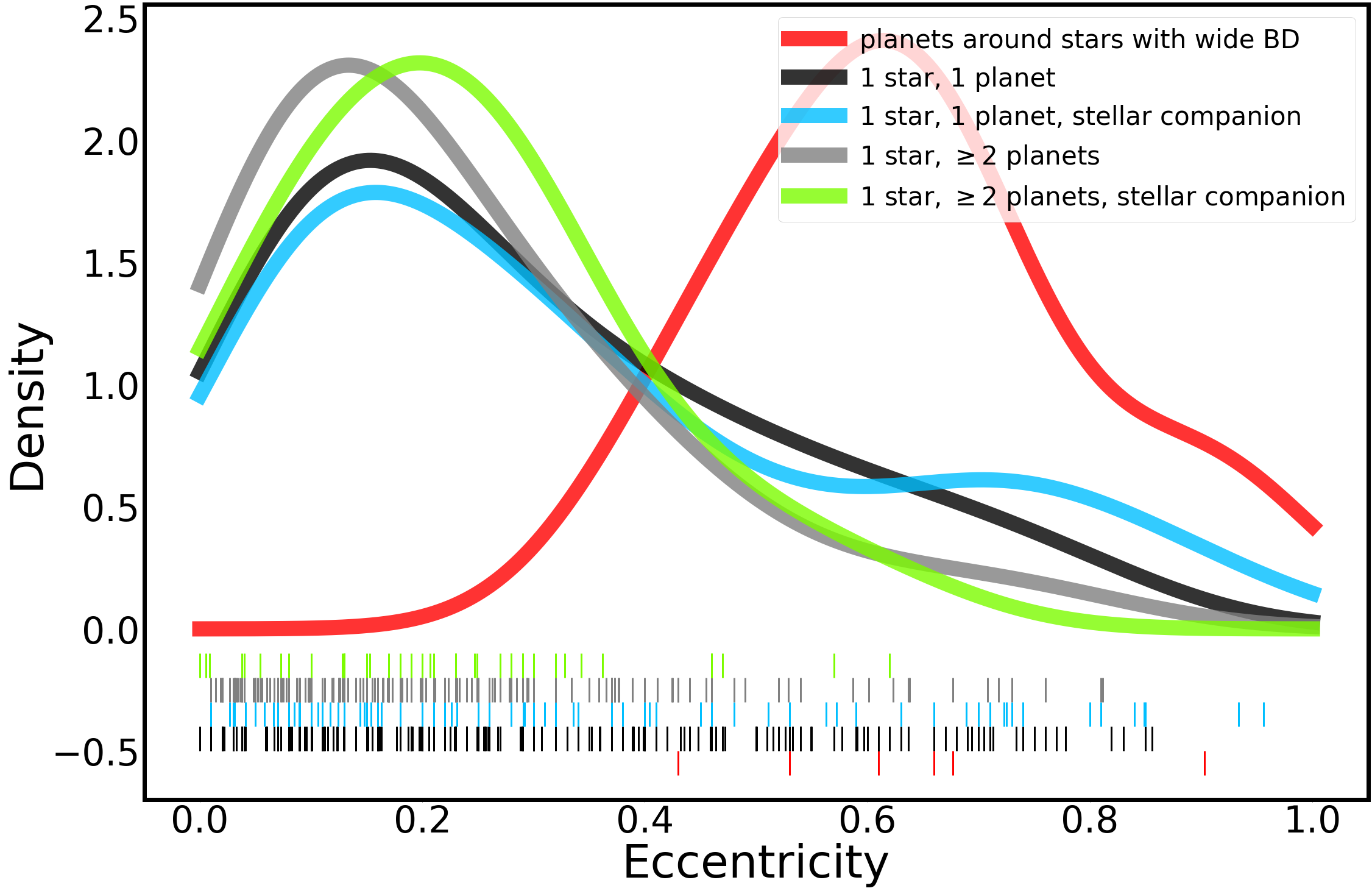}
\caption{KDEs of eccentricity comparing planets with periods larger than 50 days around single stars (black/grey), planets around stars with stellar companion (blue/green), and planets around stars with BD companion (red).} \label{fig:KDE_ecc1}
\end{figure}

As the planets in systems with wide BDs have periods larger than 50 days, we took only planets from this region to create kernel density estimates (KDEs) of the distributions of planet eccentricity (Fig.\ \ref{fig:KDE_ecc1}). We obtained the p-value of 0.0007 comparing this population to the Group\,1 (black line in Fig.\ \ref{fig:KDE_ecc1}), 0.00002 comparing to the Group\,2 (grey line), 0.002 comparing to the Group\,3 (blue line) and 0.0001 comparing to the Group\,4 (green line). It means that these planets, with a high confidence level, follow their own distribution with a maximum at $\sim0.65$. All of these planets in Fig.\ \ref{fig:family_ecc} have periods larger than 40 days and eccentricities larger than 0.4\@. One planet that does not meet these criteria is GJ\,229\,b. We reported in Appendix \ref{sec:family} the RV signal in GJ\,229 as a planetary candidate rather than a confirmed planet. So far, systems with wide BD companions do not follow the trend of the decreasing number of systems toward higher eccentricities observed for the other groups. It can suggest that wide BDs significantly affect planetary systems or that the formation mechanisms differ.

\begin{figure}[!ht]
\centering
\includegraphics[width=0.48\textwidth, trim= {0.0cm 0.0cm 0.0cm 0.0cm}]{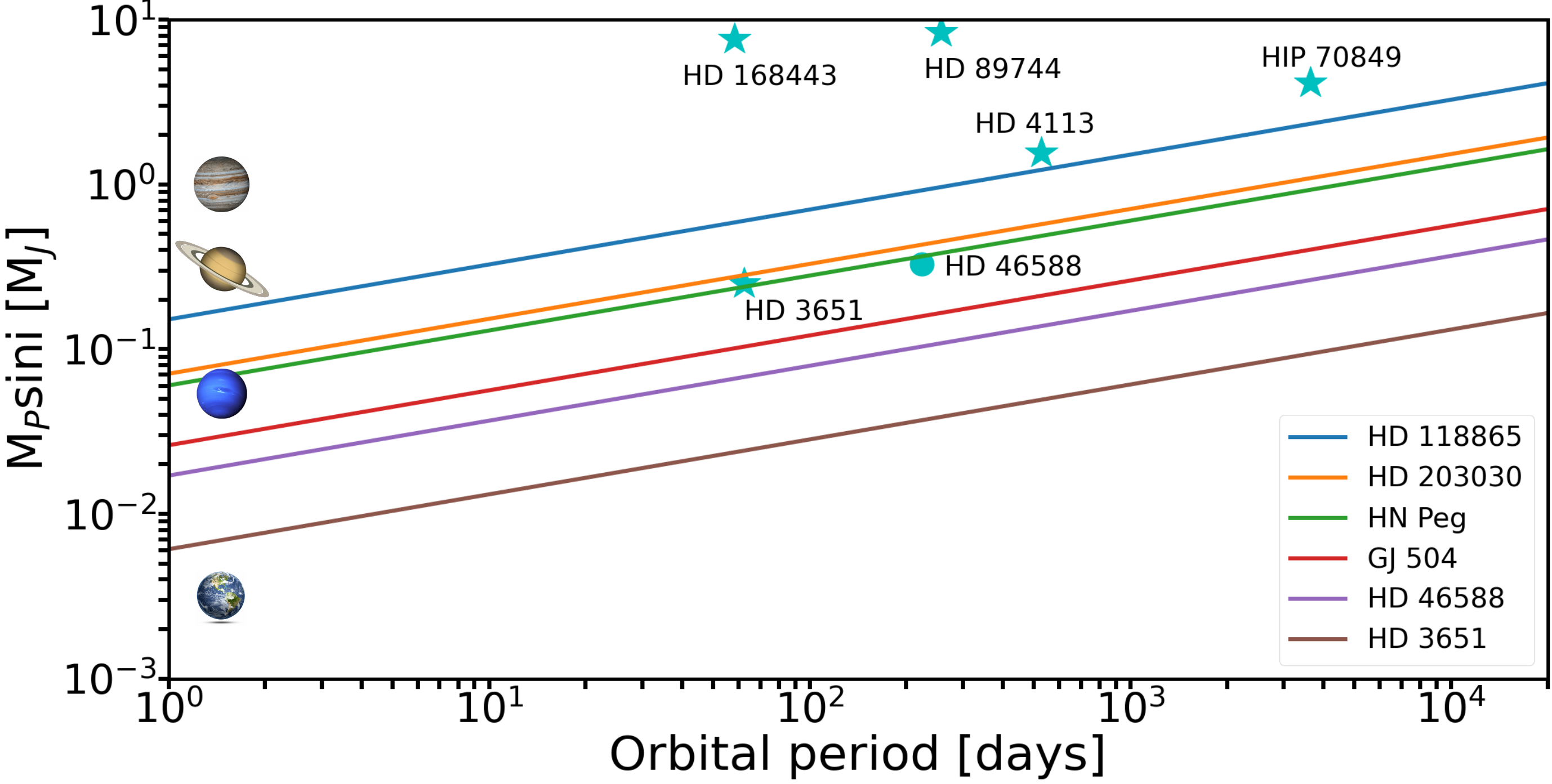}
\caption{Detection limits for stars in our sample. Cyan stars are confirmed planets in systems with a wide brown dwarf companion, and a circle represents a planetary candidate. Planets from the solar system are also plotted.} \label{fig:det_limits}
\end{figure}
\subsection{Frequency of planets in our sample}\label{freq_p}

We estimate individual detection limits before drawing any conclusions about the fraction of planets orbiting stars with wide BD companions. First, we used RV observations to measure the RMS scatter caused by stellar activity/error bars of observations for each star. Before the procedure, we subtracted signals observed in periodograms, which we linked to the activity to reduce the RMS scatter as much as possible (see Section \ref{sec:frequency}). Specifically, we subtracted the best fit sinusoidal function with the given period. Each star shows a different level of activity, mainly due to age differences, from the less active star HD\,3651 with RMS of 6\,m/s consistent with error bars of SONG data, GJ\,504 with RMS of 19\,m/s, up to HD\,118865 with RMS of 63.5\,m/s from STELLA observations. 
Second, we simulated the data with the same cadence and baseline as our observations and added Gaussian noise with a sigma equal to the RMS scatter. These data simulate our observations taking into account their error bars and the activity level of each star. Then, we inserted a sinusoidal signal into these data (higher eccentricity with sufficient phase coverage would further improve our detection limits). On such prepared datasets, we used the {\tt Exo-Striker} to perform the Keplerian fit of the inserted signal and ran MCMC analysis with 20 walkers, 1,000 burning phase steps, and 10,000 MCMC phase steps to compute the 3$\sigma$ error bars for the semi-amplitude of this signal, hence, 3$\sigma$ detection limits. We note that such detection limits would drop for specific periods based on the cadence or gaps in individual datasets; however, the cadences and timescales of our observations would enable us to detect most of the signals from several days up to several hundreds of days, so it is a reasonable simplification. Finally, we transformed the semi-amplitudes into minimum masses for all periods, as shown in Fig.\ \ref{fig:det_limits}. HD\,3651 is the less active star, so the detection limit derived for the star is equal to the instrumental limit of the SONG spectrograph for a given number of observations. We see that the activity in other systems significantly affects our detection limits. The HD\,118865 system observed with STELLA has lower RV precision than the targets observed with SONG\@. Based on our detection limits, we can often exclude more massive planets than Neptune for orbital periods of a few days and Saturn for orbital periods longer than hundreds of days, respectively. For quiet stars, we would be able to detect all planets from the literature. 

Before driving any conclusion about the frequency of planets orbiting stars with wide BD companions, we need to decide whether to include the HD\,3651 system in the statistical analysis. Our decision will greatly impact the results as it is the only system with a confirmed planet. One can argue that we should exclude HD\,3651b because this planet was known prior to our project. However, our target selection process was based on the starting criteria that this system fulfils; hence, we would include it regardless of the presence of a known planet. However, the discovery of the wide BD companion \citep{Mugrauer06} was actually triggered by the known planet companion as authors studied the multiplicity of exoplanet host stars. It is making it no longer independent from the presence of the planet in the system. However, \cite{Luhman07} later independently presented the discovery of the wide BD companion observing a large number of stars in the solar neighbourhood with the Spitzer Space Telescope. It suggests that the wide BD companion would probably be discovered regardless of the presence of a known planet; hence we decided to include HD\,3651 in our statistical analysis. In any case, we provide the frequency of planets also for the case we do not include HD\,3651.

Based on our sample of six systems, the frequency of planets orbiting stars with wide BD companions is between $17\pm23$\% (without the planetary candidate in the HD\,46588 system included) and $33\pm37$\%\footnote{The uncertainty of the frequency was estimated as \begin{equation}
    \sigma_f=\left(N_{planets}^{-1/2}+N_{stars}^{-1/2}\right)*\left(\frac{N_{planets}}{N_{stars}}\right),
\end{equation} following \citet{Bonavita07}.} (with the planetary candidate included). If we do not include HD\,3651b, then the frequency of planets orbiting stars with wide BD companions is between 0 and $20\pm29$\%. The study by \cite{Bonavita20} compares the frequency of giant planets around binaries and single stars, including different observable biases. The study did not reveal statistically significant difference with 5.1\,±\,1.57\,\% occurrence rate for binaries and 6.3\,±\,1.36\,\% occurrence rate for single stars. Given the small sample of studied systems, comparing the frequency of planets orbiting stars with wide BD companions to these frequencies has no informative value.

\section{Summary}\label{sec:summary}

We characterise a sample of five systems with a wide BD companion and search for exoplanets with the radial velocity technique. One additional system with a known exoplanet, HD\,3651, was used as a benchmark to test our detection capabilities, and we recovered the orbital solution and improved the precision of planet parameters. 

We used spectroscopic observations and available photometry to derive a complete list of physical parameters for the systems in our sample, with a special focus on the age of the systems using complementary age techniques. We found our sample to be relatively young, with three systems in order of hundreds of millions of years, one with age close to one Gyr, and two with ages older than 2.5 Gyr. 
We used ages to improve the physical parameters of the wide companions considering several substellar evolutionary models. We confirm the BD nature of all companions and place GJ\,504B in the planetary regime with a mass below the deuterium-burning limit.

In our sample of five stars, only one planetary candidate was identified in the HD\,46588 system, while the other signals are attributed to stellar activity. Our lower sensitivity limit is for Neptune-mass planets at short periods of a few days and Saturn-mass planets at longer periods of hundreds of days. For less active systems such as HD\,3651, we are sensitive to Neptune-mass planets at all periods up to 300 days, while for HD\,118865, we are sensitive only to Jupiter-mass planets at all periods. With one planetary candidate and one confirmed planet in six observed systems, we derive a frequency of planets orbiting stars with wide BD companions of below 70\% with the uncertainties included.

We summarised the properties of known planets orbiting five stars with wide BD companions. We considered planets in these systems together with the full sample of planet host stars with and without wide stellar companions to identify potential peculiarities in their parameter distributions. We detected the enhancement of planets with short periods below six days in systems with wide companions. We interpret it as a consequence of enhanced migration inwards closer to the host star. We also observe a break at the eccentricity of 0.5 with a small number of planets with higher eccentricities in multi-planetary systems. Finally, planets in systems with wide BD companions follow their own eccentricity distribution with a maximum at $\sim0.65$ and have periods larger than 40 days, masses larger than $0.1\,M_J$, and eccentricities larger than 0.4\@.

%
%
\begin{acknowledgements}
JS and PK would like to acknowledge support from MSMT grant LTT-20015.
JS and PK acknowledge a travel budget from ERASMUS+ grant 2020-1-CZ01-KA203-078200.
JS would like to acknowledge support from the Grant Agency of Charles University: GAUK No. 314421.
NL was financially supported by the Ministerio de Economia y Competitividad
and the Fondo Europeo de Desarrollo Regional (FEDER) under AYA2015-69350-C3-2-P\@.
We thank warmly Matthias Zechmeister for running the SERVAL pipeline and sending
us the CARMENES radial velocities. \\

This research has made use of the Simbad and Vizier databases, operated
at the centre de Donn\'ees Astronomiques de Strasbourg (CDS), and
of NASA's Astrophysics Data System Bibliographic Services (ADS). \\

This work has made use of data from the European Space Agency (ESA) mission {\it Gaia} (\url{https://www.cosmos.esa.int/gaia}), processed by the {\it Gaia} Data Processing and Analysis Consortium (DPAC, \url{https://www.cosmos.esa.int/web/gaia/dpac/consortium}). Funding for the DPAC
has been provided by national institutions, in particular the institutions participating in the {\it Gaia} Multilateral Agreement.

We acknowledge the use of public TESS data from pipelines at the TESS Science Office and at the TESS Science Processing Operations Center. 
Resources supporting this work were provided by the NASA High-End Computing (HEC) Program through the NASA Advanced Supercomputing (NAS) Division at Ames Research Center for the production of the SPOC data products.
This paper includes data collected with the TESS mission, obtained from the MAST data archive at the Space Telescope Science Institute (STScI). Funding for the TESS mission is provided by the NASA Explorer Program. STScI is operated by the Association of Universities for Research in Astronomy, Inc., under NASA contract NAS 5–26555.

This publication makes use of VOSA, developed under the Spanish Virtual Observatory project supported by the Spanish MINECO through grant AyA2017-84089. VOSA has been partially updated by using funding from the European Union's Horizon 2020 Research and Innovation Programme, under Grant Agreement nº 776403 (EXOPLANETS-A).

The SONG data were obtained over several semesters through programme numbers
P02-10, P03-08, P04-02, and P05-02 (PI Lodieu).
The Danish SONG telescope in Tenerife, the Hertzsprung SONG telescope, is owned and operated by Aarhus University and the University of Copenhagen in collaboration with the Astrophysics Institute of the Canary Islands (IAC). It is financed by the Villum Kann Rasmussen Foundation, Carlsberg Foundation,  the Danish Council for Independent Research | Natural Sciences (FNU), European Research Council, Danish National Research Foundation, Aarhus University, University of Copenhagen and Instituto de Astrof\'isica de Canarias. \\

Based on observations collected at the Centro Astron\'omico Hispano-Alem\'an (CAHA) at Calar Alto, operated jointly by Junta de Andaluc\'ia and Consejo Superior de Investigaciones Cient\'ificas (IAA-CSIC). The CARMENES dataset was obtained as part of programme number F19-3.5-011 (PI Lodieu).
CARMENES is an instrument for the Centro Astron\'omico Hispano-Alem\'an de Calar Alto (CAHA, Almer\'ia, Spain). CARMENES is funded by the German Max-Planck-Gesellschaft (MPG), the Spanish Consejo Superior de Investigaciones Cient\'ificas (CSIC), the European Union through FEDER/ERF FICTS-2011-02 funds, and the members of the CARMENES Consortium (Max-Planck-Institut f\"ur Astronomie, Instituto de Astrof\'isica de Andaluc\'ia, Landessternwarte K\"onigstuhl, Institut de Ciencies de l'Espai, Insitut f\"ur Astrophysik G\"ottingen, Universidad Complutense de Madrid, Th\"uringer Landessternwarte Tautenburg, Instituto de Astrof\'isica de Canarias, Hamburger Sternwarte, Centro de Astrobiolog\'ia and Centro Astron\'omico Hispano-Alem\'an), with additional contri-butions by the Spanish Ministry of Economy, the German Science Foundation through the Major Research Instrumentation Programme and DFG Research Unit FOR2544 "Blue Planets around Red Stars", the Klaus Tschira Stiftung, the states of Baden-W\"urttemberg and Niedersachsen, and by the Junta de Andaluc\'ia. \\

Based on observations made with the Italian Telescopio Nazionale Galileo (TNG) operated on the island of La Palma by the Fundaci\'on Galileo Galilei of the INAF (Istituto Nazionale di Astrofisica) at the Spanish Observatorio del Roque de los Muchachos of the Instituto de Astrof\'isica de Canarias. Part of the HARPS-N data used in this work have been downloaded from the TNG archive.

Based on observations collected at the European Organisation for Astronomical Research in the Southern Hemisphere under ESO programmes 072.C-0488, 183.C-0972, 085.C-0019, 087.C-0831, 183.C-0972, 089.C-0732, 090.C-0421, 091.C-0034, 093.C-0409, 095.C-0551, 096.C-0460, 196.C-1006, 098.C-0366, 099.C-0458, 0100.C-0097, 0101.C-0379, 0102.C-0558, 0103.C-0432.
\end{acknowledgements}

\clearpage
\bibliographystyle{aa}
\bibliography{astro_citations}
\twocolumn
\appendix
\renewcommand\thefigure{\thesection.\arabic{figure}}    
\section{Stellar and brown dwarf parameters}

\subsection{Stellar parameters with {\tt iSpec}}
\label{sec:iSpec}

We used the {\tt iSpec} framework \citep{Blanco14,Blanco19} to derive the parameters of the host stars. We used the Spectroscopy Made Easy radiative transfer code \citep[{\tt SME};][]{Valenti96,Piskunov17} incorporated into the {\tt iSpec} framework, and complementary to it, we use the MARCS atmosphere models \citep{Gustafsson08}, and version 5 of the GES atomic line list \citep{Heiter15}. The spectral fitting technique minimizes the $\chi^2$ value between the calculated synthetic spectrum and the observed spectrum. 

We used {\tt SERVAL} to co-add all CARMENES spectra to the very high signal-to-noise final spectrum. We then used these final spectra for each system from our sample to determine the effective temperature $T_{\rm eff}$, metallicity $\rm [Fe/H]$, surface gravity $\log{g}$, the projected stellar equatorial velocity $v\sin{i}$, and the lithium equivalent width $eqw_{Li}$ of the Li\,I line at 670.8\,nm following the same procedure as in \cite{Fridlund17}. In our analysis, we computed values for the micro-turbulence and macro-turbulence velocities (V$_{mic}$, V$_{mac}$) from empirical relations incorporated into the {\tt iSpec} framework. To determine $T_{\rm eff}$, we fit the wings of H$\alpha$ Balmer line \citep{Cayrel11}. We then used this temperature to fit the line wings of the Ca\,I triplet (610.27\,nm) to derive a value of $\log{g}$. We also used a large sample of clean and unblended Fe\,I lines in the interval from 597\,nm to 643\,nm to derive $\rm [Fe/H]$ and $v\sin{i}$. We report the stellar parameters for the stars of our sample in Table \ref{tab5}.

To estimate stellar mass and radius, we used the Bayesian parameter estimation code {\tt PARAM 1.5}\footnote{\url{http://stev.oapd.inaf.it/cgi-bin/param_1.3}} \citep{DaSilva06,Rodrigues14,Rodrigues17} based on the PARSEC isochrones. As input parameters, we set $Gaia$ eDR3 parallaxes, $Gaia$ DR2 magnitudes, luminosities from {\tt VOSA}, and $T_{\rm eff}$ with $\rm [Fe/H]$ derived from {\tt iSpec}. We report all derived parameters in Table \ref{tab5}.

\subsection{SED analysis with {\tt VOSA}}
\label{sec:VOSA}

As an independent check, we used the Virtual Observatory SED Analyser \citep[{\tt VOSA}\footnote{\url{http://svo2.cab.inta-csic.es/theory/vosa/}};][]{Bayo08} to determine stellar parameters. To model the spectral energy distribution of the stars in our sample we used grids of five different models: BT-Settl-AGSS2009 \citep{Barber06,Asplund09,Allard12}, BT-Settl-CIFIST \citep{Barber06,Caffau11,Allard12}, BT-NextGen GNS93 \citep{Grevesse93,Barber06,Allard12}, BT-NextGen AGSS2009 \citep{Barber06,Asplund09,Allard12}, and Coelho Synthetic stellar library \citep{Coelho14} and performed the ${\chi}^2$ minimization procedure to compare theoretical models with the observed photometry. We used the available photometric measurements, specifically, the Str{\"o}mgren-Crawford uvby{\ensuremath{\beta}} \citep{Paunzen15}, Tycho \citep{Hog00}, $Gaia$ DR2 \citep{Gaia18}, $Gaia$ eDR3 \citep{Gaia21}, 2MASS \citep{Cutri03}, AKARI \citep{Ishihara10}, and WISE \citep{Cutri14} photometry. We set priors for the T$_{\rm eff}$\,=\,4000--7000\,K, $\log$(g)\,=\,4.0--5.0 dex, and [Fe/H]\,=\,$-$0.5--0.5 based on results from {\tt iSpec}. For each model, we take only results with the lowest ${\chi}^2$, and use them to create the final intervals of derived parameters. We report the final intervals in Table \ref{tab5}. In Table \ref{tab5}, we also compare the stellar radii computed from stellar isochrones with ones from {\tt VOSA} determined via the Stefan–Boltzmann law.

\subsection{Revised parameters of the wide BD companions}
\label{phys_params_compan}
We used the SpeX Prism Library Analysis Toolkit ({\tt SPLAT}) \citep{Burgasser17} to derive the parameters of the wide BD companions in our sample. SPLAT is a python-based package designed to interface with the SpeX Prism Library ({\tt SPL}) \citep{Burgasser14}, which is an online repository of over 2000 low-resolution, near-infrared spectra of stars and BDs. The package enables conversion between observable (temperature, luminosity, surface gravity) and physical parameters (mass, radius, age) of BDs using published evolutionary model grids. First, we input the age of the stars in our sample as derived in Section \ref{sec:age}. We used the BD luminosities published in the literature and compiled them in Table \ref{tab3}. The {\tt SPLAT} output parameters include mass, temperature, surface gravity, and radius.
We considered different BD evolutionary models from \cite{Burrows01,baraffe03,saumon08} incorporated in the package. The models of \citet{Burrows01} and \citet{baraffe03} models are only for the solar metallicity, while models of \citet{saumon08} incorporate the solar metallicity and $-$0.3, $+$0.3 dex. As several stars in our sample have metallicities of $\sim$0.2 dex, we provide two solutions for them: solar metallicity and $+$0.3 dex based on the \citet{saumon08} models. We do not consider the metallicity of $-$0.3 dex because none of our targets has metallicity below solar (the ones with slightly negative values are still consistent with solar in terms of error bars). The results are summarized in Table \ref{tab3}. 
The minimum and maximum values for each parameter (temperature, gravity, mass, radius) take into account the full interval of errors in age and luminosities as well as the discrepancies between models
The ranges of the physical parameters of the BDs quoted in Table \ref{tab3}; hence, our conclusions do not strongly depend on the models. According to the derived mass intervals, all companions are BDs, except GJ\,504B, whose mass estimates place it in the planetary mass rather than the substellar regime. \citet{Bonnefoy18} discussed that two possible age scenarios (21$\pm$2\,Myr and 4.0$\pm$1.8\,Gyr) make that companion either a BD or a planet. Our analysis favours the young case over the older option, hence, the lowest mass.

\section{Age analysis}\label{sec:age_app}

\subsection{Stellar isochrones}

We use the {\tt PARAM 1.5} code \citep{DaSilva06,Rodrigues14,Rodrigues17} code to infer stellar ages. It is based on the PARSEC and MIST stellar evolution tracks \citep{Bressan12,Choi16}. {\tt PARAM 1.5} derives stellar parameters by comparing a variety of observational inputs, such as photometric colours, spectroscopic properties or asteroseismic parameters, to interpolated model values. We use the PARSEC isochrones, and as inputs for modelling, we use $Gaia$ eDR3 parallax, the $Gaia$ DR2 colours, luminosities derived from {\tt VOSA} modelling, together with Teff, log g, and [Fe/H] derived from {\tt iSpec} modelling from Table \ref{tab5}. Results are listed in Table \ref{tab:age}.

\begin{figure*}[!ht]
\centering
\includegraphics[width=1.0\textwidth,height=0.35\textwidth]{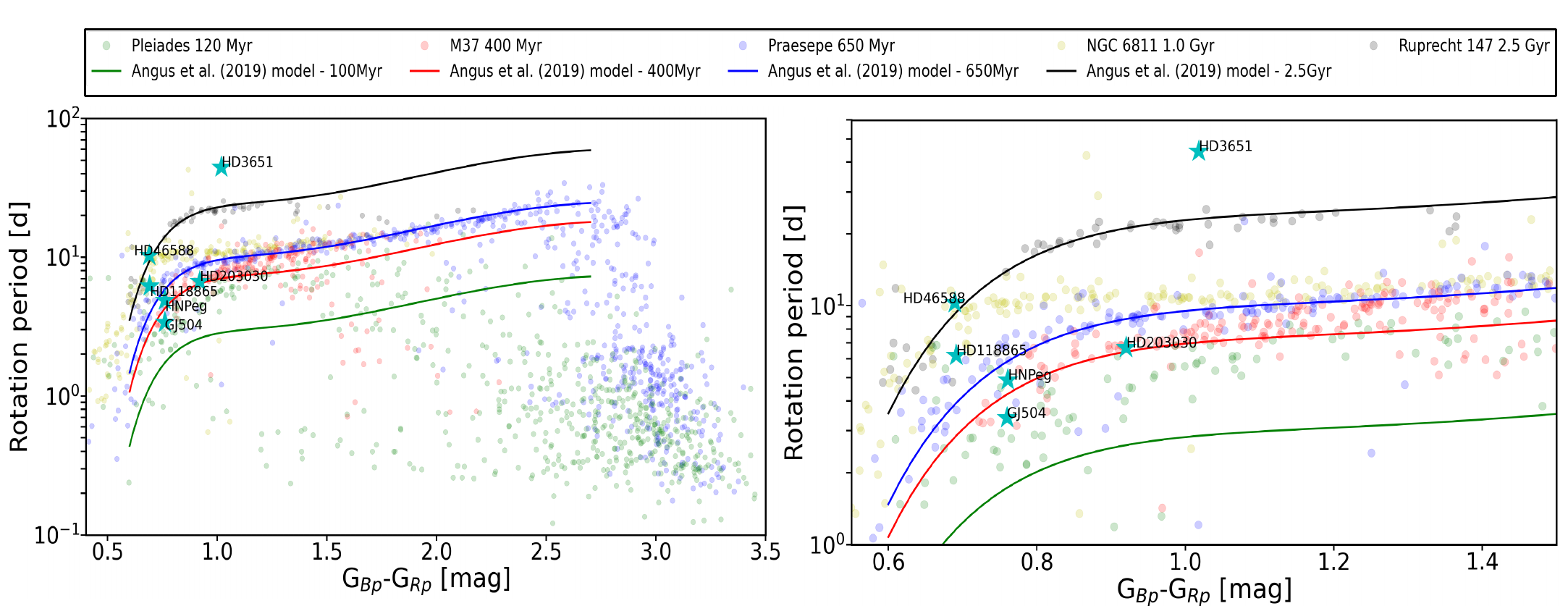}
\caption{Color-period diagram of the stars from our sample (magenta stars) together with members of well-studied clusters: Pleiades cluster, M37 cluster, Praesepe cluster, NGC\,6811 cluster, Ruprecht\,147 cluster and NGC\,6819 cluster. Lines represent the 100, 400, 650, and 2500\,Myr curves compute from the empirical relation from \cite{Angus2019}.} \label{fig:rot_per}
\end{figure*}

\subsection{Gyrochronology}

As described in \cite{Angus2019}, the gyrochronology relation used in the {\tt stardate} code is derived based on the Praesepe cluster fitting a broken power-law to observed rotation periods of stars from this cluster in form: 

\begin{equation}
      log{P_{rot}} = 
         c_{a}log{(t)} + \sum_{n=0}^{4} c_{n}[log(G_{Bp}-G_{Rp})]^{n} \,,
\end{equation}

where $t$ defines the age of a cluster. To study the ages of the stars in our sample, we used this empirical relation, with coefficients derived on the Praesepe cluster, to compute curves for ages 100, 400, 650, and 2500\,Myr and plot them together with the members of the well-defined clusters studied in \cite{Godoy21}: Pleiades cluster ($\sim$120 Myr), M37 cluster ($\sim$400 Myr), Praesepe cluster ($\sim$650 Myr), NGC\,6811 cluster ($\sim$1 Gyr), together with the Ruuprecht\,147\,+\,NGC\,6819 clusters studied in \cite{Curtis20}. We calculated reddening E(B-V) for each star in the clusters using the {\tt dustmaps} code \citep{Green18} and three-dimensional Bayestar dust maps \citep{Green19}. We then followed the approach from \citet{Gaia2018} to compute reddening in the Gaia colours. In Fig.\ \ref{fig:rot_per}, we overplot our sample highlighted with cyan star symbols, specifically, rotation periods derived from TESS SPOC LCs in Section \ref{surface_rotation}, and $Gaia$ eDR3 $G_{Bp}$, $G_{Rp}$ magnitudes. Using empirical relations together with cluster members, we are able to make assumptions for the ages of our systems. 

We conclude that HD\,3651 is older than Ruprecht\,147. HD\,46588 looks to have an age between NGC\,6811 and Ruprecht\,147. HD\,118865 has an age between Praesepe and NGC\,6811. HN\,Peg, and HD\,203030 are probably younger than Praesepe and have ages consistent with M37. Finally, GJ\,504, has an age between Pleiades and M37. It looks like the youngest system in our sample; however, we cannot exclude the older scenario previously discussed.

\subsection{Lithium EW}


We used the lithium line at 6708\,\AA\,to measure the equivalent width with {\tt iSpec}. We fitted the line with a Gaussian profile, and the EW corresponds to the area within the gaussian fit. Similarly, as with the rotation period, we compare the EW of Li vs colour to members of well-studied clusters. We use the Tuc-Hor young moving group \citep[$\sim$45 Myr;][]{Mentuch08}, the Pleiades \citep[$\sim$120 Myr;][]{Soderblom93}, M34 \citep[$\sim$220 Myr;][]{Jones97}, Ursa Major Group \citep[$\sim$400 Myr;][]{Soderblom1993}, Praesepe \citep[$\sim$650 Myr;][]{Soderblom993}, Hyades \citep[$\sim$650 Myr;][]{Soderblom90}, and M67 clusters \citep[$\sim$4 Gyr;][]{Jones99}. If needed, we deredden clusters using E(B-V) values from \citet{Gaia2018}. According to Li EW, two of our systems look younger than the rest. They are HN\,Peg, and GJ\,504, with ages between the M34 and Praesepe/Hyades. HD\,46588 and HD\,118865 are consistent with Praesepe or older. HD\,203030 is consistent with Praesepe. Finally, in HD\,3651, we plot the upper limit as we did not detect any lithium, setting a minimum age of 220 Myr (age of M34).

\begin{figure}[!ht]
\centering
\includegraphics[width=0.46\textwidth, trim= {0.0cm 0.0cm 0.0cm 0.0cm}]{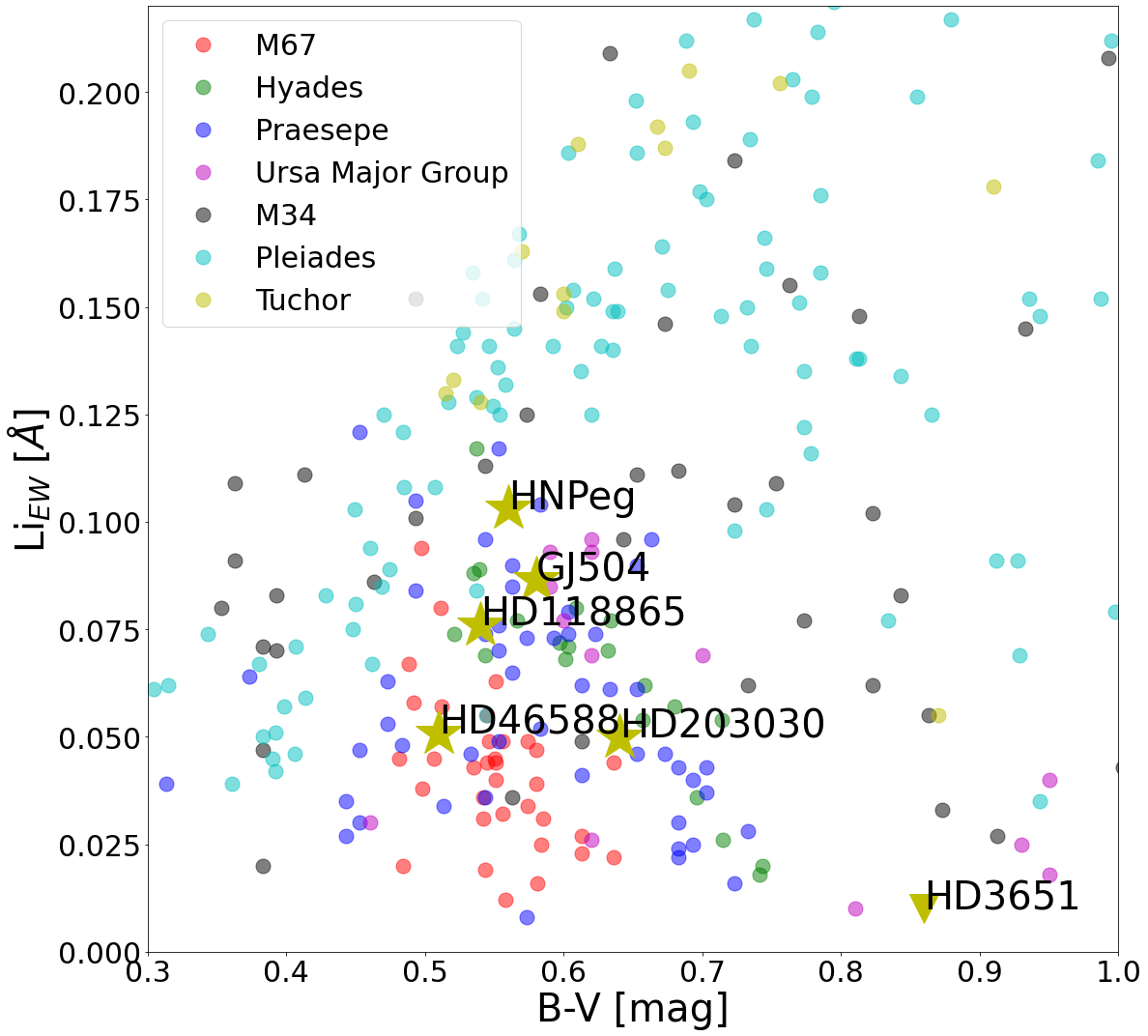}
\caption{Color vs. EW of lithium line Li\,6708 \AA{} of the stars from our sample (gold stars) together with members of well-studied clusters: Tuc-Hor young moving group \citep[$\sim$45 Myr;][]{Mentuch08}, the Pleiades \citep[$\sim$120 Myr;][]{Soderblom93}, M34 \citep[$\sim$220 Myr;][]{Jones97}, Ursa Major Group \citep[$\sim$400 Myr;][]{Soderblom1993}, Praesepe \citep[$\sim$650 Myr;][]{Soderblom993}, Hyades \citep[$\sim$650 Myr;][]{Soderblom90}, and M67 clusters \citep[$\sim$4 Gyr;][]{Jones99}.} \label{fig:Li_ew}
\end{figure}

\subsection{X-ray luminosity}

\begin{figure}[!ht]
\centering
\includegraphics[width=0.46\textwidth, trim= {0.0cm 0.0cm 0.0cm 0.0cm}]{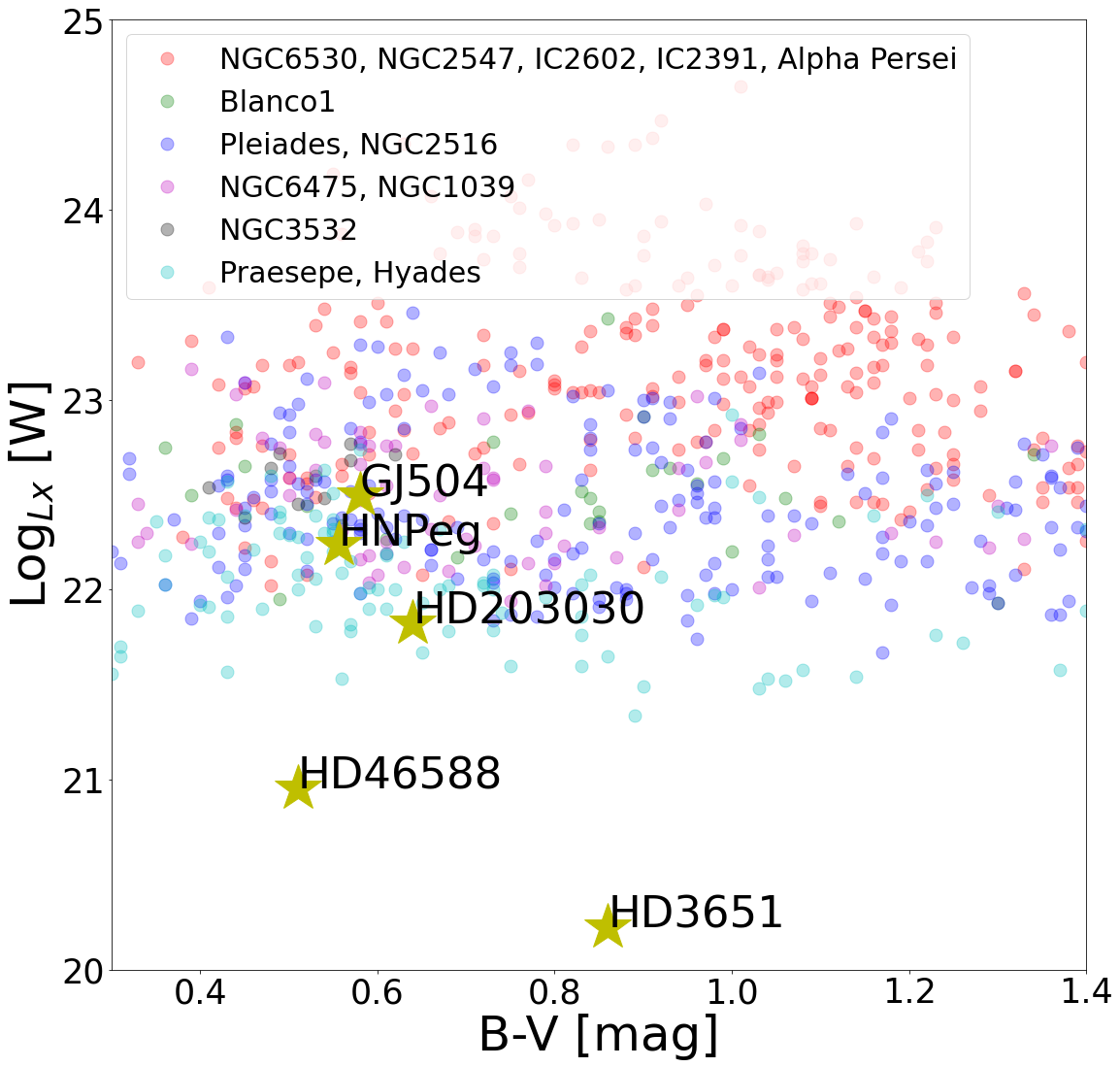}
\caption{X-ray luminosity vs. color of the stars from our sample (gold stars) together with members of well-defined clusters from \cite{Jackson12}.} \label{fig:Xray}
\end{figure}

We used clusters listed in \cite{Jackson12} and ROSAT observations\footnote{\url{https://heasarc.gsfc.nasa.gov/cgi-bin/W3Browse/w3browse.pl}} \citep{Voges99} of our systems to compare the X-ray luminosity of our sample with clusters of known ages. We found ROSAT data available for all systems except for HD\,118865, which is missing in this analysis. The X-ray luminosities are listed in Table \ref{tab5}. The final plot can be found in Fig.\ \ref{fig:Xray}. According to this plot, systems HD\,3651 and HD\,46588 are older than the Praesepe and Hyades. HD\,203030 seems to have an age similar to the Praesepe and Hyades. HN\,Peg can have any age between the Pleiades and Praesepe/Hyades included, while GJ\,504 is probably younger than the Praesepe/Hyades.

\begin{figure*}[!ht]
\centering
\includegraphics[width=1.0\textwidth,height=0.35\textwidth]{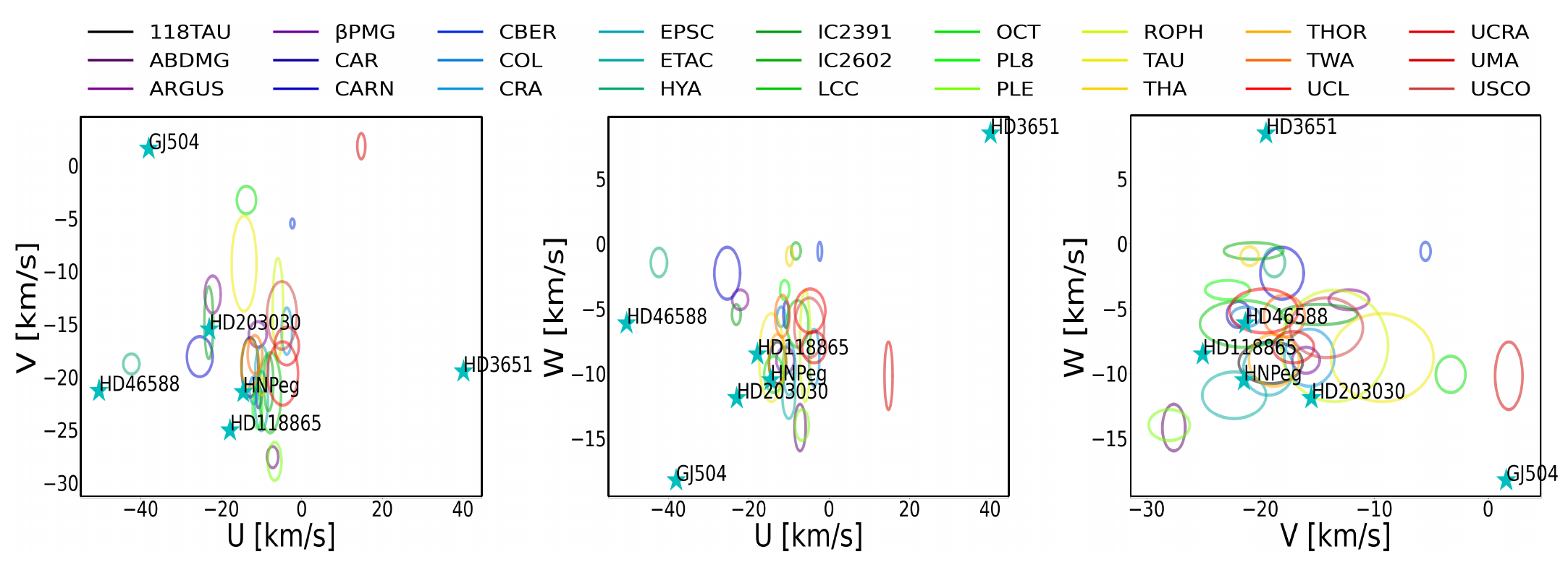}
\caption{Membership to young associations. Ellipses represent the 1-sigma position of young stellar associations in space velocities U, V, W taken from \cite{Gagne18}. Our sample of stars is plotted as magenta stars.} \label{fig:associations}
\end{figure*}

\subsection{Membership to young associations}

We used the {\tt BANYAN} code \citep{Gagne18} to derive membership probabilities for our sample of stars. {\tt BANYAN} includes 27 young associations with ages between 1--800\,Myr. As an independent check, we also used the LocAting Constituent mEmbers In Nearby Groups code \citep[{\tt LACEwING};][]{Riedel17}. {\tt LACEwING} includes 13 YMGs and three open clusters within 100\,pc. All YMGs are in common with \cite{Gagne18}. The $Gaia$ eDR3 astrometric data, which can be found in Table \ref{tab1}, were used as input parameters. {\tt BANYAN} code does not confirm membership to any association and derives that all stars in our sample are field stars. The {\tt LACEwING} code found 23\% probability for HD\,203030 and 10\% probability for HN\,Peg to be the member of Argus association \citep[$\sim$50 Myr;][]{Zuckerman19}. Given the age indicators from previous subsections, we consider these stars to be older than the Argus association and hence unlikely members of this YMG. The results for the rest of the stars are in agreement with the {\tt BANYAN} code, and they appear to be field stars belonging to the Thin disk \citep{Leggett92,Bensby03}. \citet{Gaidos98} discovered that HN\,Peg is a member of the Her--Lyr moving group, which is not included in {\tt BANYAN} and {\tt LACEwING}.

In Fig.\ \ref{fig:associations}, we plot the 1-sigma position of young stellar associations in space velocities U, V, W taken from \cite{Gagne18} together with our sample of stars. To determine U, V, W velocities for our systems used in this figure, we used the python package {\tt PyAstronomy}, specifically the {\tt gal\_uvw}\footnote{\url{https://pyastronomy.readthedocs.io/en/latest/pyaslDoc/aslDoc/gal_uvw.html}} function.

\section{The family of planet host stars with BD companions}
\label{sec:family}
Besides HD\,3651, eight systems are known with exoplanet-host stars and wide BD companions: HIP\,70849, HD\,89744, HD\,168443, GJ\,229, HD\,4113, HD\,65216, HD\,41004 and $\epsilon$ Indi. We looked at original discoveries papers, and in this section, we summarized the parameters derived by other authors. For HIP\,70849, we have revised stellar parameters and updated orbital parameters as we found a large number of available HARPS spectra not previously reported, which significantly improved the parameters of the system.




\subsection{HIP\,70849}
We processed all 54 HARPS archival spectra observed between April 2006 and August 2019 with {\tt SERVAL} \citep{Zechmeister18} to derive stellar parameters, refine the orbital solution of the planet, and discuss the star's activity level. We co-added all HARPS spectra to produce a very high signal-to-noise final spectrum input in the {\tt iSpec}/{\tt PARAM} 1.5 tools to determine a stellar mass of 0.53$\pm$0.04\,$\rm \mst$, a radius of $R_\star$\,=\,0.47$\pm$0.04\,$\rm \rst$, an effective temperature of $T_{\rm eff}$\,=\,4112$\pm$72\,K, and a metallicity of $-$0.31$\pm$0.08\,dex. \citet{Segransan11} reported an age of 3$\pm$2\,Gyr, while we derived an age of 0.6--3.0\,Gyr (Section \ref{sec:age}). A T4.5 dwarf companion orbits the star at the separation of 6.3\,arcmins ($\sim$ 9000\,au) \citep{Lodieu14}. \citet{Segransan11} reported an additional long-period companion with a period between 5 and 75 years, a minimum mass of $msini$ between 3.5--15 $M_J$, and an eccentricity of 0.4--0.98 using data from the HARPS Echelle spectrograph. The large uncertainties of the derived parameters are due to the (1) small number of RV measurements with only one observation close to the periastron and (2) large eccentricity, causing the period to be insufficiently covered. Hence, we downloaded the 18 archival HARPS spectra used in \citet{Segransan11} as well as the additional spectra obtained since then, not yet reported in the literature.

\begin{figure}[!ht]
\centering
\includegraphics[width=0.48\textwidth, trim= {0.0cm 0.0cm 0.0cm 0.0cm}]{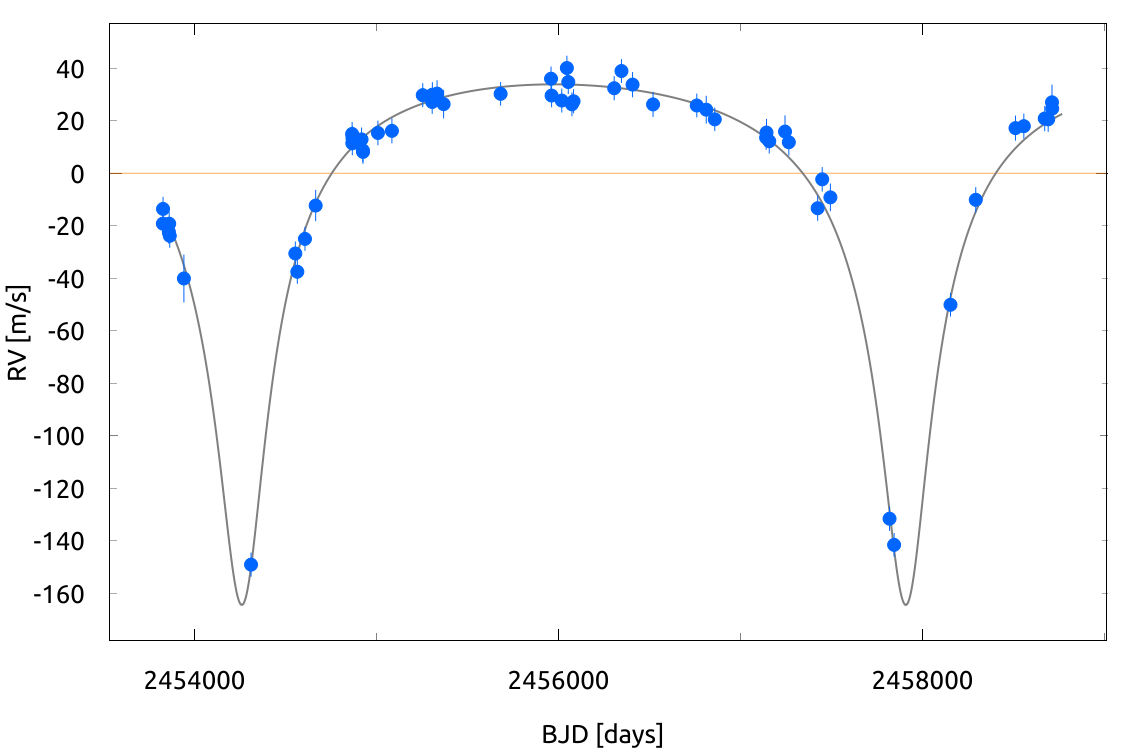}
\caption{Orbital solution for HIP\,70849 showing the HARPS RVs (blue points) and the RV model in black.} \label{fig:hip70849}
\end{figure}

With this enhanced dataset, we observe an RVs signal at 3648$\pm$12\,days in the periodogram. To confirm the planetary nature of this signal, we used {\tt SERVAL} to derive various activity indicators. We can not see any counterpart at this period in any periodogram of activity indicators supporting the planetary origin of the signal. 
We used {\tt Exo-Striker} to fit the orbital solution with the revised stellar parameters. We set 14 walkers and ran 1,000 burning phase steps and 10,000 MCMC phase steps. Two more observations were obtained close to the periastron, significantly improving the orbital solution given HARPS precision and low RMS scatter. We summarize the updated planetary parameters in Table \ref{table:hip70849b}. The orbital solution is plotted in Fig.\ \ref{fig:hip70849} and the plot of correlations between parameters together with the derived posterior probability distributions from MCMC is in Fig.\ \ref{fig:hip70849-PPD} in Appendix. We do not detect additional signals in RVs.

\begin{table}
 \centering
 \caption[]{HIP\,70849b: updated parameters
 }
 \label{table:hip70849b}
\scalebox{0.9}{
 \begin{tabular}{@{\hspace{0mm}}l c c c@{\hspace{0mm}}}
 \hline
$msini$ ($\ensuremath{\,{\rm M_J}}$) & $3.7\pm0.2$ \cr
$K$ (m/s) & $99\pm3$ \cr
$T_0$ (days) & $2457699\pm8$ \cr
$P$ (days) & $3649\pm12$ \cr
$e$ & $0.66\pm0.01$ \cr
$\omega$ (deg) & $183\pm1$ \cr
\hline

 \end{tabular}
}
\end{table}

\subsection{HD\,89744}
HD\,89744 is an F7V star with a mass of $M_\star$\,=\,1.56$\pm$0.05\,$\rm \mst$, radius of $R_\star$\,=\,2.14$\pm$0.20\,$\rm \rst$, and age of $1.9\pm0.2$\,Gyr \citep{Takeda07}. The star has an effective temperature of $T_{\rm eff}$\,=\,6300$\pm$30\,K, and a metallicity of 0.24$\pm$0.03\,dex \citep{Tsantaki14}. \citet{Wilson01} reported a wide massive brown dwarf candidate of spectral type L0V based on spectroscopic observations. \citet{Mugrauer04} then confirmed this companion at a separation of 63 arcsec ($\sim$2500\,au) with the mass between 0.072--0.081 $\rm \mst$, and temperature of $\sim$2200K at the stellar/substellar boundary. \citet{Korzennik00} earlier reported a massive planet using the Advanced Fiber-Optic Echelle spectrograph at the Whipple Observatory. The solution was later revised by \citet{Wittenmyer19} using four different instruments to $msini$\,=\,8.4$\pm$0.2\,$\ensuremath{\,{\rm M_J}}$ with the period of $P$\,=\,256.78$\pm$0.02\,days and eccentricity of $e$\,=\,0.677$\pm$0.003. The discovery paper and later papers about this system agree that the semi-amplitude of 270m/s is too large to be solely associated with the stellar activity of a late-F star.

\subsection{HD\,168443}
HD\,168443 is an G6V star with a mass of $M_\star$\,=\,1.00$\pm$0.02\,$\rm \mst$, radius of $R_\star$\,=\,1.52$\pm$0.07\,$\rm \rst$, and age of $11.1\pm0.7$\,Gyr \citep{Takeda07}. The star has an effective temperature of $T_{\rm eff}$\,=\,5530$\pm$97\,K, and a metallicity of 0.08$\pm$0.06\,dex \citep{Rosenthal21}. \citet{Marcy01} reported two companions observed in 4.4 yr long period of the Keck/HIRES observations. The wide BD companion has $msini$\,=\,18.0$\pm$1.8\,$\ensuremath{\,{\rm M_J}}$, the period of $P$\,=\,1753.1$\pm$0.8\,days and eccentricity of $e$\,=\,0.210$\pm$0.003. The giant planet has $msini$\,=\,7.6$\pm$0.6\,$\ensuremath{\,{\rm M_J}}$, the period of $P$\,=\,58.1119$\pm$0.0002\,days and eccentricity of $e$\,=\,0.530$\pm$0.002.

\subsection{GJ\,229}
GJ\,229 is an M1.5V dwarf with a mass of $M_\star = 0.58 \mst$, a radius of $R_\star = 0.46 \rst$, and a temperature of $T_{\rm eff}$\,=\,3564K \citep{Tuomi14}. The star hosts a BD companion detected through direct imaging \citep{Nakajima95}. In a new analysis of the system, \citet{Brandt20} derived a mass of 70$\pm$5\,M$_{J}$ for the BD with a semi-major axis of 34.7$\pm$1.9\,au, and eccentricity of 0.846$\pm$0.015 combining Keck/HIRES RVs, Hipparcos and $Gaia$ DR2 astrometry, and HiCIAO--Subaru and the Hubble Space Telescope imaging. 

\citet{Tuomi14} reported a two-sigma detection of a possible additional planet with $msini\,=\,0.1 \ensuremath{\,{\rm M_J}}$, a period of 471 days, and an eccentricity of 0.1 using the HARPS and UVES observations. However, \citet{Brandt20} did not find evidence for this planet in 47 Keck/HIRES RVs. \citet{Feng20} discuss the system using all data available: 47 Keck/HIRES, 74 UVES, 200 HARPS RVs. They report two planets in the system, one with mass $msini\,=\,0.023 \ensuremath{\,{\rm M_J}}$, the period of $P\,=\,122$\,days and eccentricity of $e\,=\,0.19$ and the second with $msini\,=\,0.027 \ensuremath{\,{\rm M_J}}$, $P\,=\,526$\,days and $e\,=\,0.10$. They also provide different activity indicators for each instrument besides UVES (see Fig.\ 10 in \citet{Feng20}).

We have visually investigated the plots and results of the work by \cite{Feng20}. The periodograms of HARPS activity indicators with the largest number of observations reveal significant signals close to the 526-day peak reported as a planet. The authors also reported activity signal at $\sim$278 days, which can be just the first harmonic. Hence, we argue that the 526-day peak seen in RVs might be associated with stellar activity. The signal with a period of 122 days has a better chance of being the planet; however, it can also be the higher-order harmonics. For example, this peak is visible in the HARPS R$_{HK}$ activity indicator. Furthermore, the semi-amplitudes of both signals are below 2m/s, smaller than the RMS scatter of both datasets. Because of our scepticism regarding the planetary nature of these signals, we decided not to include them in our analyses.

\subsection{HD\,4113}
HD\,4113 is an G5V star with a mass of $M_\star$\,=\,1.05$\pm$0.10\,$\rm \mst$, a radius of $R_\star$\,=\,1.09$\pm$0.09\,$\rm \rst$, a temperature of $T_{\rm eff}$\,=\,5646$\pm$70\,K, and a metallicity of 0.19$\pm$0.05\,dex \citep{Ghezzi10}. \citet{Tamuz08} reported a massive planet in the system with $msini$\,=\,1.56$\pm$0.04\,$\ensuremath{\,{\rm M_J}}$, a period of $P$\,=\,526.6$\pm$0.3\,days, and an eccentricity of $e$\,=\,0.903$\pm$0.005 using the CORALIE spectrograph. These authors also observed a long trend in RVs, predicting a possible second companion with a minimum mass of 10\ensuremath{\,{\rm M_J}} and minimum semi-major axis of 8\,au. \cite{Mugrauer04} later reported a wide M-dwarf companion at a separation of 49 arcsec ($\sim$2000\,au) with the mass of $0.56$ $\rm \mst$, too far away to explain the long trend in RVs. \citet{Cheetham18} finally detected a third companion responsible for this trend, using CORALIE and KECK/HIRES RVs, and high-contrast imaging with SPHERE: it is a BD with a mass of 66$\pm$5\,M$_{J}$, period of 105$\pm$29\,yr, and an eccentricity of 0.38$\pm$0.08\@.

\subsection{HD\,65216, HD\,41004 and $\epsilon$ Indi}
All three systems have different architectures from the previous ones. HD\,65216 is a G5V star with a mass of $M_\star = 0.88 \pm 0.04$ $\rm \mst$, a radius of $R_\star = 0.88 \pm 0.01$ $\rm \rst$, an effective temperature of $T_{\rm eff}$\,=\,5645$\pm$20\,K, and a metallicity of $-$0.16$\pm$0.02 dex \citep{Stassun17}. The star forms a hierarchical triple system with the binary companion at a separation of 253 au. The two wide components, HD\,65216BC, are separated by 6\,au and consist of an M2--M3$+$L2--L3 pair \citep{Mugrauer07}. The star also hosts a planet with a period of 613 days and an eccentricity of 0.41 \citep{Mayor04} using 52 spectra obtained with CORALIE spectrograph. Another 24 HARPS spectra were added to the analysis by \citet{Wittenmyer19} to report a second potential companion with a period of 5370 days and eccentricity 0.17, which would refine the orbital parameters of the inner planet to P\,=\,578 days and e\,=\,0.27.

HD\,41004\,A is a K1V star with a mass of $M_\star$\,=\,0.95$\pm$0.10\,$\rm \mst$, a radius of $R_\star$\,=\,0.85$\pm$0.07\,$\rm \rst$, an effective temperature of $T_{\rm eff}$\,=\,5310$\pm$65\,K, and a metallicity of 0.23$\pm$0.04\,dex \citep{Ghezzi10}. This star hosts an exoplanet and is the primary of a system with a wide companion hosting a BD. \cite{Zucker04} analysed 149 CORALIE spectra to derive the parameters of the system by separating the RVs of the two stellar components. The primary star host an exoplanet with $msini$\,=\,2.5$\pm$0.7\,$\ensuremath{\,{\rm M_J}}$, a period of 963$\pm$38\,days, and an eccentricity of 0.74$\pm$0.20\@. The secondary, HD\,41004\,B (M2), is separated by 0.5 arcsec ($\sim$21\,au) from HD\,41004\,A has a mass of 0.4\,$\rm \mst$, and host the BD companion with $msini$\,=\,18.37$\pm$0.22\,$\ensuremath{\,{\rm M_J}}$, a period of 1.328300$\pm$0.000012\,days, and eccentricity of 0.081$\pm$0.012 \citep{Zucker04}.

$\epsilon$ Indi is a nearby triple system with a K5V spectral type primary exoplanet-host star and wide BD binary at a separation of 402.3\,arcsec, whose components have a projected separation of 0.7\,arcsec \citep{Scholz03,Mccaughrean04}. The primary star has a mass of $M_\star$\,=\,0.73$\pm$0.09\,$\rm \mst$, a radius of $R_\star$\,=\,0.74$\pm$0.07\,$\rm \rst$, an effective temperature of $T_{\rm eff}$\,=\,4611$\pm$157\,K, and a metallicity of $-$0.13$\pm$0.03\,dex \citep{Stassun19}. \cite{Dieterich18} determined dynamical masses of the BD binary, which consists of T1.5 dwarf with a mass of 75.0$\pm$0.8\,$\ensuremath{\,{\rm M_J}}$ and a T6 secondary with a mass of 70.1$\pm$0.7\,$\ensuremath{\,{\rm M_J}}$. Both objects have masses very close to the transition between low-mass stars and BDs. \cite{Endl02} reported a linear trend in RVs of the primary star using the Coude Echelle Spectrometer (CES) at ESO La Silla. \cite{Feng19} used spectroscopic data from CES, HARPSpre, HARPSpost, and UVES together with astrometric data from Hipparcos and $Gaia$ to derive the parameters of the planet to $M_P$\,=\,3.25\,$\ensuremath{\,{\rm M_J}}$ with a period of 45.2 years and e\,=\,0.26\@.

\clearpage

\begin{figure*}[!ht]
\centering
\includegraphics[width=0.9\textwidth,height=0.65\textwidth]{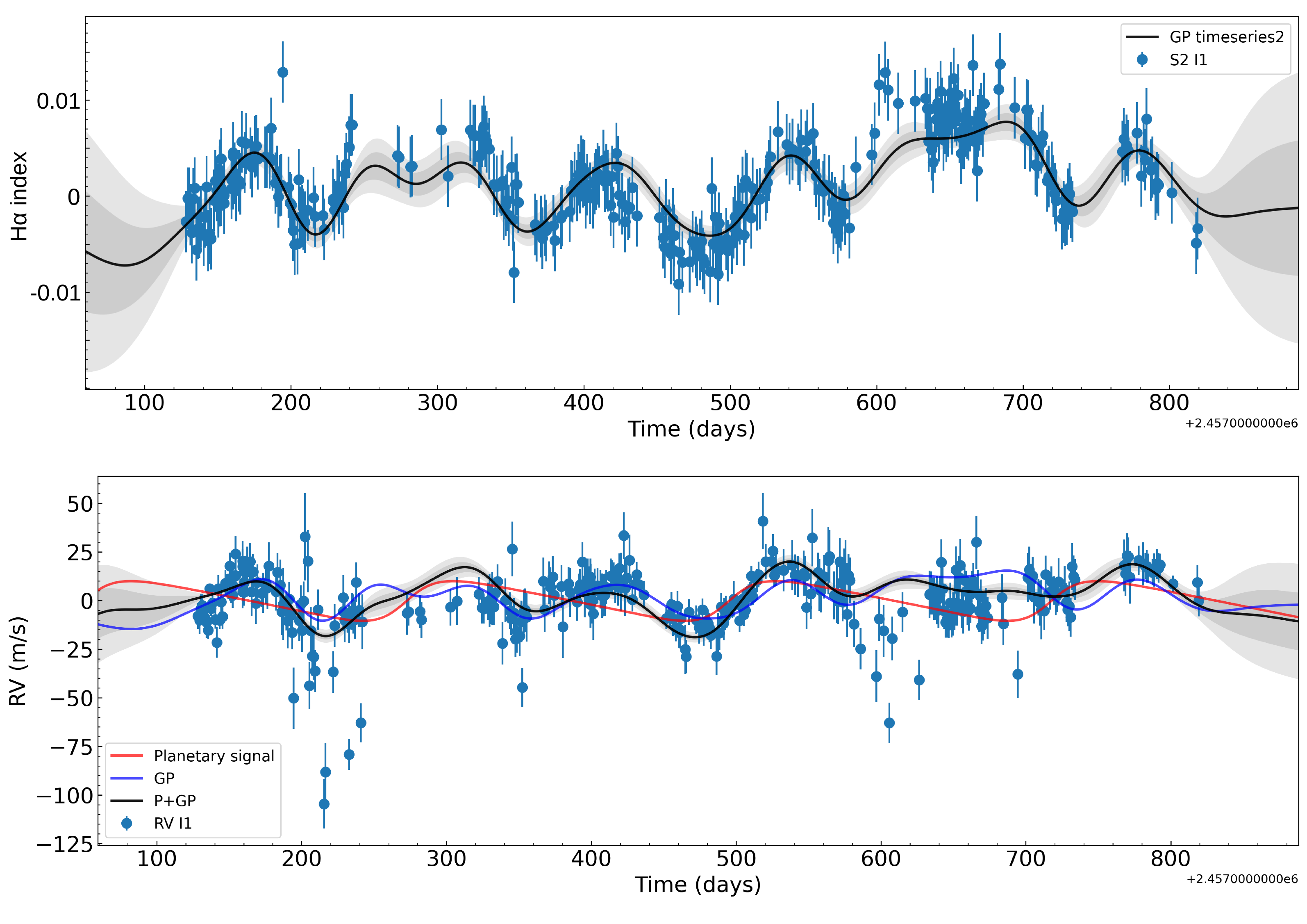}
\caption{H$\alpha$ and radial velocity time series of HD\,46588. Top plot shows the inferred stellar model (black). Bottom panel shows the inferred stellar model (blue), planetary model (red) and stellar+planetary model (black).} 
\label{fig:gp_hd46588}
\end{figure*}

\clearpage

\begin{figure*}[b]
\centering
\includegraphics[width=1.0\textwidth]{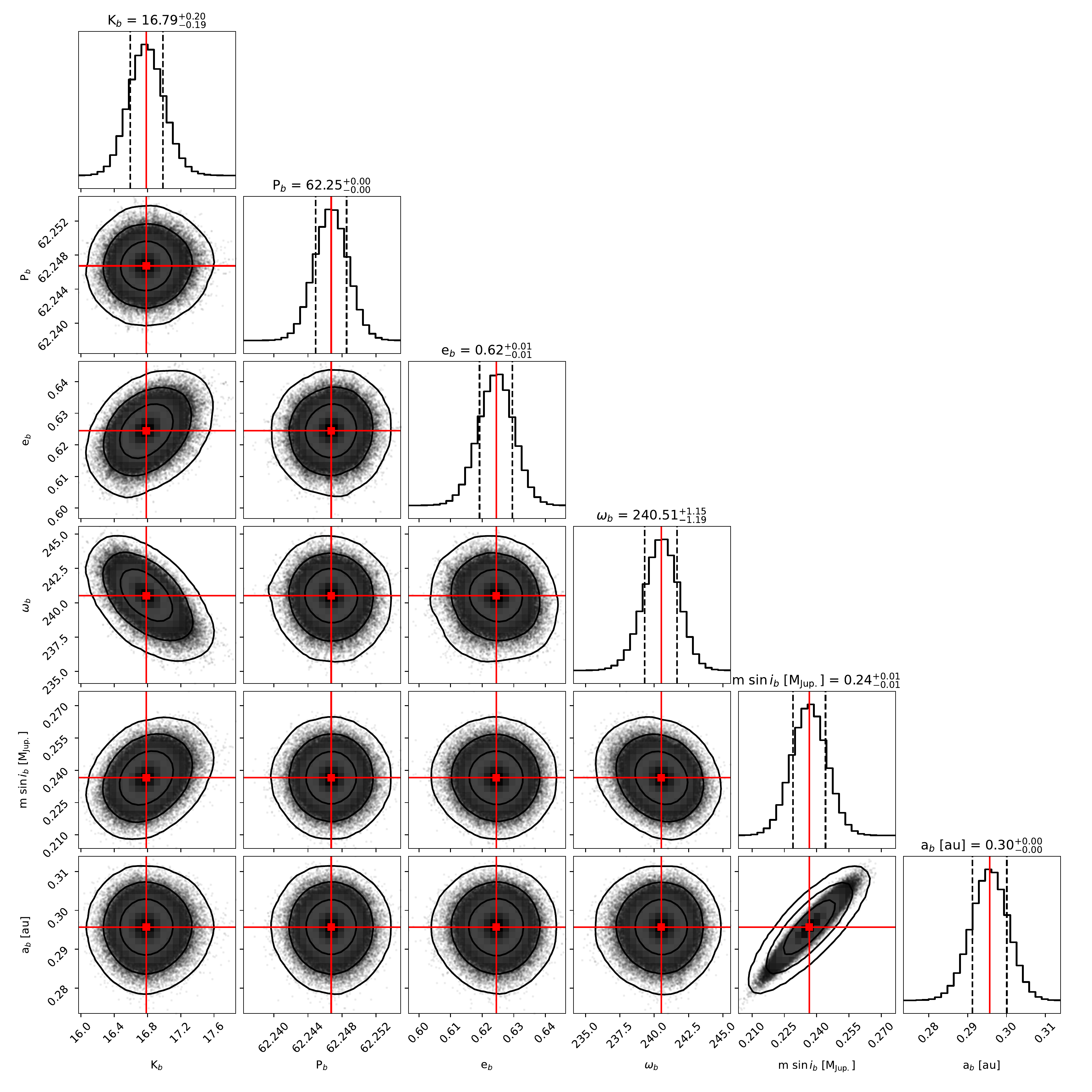}
\caption{Correlations between the free parameters of the RV model from the MCMC analysis using the {\tt Exo-Striker} package for the HD\,3651. At the end of each row is shown the derived posterior probability distribution.} \label{fig:HD3651-PPD}
\end{figure*}

\clearpage

\begin{figure*}[b]
\centering
\includegraphics[width=1.0\textwidth]{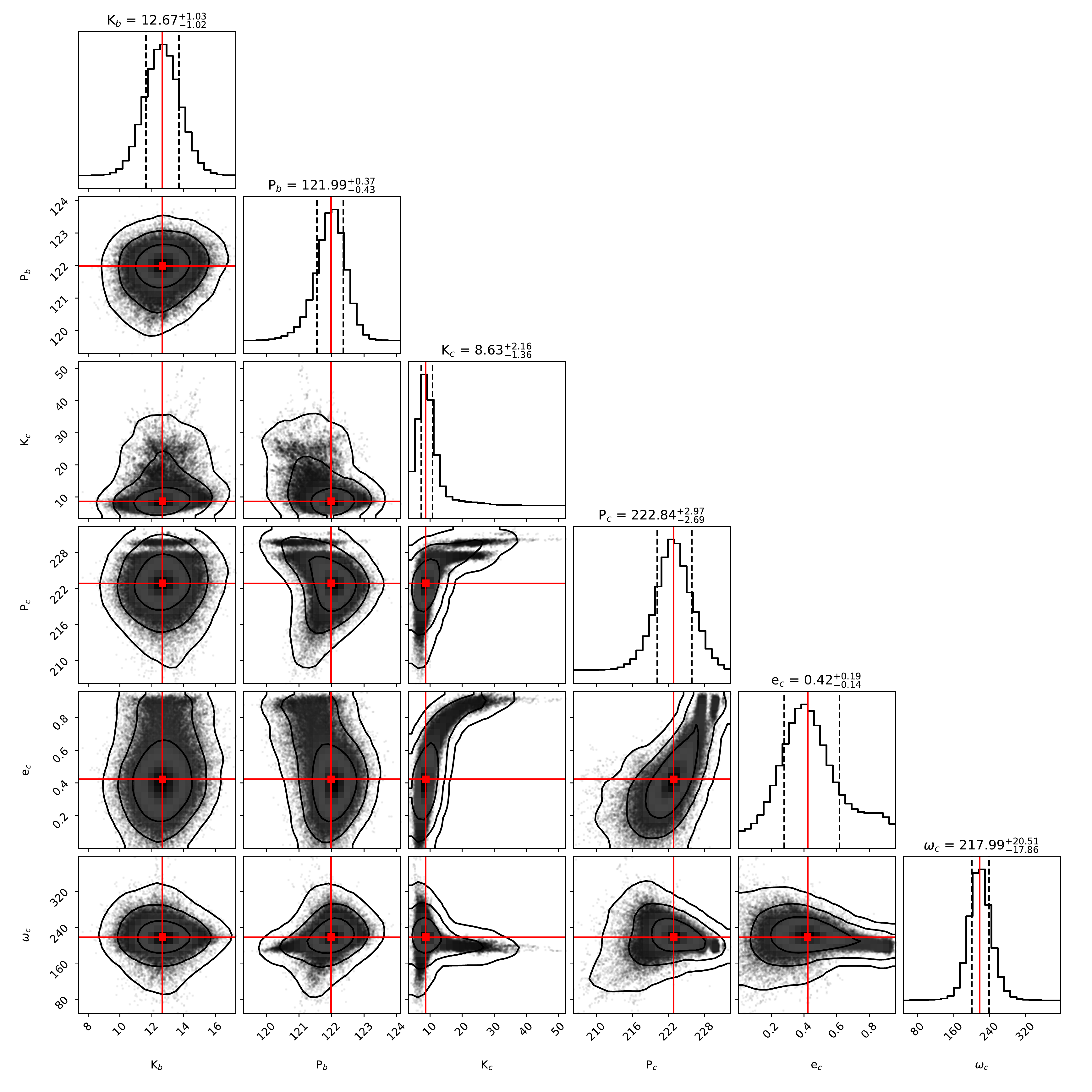}
\caption{Correlations between the free parameters of the RV model from the MCMC analysis using the {\tt Exo-Striker} package for the HD\,46588. At the end of each row is shown the derived posterior probability distribution. Index b represents a stellar activity signal, and index c is a planetary candidate.} \label{fig:HD46588-PPD}
\end{figure*}

\clearpage

\begin{figure*}[hb!]
\centering
\includegraphics[width=1.0\textwidth]{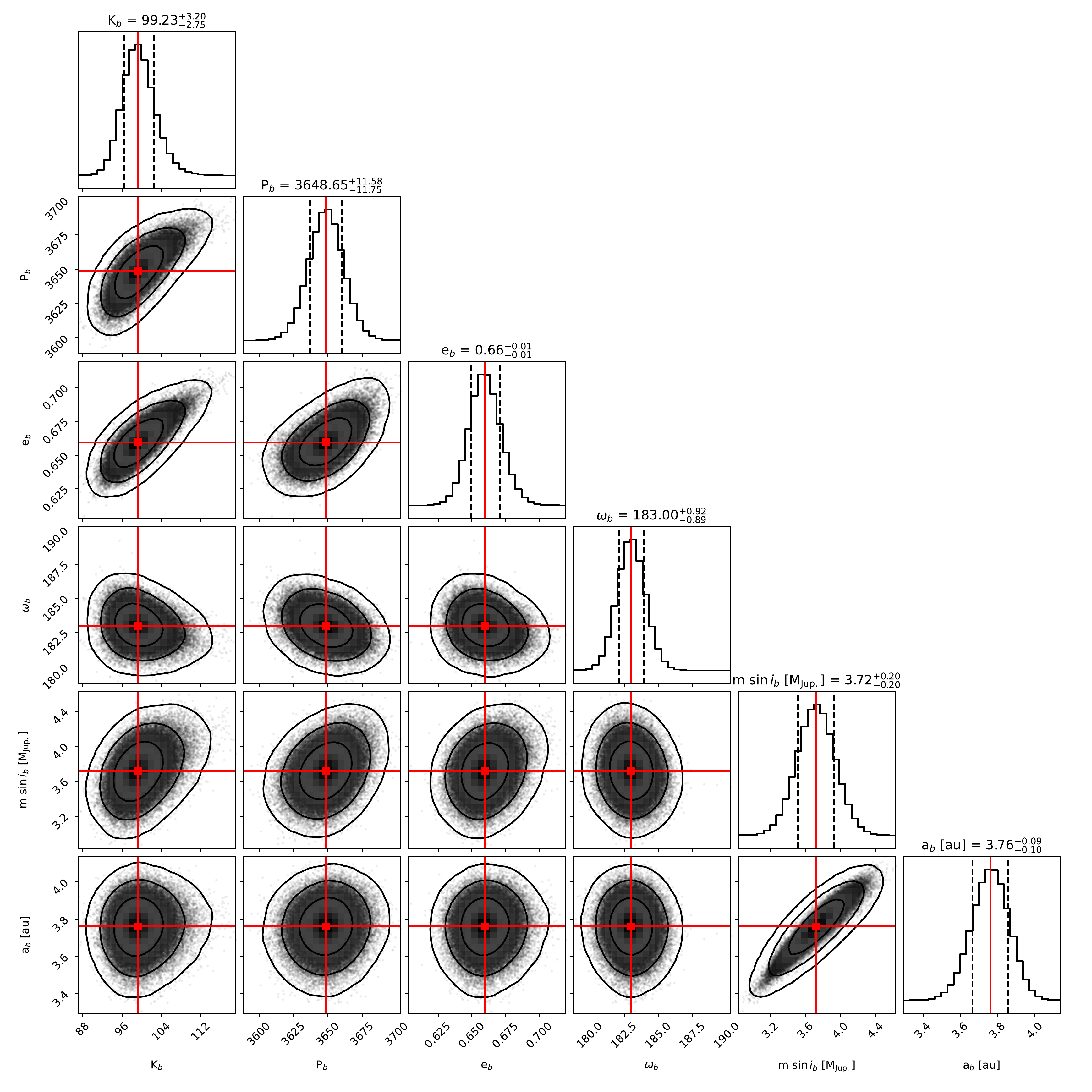}
\caption{Correlations between the free parameters of the RV model from the MCMC analysis using the {\tt Exo-Striker} package for the HIP\,70849. At the end of each row is shown the derived posterior probability distribution.} \label{fig:hip70849-PPD}
\end{figure*}

\end{document}